\documentclass{article}

\usepackage{pdfpages}
\usepackage{fullpage}
\usepackage{textcomp}

\usepackage{subcaption}
\usepackage{enumerate}
\usepackage{amsmath}
\usepackage{amsfonts}
\usepackage{graphicx}
\usepackage{amssymb}
\usepackage{multirow}

\newtheorem{lemma}{Lemma}
\newtheorem{theorem}[lemma]{Theorem}
\newtheorem{corollary}[lemma]{Corollary}

\usepackage{wrapfig}
\usepackage{bm}

\title{New bounds for range closest-pair problems}
\author{
  Jie Xue\footnote{University of Minnesota - Twin Cities, MN, USA; \texttt{\{xuexx193,janardan\}@umn.edu}.}
  \and
  Yuan Li\footnote{Facebook Inc., Seattle, WA, USA; \texttt{lydxlx@fb.com}.}
  \and
  Saladi Rahul\footnote{University of Illinois at Urbana-Champaign, IL, USA; \texttt{saladi.rahul@gmail.com}.}
  \and
  Ravi Janardan\footnotemark[1]
}
\date{}

\begin{document}

\maketitle

\begin{abstract}
Given a dataset $S$ of points in $\mathbb{R}^2$, the range closest-pair (RCP) problem aims to preprocess $S$ into a data structure such that when a query range $X$ is specified, the closest-pair in $S \cap X$ can be reported efficiently.
The RCP problem can be viewed as a range-search version of the classical closest-pair problem, and finds applications in many areas.
Due to its non-decomposability, the RCP problem is much more challenging than many traditional range-search problems.
This paper revisits the RCP problem, and proposes new data structures for various query types including quadrants, strips, rectangles, and halfplanes.
Both worst-case and average-case analyses (in the sense that the data points are drawn uniformly and independently from the unit square) are applied to these new data structures, which result in new bounds for the RCP problem.
Some of the new bounds significantly improve the previous results, while the others are entirely new.
\end{abstract}

\section{Introduction}
The closest-pair problem is one of the most fundamental problems in computational geometry
and finds many applications, e.g., collision detection, similarity search, traffic control, etc.
In this paper, we study a range-search version of the closest-pair problem called the \textit{range closest-pair} (RCP) problem.
Let $\mathcal{X}$ be a certain collection of ranges called \textit{query space}.
The RCP problem with query space $\mathcal{X}$ (or the $\mathcal{X}$-RCP problem for short) aims to preprocess a given dataset $S$ of points into a low-space data structure such that when a query range $X \in \mathcal{X}$ is specified, the closest-pair in $S \cap X$ can be reported efficiently.
The motivation for the RCP problem is clear and similar to that of range search: in many situations, one is interested in local information (i.e., local closest-pairs) inside specified ranges rather than global information (i.e., global closest-pair) of the dataset.

The RCP problem is quite challenging due to a couple of reasons.
First, in the RCP problem, the objects of interest are in fact point-pairs instead of single points, and in a dataset there is a quadratic number of point-pairs to be dealt with.
Moreover, the RCP problem is non-decomposable in the sense that even if the query range $X \in \mathcal{X}$ can be written as $X = X_1 \cup X_2$, the closest-pair in $S \cap X$ cannot be computed from the closest-pairs in $S \cap X_1$ and $S \cap X_2$.
The non-decomposability makes many traditional range-search techniques inapplicable to the RCP problem, and thus makes the problem much more challenging.

The RCP problem in $\mathbb{R}^2$ has been studied in prior work over the last fifteen years, e.g., \cite{abam2009power,gupta2006range,gupta2014data,shan2003spatial,sharathkumar2007range}.
In this paper, we revisit this problem and make significant improvements to the existing solutions.
Following the existing work, the query types considered in this paper are orthogonal queries (specifically, quadrants, strips, rectangles) and halfplane query.


\subsection{Our contributions, techniques, and related work} \label{sec-contribution}
The closest-pair problem and range search are both classical topics in computational geometry; see \cite{agarwal1999geometric:range_search,smid1995closest} for references.
The RCP problem is relatively new.
The best existing bounds in $\mathbb{R}^2$ and our new results are summarized in Table~\ref{tab-results} ($\mathsf{Space}$ refers to space cost and $\mathsf{Qtime}$ refers to query time), and we give a brief explanation below.
\smallskip

\begin{table}[h]
    \centering
    \begin{tabular}{|c|c|c|c|c|c|}
        \hline
        \multirow{2}{*}{Query} & \multirow{2}{*}{Source} & \multicolumn{2}{c|}{Worst-case} & \multicolumn{2}{c|}{Average-case} \\
        \cline{3-6}
        & & \textsf{Space} & \textsf{Qtime} & \textsf{Space} & \textsf{Qtime} \\
        \hline
        \multirow{2}{*}{Quadrant} & \cite{gupta2014data} & $O(n \log n)$ & $O(\log n)$ & - & - \\
        \cline{2-6}
        & \textbf{Theorem~\ref{thm-quadrant}} & \boldmath $O(n)$ & \boldmath $O(\log n)$ & \boldmath $O(\log^2 n)$ & \boldmath $O(\log \log n)$  \\
        \hline
        \multirow{2}{*}{Strip} & \cite{sharathkumar2007range} & $O(n \log^2 n)$ & $O(\log n)$ & - & - \\
        \cline{2-6}
        & \textbf{Theorem~\ref{thm-strip}} & \boldmath $O(n \log n)$ & \boldmath $O(\log n)$ & \boldmath $O(n)$ & \boldmath $O(\log n)$  \\
        \hline
        \multirow{4}{*}{Rectangle} & \cite{gupta2014data} & $O(n \log^5 n)$ & $O(\log^2 n)$ & - & - \\
        \cline{2-6}
        & \cite{sharathkumar2007range} & $O(n \log^3 n)$ & $O(\log^3 n)$ & - & - \\
        \cline{2-6}
        & \cite{gupta2014data} & - & - & $O(n \log^4 n)$ & $O(\log^4 n)$ \\
        \cline{2-6}
        & \textbf{Theorem~\ref{thm-rectangle}} & \boldmath $O(n \log^2 n)$ & \boldmath $O(\log^2 n)$ & \boldmath $O(n \log n)$ & \boldmath $O(\log n)$ \\
        \hline
        \multirow{2}{*}{Halfplane} & \cite{abam2009power} & $O(n \log n)$ & $O(n^{0.5+\varepsilon})$ & - & - \\
        \cline{2-6}
        & \textbf{Theorem~\ref{thm-halfplane}} & \boldmath $O(n)$ & \boldmath $O(\log n)$ & \boldmath $O(\log^2 n)$ & \boldmath $O(\log\log n)$ \\
        \hline
    \end{tabular}
    \caption{Summary of the best existing bounds and our new results for the RCP problem in $\mathbb{R}^2$ (each row corresponds to an RCP data structure for the corresponding query space).}
    \label{tab-results}
\end{table}

\noindent
$\bullet$ \textbf{Related work.}
The RCP problem for orthogonal queries was studied in \cite{gupta2014data,shan2003spatial,sharathkumar2007range}.
The best known solution for quadrant query was given by \cite{gupta2014data}, while \cite{sharathkumar2007range} gave the best known solution for strip query.
For rectangle query, there are two best known solutions (in terms of worst-case bounds) given by \cite{gupta2014data} and \cite{sharathkumar2007range} respectively.
The above results only considered worst-case performance of the data structures.
The authors of \cite{gupta2014data} for the first time applied average-case analysis to RCP data structures in the model where the data points are drawn independently and uniformly from the unit square.
Unfortunately, \cite{gupta2014data} only gave a rectangle RCP data structure with low average-case \textit{preprocessing} time, while its average-case space cost and query time are even higher than the worst-case counterparts of the data structure given by \cite{sharathkumar2007range} (even worse, its worst-case space cost is super-quadratic).
In fact, in terms of space cost and query time, no nontrivial average-case bounds were known for any kind of query before this paper.
The RCP problem for halfplane query was studied in \cite{abam2009power}.
Two data structures were proposed.
We only present the first one in Table~\ref{tab-results}.
The second one (not in the table), while having higher space cost and query time than the first one, can be built in $O(n \log^2 n)$ time.
Both data structures require (worst-case) super-linear space cost and polynomial query time.
\smallskip

\noindent
$\bullet$ \textbf{Our contributions.}
In this paper, we improve all the above results by giving new RCP data structures for various query types.
The improvements can be seen in Table~\ref{tab-results}.
In terms of worst-case bounds, the highlights are our rectangle RCP data structure which simultaneously improves the two best known results (given by \cite{gupta2014data} and \cite{sharathkumar2007range}) and our halfplane RCP data structure which is \textit{optimal} and significantly improves the bounds in \cite{abam2009power}.
Furthermore, by applying average-case analysis to our new data structures, we establish the first nontrivial average-case bounds for all the query types studied.
Our average-case analysis applies to datasets generated in not only the unit square but also an arbitrary axes-parallel rectangle.
These average-case bounds demonstrate that our new data structures might have much better performance in practice than one can expect from the worst-case bounds.
Finally, we also give an $O(n \log^2 n)$-time algorithm to build our halfplane RCP data structure, matching the preprocessing time in \cite{abam2009power}.
The preprocessing for our orthogonal RCP data structures is not considered in this paper; we are still in the process of investigating this.
\smallskip

\noindent
$\bullet$ \textbf{Our techniques.}
An important notion in our techniques is that of a \textit{candidate pair}, i.e., a pair of data points that is the answer to some RCP query.
Our solutions for the quadrant and strip RCP problems use the candidate pairs to construct a planar subdivision and take advantage of point-location techniques to answer queries.
The data structures themselves are simple, and our main technical contribution here occurs in the average-case analysis of the data structures.
The analysis requires a nontrivial study of the expected number of candidate pairs in a random dataset, which is of both geometric and combinatorial interest.
Our data structure for the rectangle RCP problem is subtle; it is constructed by properly combining two simpler data structures, each of which partially achieves the desired bounds.
The high-level framework of the two simpler data structures is identical: it first ``decomposes'' a rectangle query into four quadrant queries and then simplifies the problem via some geometric observations similar to those in the standard divide-and-conquer algorithm for the classical closest-pair problem.
Also, the analysis of the data structures is technically interesting.
Our solution for the halfplane RCP problem applies the duality technique to map the candidate pairs to wedges in the dual space and form a planar subdivision, which allows us to solve the problem by using point-location techniques on the subdivision, similarly to the approach for the quadrant and strip RCP problems.
However, unlike the quadrant and strip cases, to bound the complexity of the subdivision here is much more challenging, which requires non-obvious observations made by properly using the properties of duality and the problem itself.
The average-case bounds of the data structure follow from a technical result bounding the expected number of candidate pairs, which also involves a nontrivial proof.
\smallskip

\noindent
$\bullet$ \textbf{Organization.}
Section~\ref{sec-notation} presents the notations and preliminaries that are used throughout the paper.
We suggest that the reader reads this section carefully before moving on.
Our solutions for quadrant, strip, rectangle, and halfplane queries are presented in Section~\ref{sec-quadrant}, \ref{sec-strip}, \ref{sec-rectangle}, and \ref{sec-halfplane}, respectively.
In Section~\ref{sec-conclusion}, we conclude our results and give some open questions for future work.
To make the paper more readable, some technical proofs are deferred to Appendix~\ref{appx-prfs}.
Some implementation details of the preprocessing algorithms for our halfplane RCP data structure are deferred to Appendix~\ref{appx-implement}.


\subsection{Notations and Preliminaries} \label{sec-notation}
We introduce the notations and preliminaries that are used throughout the paper.
\smallskip

\noindent
$\bullet$ \textbf{Query spaces.}
The following notations denote various query spaces (i.e., collections of ranges in $\mathbb{R}^2$): $\mathcal{Q}$ quadrants, $\mathcal{P}$ strips, $\mathcal{U}$ 3-sided rectangles, $\mathcal{R}$ rectangles, $\mathcal{H}$ halfplanes (quadrants, strips, 3-sided rectangles, rectangles under consideration are all axes-parallel).
Define $\mathcal{Q}^\nearrow = \{[x,\infty) \times [y,\infty): x,y \in \mathbb{R}\} \subseteq \mathcal{Q}$ as the sub-collection of all northeast quadrants, and define $\mathcal{Q}^\nwarrow,\mathcal{Q}^\searrow,\mathcal{Q}^\swarrow$ similarly.
Define $\mathcal{P}^\text{v} = \{[x_1,x_2] \times \mathbb{R}: x_1,x_2 \in \mathbb{R}\} \subseteq \mathcal{P}$ as the sub-collection of all vertical strips, and similarly $\mathcal{P}^\text{h}$ horizontal strips.
If $l$ is a vertical (resp., horizontal) line, an $l$-\textit{anchored} strip is a vertical (resp., horizontal) strip containing $l$; define $\mathcal{P}_l \subseteq \mathcal{P}$ as the sub-collection of all $l$-anchored strips.
Define $\mathcal{U}^\downarrow = \{[x_1,x_2] \times (-\infty,y]: x_1,x_2,y \in \mathbb{R}\} \subseteq \mathcal{U}$ as the sub-collection of all bottom-unbounded rectangles, and define $\mathcal{U}^\uparrow,\mathcal{U}^\leftarrow,\mathcal{U}^\rightarrow$ similarly.
If $l$ is a non-vertical line, denote by $l^\uparrow$ (resp., $l^\downarrow$) the halfplane above (resp., below) $l$; define $\mathcal{H}^\uparrow \subseteq \mathcal{H}$ (resp., $\mathcal{H}^\downarrow \subseteq \mathcal{H}$) as the sub-collection of all such halfplanes.
\smallskip

\noindent
$\bullet$ \textbf{Candidate pairs.}
For a dataset $S$ and query space $\mathcal{X}$, a \textit{candidate pair} of $S$ with respect to $\mathcal{X}$ refers to a pair of points in $S$ which is the closest-pair in $S \cap X$ for some $X \in \mathcal{X}$.
We denote by $\varPhi(S,\mathcal{X})$ the set of the candidate pairs of $S$ with respect to $\mathcal{X}$.
If $l$ is a line, we define $\varPhi_l(S,\mathcal{X}) \subseteq \varPhi(S,\mathcal{X})$ as the subset consisting of the candidate pairs that cross $l$ (i.e., whose two points are on opposite sides of $l$).
\smallskip

\noindent
$\bullet$ \textbf{Data structures.}
For a data structure $\mathcal{D}$, we denote by $\mathcal{D}(S)$ the data structure instance of $\mathcal{D}$ built on the dataset $S$.
The notations $\mathsf{Space}(\mathcal{D}(S))$ and $\mathsf{Qtime}(\mathcal{D}(S))$ denote the space cost and query time (i.e., the maximum time for answering a query) of $\mathcal{D}(S)$, respectively.
\smallskip

\noindent
$\bullet$ \textbf{Random datasets.}
If $X$ is a region in $\mathbb{R}^2$ (or more generally in $\mathbb{R}^d$), we write $S \propto X^n$ to mean that $S$ is a dataset of $n$ random points drawn independently from the uniform distribution $\mathsf{Uni}(X)$ on $X$.
More generally, if $X_1,\dots,X_n$ are regions in $\mathbb{R}^2$ (or more generally in $\mathbb{R}^d$), we write $S \propto \prod_{i=1}^n X_i$ to mean that $S$ is a dataset of $n$ random points drawn independently from $\mathsf{Uni}(X_1),\dots,\mathsf{Uni}(X_n)$ respectively.
\smallskip

\noindent
$\bullet$ \textbf{Other notions.}
For a point $a \in \mathbb{R}^2$, we denote by $a.x$ and $a.y$ the $x$-coordinate and $y$-coordinate of $a$, respectively.
For two points $a,b \in \mathbb{R}^d$, we use $\text{dist}(a,b)$ to denote the Euclidean distance between $a$ and $b$, and use $[a,b]$ to denote the segments connecting $a$ and $b$ (in $\mathbb{R}^1$ this coincides with the notation for a closed interval).
We say $I_1,\dots,I_n$ are vertical (resp., horizontal) \textit{aligned} segments in $\mathbb{R}^2$ if there exist $r_1,\dots,r_n,\alpha,\beta \in \mathbb{R}$ such that $I_i = \{r_i\} \times [\alpha,\beta]$ (resp., $I_i = [\alpha,\beta] \times \{r_i\}$).
The \textit{length} of a pair $\phi = (a,b)$ of points is the length of the segment $[a,b]$.
For $S \subseteq \mathbb{R}^2$ of size at least 2, the notation $\kappa(S)$ denotes the \textit{closest-pair distance} of $S$, i.e., the length of the closest-pair in $S$.
\smallskip

\noindent
The following result regarding the closest-pair distance of a random dataset will be used to bound the expected number of candidate pairs with respect to various query spaces.
\begin{lemma} \label{lem-kappa}
Let $R$ be a rectangle of size $\Delta \times \Delta'$ where $\Delta \leq \Delta'$, and $A \propto R^m$.
Then
\begin{equation*}
    \mathbb{E}[\kappa^p(A)] = \Theta\left(\max \left\{(\Delta'/m^2)^p, (\sqrt{\Delta \Delta'}/m)^p \right\} \right) \text{ for any constant } p>1.
\end{equation*}
In particular, if $R$ is a segment of length $\ell$, then $\mathbb{E}[\kappa^p(A)] = \Theta((\ell/m^2)^p)$.
\end{lemma}

\section{Quadrant query} \label{sec-quadrant}
\begin{figure}[h]
	\begin{center}
		\includegraphics[height=3cm]{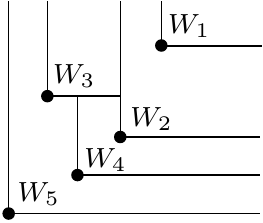}
	\end{center}
	\caption{The subdivision induced by successively overlaying the quadrants.}
	\label{figure:weighted quadrant}
\end{figure}

We consider the RCP problem for quadrant queries, i.e., the $\mathcal{Q}$-RCP problem.
In order to solve the $\mathcal{Q}$-RCP problem, it suffices to consider the $\mathcal{Q}^\swarrow$-RCP problem.
Let $S \subseteq \mathbb{R}^2$ be a dataset of size $n$.
Suppose $\varPhi(S,\mathcal{Q}^\swarrow) = \{\phi_1,\dots,\phi_m\}$ where $\phi_i = (a_i,b_i)$, and assume $\phi_1,\dots,\phi_m$ are sorted in increasing order of their lengths.
It was shown in \cite{gupta2014data} that $m = O(n)$.
We construct a mapping $\varPhi(S,\mathcal{Q}^\swarrow) \rightarrow \mathbb{R}^2$ as $\phi_i \mapsto w_i$ where $w_i = (\max\{a_i.x, b_i.x\},\max\{a_i.y, b_i.y\})$, and observe that for a query range $Q \in \mathcal{Q}^\swarrow$, $\phi_i$ is contained in $Q$ iff $w_i \in Q$.
Let $W_i$ be the northeast quadrant with vertex $w_i$.
Then we further have $w_i \in Q$ iff $q \in W_i$ where $q$ is the vertex of $Q$.
As such, the closest-pair in $S \cap Q$ to be reported is $\phi_\eta$ for $\eta = \min\{i: q \in W_i\}$.
We create a planar subdivision $\varGamma$, by successively overlaying $W_1,\dots,W_m$ (see Figure~\ref{figure:weighted quadrant}).  
Note that the complexity of $\varGamma$ is $O(m)$, since overlaying each quadrant creates at most two vertices of $\varGamma$.
By the above observation, the answer for $Q$ is $\phi_i$ iff $q$ is in the cell $W_i \backslash \bigcup_{j=1}^{i-1} W_j$.
Thus, we can use the optimal planar point-location data structures (e.g., \cite{edelsbrunner1986optimal,sarnak1986planar}) to solve the problem in $O(m)$ space with $O(\log m)$ query time.
Since $m = O(n)$, we obtain a $\mathcal{Q}$-RCP data structure using $O(n)$ space with $O(\log n)$ query time in worst-case.

Next, we analyze the average-case performance of the above data structure.
In fact, it suffices to bound the expected number of the candidate pairs.
Surprisingly, we have the following poly-logarithmic bound.
\begin{lemma} \label{lem-quadcand}
    For a random dataset $S \propto R^n$ where $R$ is an axes-parallel rectangle, $\mathbb{E}[|\varPhi(S,\mathcal{Q})|] = O(\log^2 n)$.
\end{lemma}
Using the above lemma, we can immediately conclude that our data structure uses $O(\log^2 n)$ space in average-case.
The average-case query time is in fact $O(\mathbb{E}[\log |\varPhi(S,\mathcal{Q})|])$.
Note that $\mathbb{E}[\log x] \leq \log \mathbb{E}[x]$ for a positive random variable $x$, thus $\mathbb{E}[\log |\varPhi(S,\mathcal{Q})|] = O(\log \log n)$.
\begin{theorem} \label{thm-quadrant}
There exists a $\mathcal{Q}$-RCP data structure $\mathcal{A}$ such that \\
$\bullet$ For any $S \subseteq \mathbb{R}^2$ of size $n$, $\mathsf{Space}(\mathcal{A}(S)) = O(n)$ and $\mathsf{Qtime}(\mathcal{A}(S)) = O(\log n)$. \\
$\bullet$ For a random $S \propto R^n$ where $R$ is the unit square or more generally an arbitrary axes-parallel rectangle, $\mathbb{E}[\mathsf{Space}(\mathcal{A}(S))] = O(\log^2 n)$ and $\mathbb{E}[\mathsf{Qtime}(\mathcal{A}(S))] = O(\log\log n)$.
\end{theorem}

\section{Strip query} \label{sec-strip}
We consider the RCP problem for strip queries, i.e., the $\mathcal{P}$-RCP problem.
In order to solve the $\mathcal{P}$-RCP problem, it suffices to consider the $\mathcal{P}^\text{v}$-RCP problem.
Let $S \subseteq \mathbb{R}^2$ be a dataset of size $n$.
Suppose $\varPhi(S,\mathcal{P}^\text{v}) = \{\phi_1,\dots,\phi_m\}$ where $\phi_i = (a_i,b_i)$, and assume $\phi_1,\dots,\phi_m$ are sorted in increasing order of their lengths.
It was shown in \cite{sharathkumar2007range} that $m = O(n \log n)$.
We construct a mapping $\varPhi(S,\mathcal{P}^\text{v}) \rightarrow \mathbb{R}^2$ as $\phi_i \mapsto w_i$ where $w_i = (\min\{a_i.x, b_i.x\},\max\{a_i.x, b_i.x\})$, and observe that for a query range $P = [x_1,x_2] \times \mathbb{R} \in \mathcal{P}^\text{v}$, $\phi_i$ is contained in $P$ iff $w_i$ is in the southeast quadrant $[x_1,\infty) \times (-\infty,x_2]$.
Let $W_i$ be the northwest quadrant with vertex $w_i$.
Then we further have $w_i \in [x_1,\infty) \times (-\infty,x_2]$ iff $p \in W_i$ where $p = (x_1,x_2)$.
As such, the closest-pair in $S \cap P$ is $\phi_\eta$ for $\eta = \min\{i: p \in W_i\}$.
Thus, as in Section~\ref{sec-quadrant}, we can successively overlay $W_1,\dots,W_m$ to create a planar subdivision, and use point-location to solve the problem in $O(m)$ space and $O(\log m)$ query time.
Since $m = O(n \log n)$ here, we obtain a $\mathcal{P}$-RCP data structure using $O(n \log n)$ space with $O(\log n)$ query time in worst-case.

Next, we analyze the average-case performance of our data structure.
Again, it suffices to bound the expected number of the candidate pairs.
For later use, we study here a more general case in which the candidate pairs are considered with respect to 3-sided rectangle queries.
\begin{lemma} \label{lem-probij}
Let $S \propto \prod_{i=1}^n I_i$ where $I_1,\dots,I_n$ are distinct vertical \textnormal{(}resp., horizontal\textnormal{)} aligned segments sorted from left to right \textnormal{(}resp., from bottom to top\textnormal{)}.
Suppose $a_i \in S$ is the point drawn on $I_i$.
Then for $i,j \in \{1,\dots,n\}$ with $i<j$ and $\mathcal{X} \in \{\mathcal{U}^\downarrow,\mathcal{U}^\uparrow\}$ \textnormal{(}resp., $\mathcal{X} \in \{\mathcal{U}^\leftarrow,\mathcal{U}^\rightarrow\}$\textnormal{)},
\begin{equation*}
    \Pr[(a_i,a_j) \in \varPhi(S,\mathcal{X})] = O\left(\frac{\log (j-i)}{(j-i)^2}\right).
\end{equation*}
\end{lemma}
From the above lemma, a direct calculation gives us the following corollary.
\begin{corollary} \label{cor-3sidedcand}
    For a random dataset $S \propto R^n$ where $R$ is an axes-parallel rectangle, $\mathbb{E}[|\varPhi(S,\mathcal{U})|] = \Theta(n)$ and $\mathbb{E}[|\varPhi(S,\mathcal{P})|] = \Theta(n)$.
\end{corollary}
\textit{Proof.}
Without loss of generality, assume $R = [0,1] \times [0,\Delta]$.
Since $\varPhi(S,\mathcal{P}) \subseteq \varPhi(S,\mathcal{U})$ for any $S$, it suffices to show $\mathbb{E}[|\varPhi(S,\mathcal{P})|] = \Omega(n)$ and $\mathbb{E}[|\varPhi(S,\mathcal{U})|] = O(n)$.
The former is clear, since every pair of $x$-adjacent or $y$-adjacent points in $S$ is a candidate pair with respect to $\mathcal{P}$.
The latter can be shown using Lemma~\ref{lem-probij} as follows.
We only need to bound $\mathbb{E}[|\varPhi(S,\mathcal{U}^\downarrow)|]$.
We first show that if $S \propto \prod_{i=1}^n I_i$ where $I_1,\dots,I_n$ are vertical aligned segments sorted from left to right, then $\mathbb{E}[|\varPhi(S,\mathcal{U}^\downarrow)|] = O(n)$.
In fact, this follows directly from Lemma~\ref{lem-probij}.
Let $a_i$ be the random point drawn on $I_i$.
Then
\begin{equation*}
    \mathbb{E}[|\varPhi(S,\mathcal{U}^\downarrow)|] = \sum_{i=1}^{n-1} \sum_{j=i+1}^n \Pr[(a_i,a_j) \in \varPhi(S,\mathcal{U}^\downarrow)].
\end{equation*}
We plug in the bound $\Pr[(a_i,a_j) \in \varPhi(S,\mathcal{U}^\downarrow)] = O(\log(j-i)/(j-i)^2)$ shown in Lemma~\ref{lem-probij} to the above equation.
Noting the fact that $\sum_{t=1}^\infty \log t/t^2 = O(1)$, a direct calculation then gives us $\mathbb{E}[|\varPhi(S,\mathcal{U}^\downarrow)|] = O(n)$.
Now assume $S \propto R^n$.
Define a random multi-set $X = \{a.x: a \in S\}$, which consists of the $x$-coordinates of the $n$ random points in $S$.
We shall show that for all $x_1,\dots,x_n \in [0,1]$ such that $x_1<\cdots<x_n$,
\begin{equation} \label{eq-expect}
    \mathbb{E}\left[\left.|\varPhi(S,\mathcal{U}^\downarrow)|\right|X = \{x_1,\dots,x_n\}\right] = O(n),
\end{equation}
which implies that $\mathbb{E}[|\varPhi(S,\mathcal{U}^\downarrow)|] = O(n)$, because the random points in $S$ have distinct $x$-coordinates with probability 1.
Let $I_i = x_i \times [0,\Delta]$ for $i \in \{1,\dots,n\}$, then $I_1,\dots,I_n$ are vertical aligned segments sorted from left to right.
Note that, under the condition $X = \{x_1,\dots,x_n\}$, the $n$ random points in $S$ can be viewed as independently drawn from the uniform distributions on $I_1,\dots,I_n$, respectively.
Thus, Equation~\ref{eq-expect} follows directly from our previous argument for the case $S \propto \prod_{i=1}^n I_i$.
As a result, $\mathbb{E}[|\varPhi(S,\mathcal{U}^\downarrow)|] = O(n)$.
\hfill $\Box$
\medskip

\noindent
Using the above argument and our previous data structure, we conclude the following.
\begin{theorem} \label{thm-strip}
There exists a $\mathcal{P}$-RCP data structure $\mathcal{B}$ such that \\
$\bullet$ For any $S \subseteq \mathbb{R}^2$ of size $n$, $\mathsf{Space}(\mathcal{B}(S)) = O(n \log n)$ and $\mathsf{Qtime}(\mathcal{B}(S)) = O(\log n)$. \\
$\bullet$ For a random $S \propto R^n$ where $R$ is the unit square or more generally an arbitrary axes-parallel rectangle, $\mathbb{E}[\mathsf{Space}(\mathcal{B}(S))] = O(n)$ and $\mathbb{E}[\mathsf{Qtime}(\mathcal{B}(S))] = O(\log n)$.
\end{theorem}

\section{Rectangle query} \label{sec-rectangle}
We consider the RCP problem for rectangle queries, i.e., the $\mathcal{R}$-RCP problem.
Interestingly, our final solution for the $\mathcal{R}$-RCP problem is a combination of two simpler solutions, each of which partially achieves the desired bounds.

\begin{figure}[h]
    \centering
	\begin{subfigure}[b]{6.5cm}
	    \includegraphics[width=5.5cm]{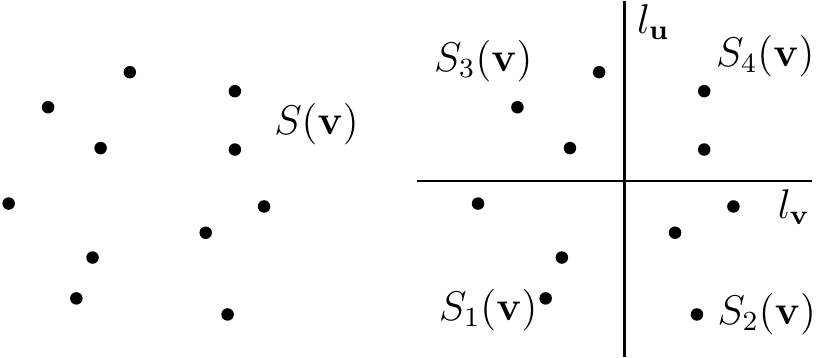}
	    \caption{Illustrating the subsets $S_1(\mathbf{v}),\dots,S_4(\mathbf{v})$.}
		\label{figure:splitting1}
	\end{subfigure}
	\hspace{0.5cm}
	\begin{subfigure}[b]{6.5cm}
	    \includegraphics[width=5.5cm]{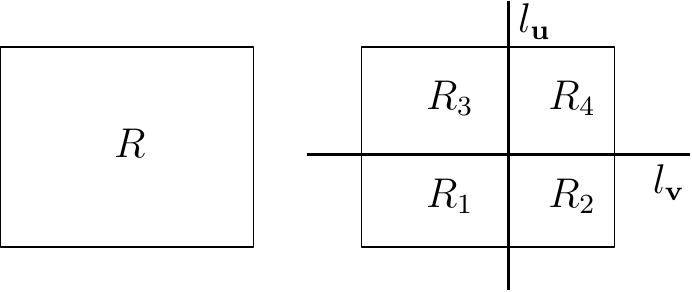}
	    \caption{Illustrating the rectangles $R_1,\dots,R_4$.}
		\label{figure:splitting2}
	\end{subfigure}
    \caption{Illustrating $S_1(\mathbf{v}), \dots, S_4(\mathbf{v})$ and $R_1, \dots, R_4$.}
\end{figure}

We first describe the common part of our two solutions.
Let $S \subseteq \mathbb{R}^2$ be a dataset of size $n$.
The common component of our two data structures is a standard 2D range tree built on $S$ \cite{de2000computational}.
The main tree (or primary tree) $\mathcal{T}$ is a range tree built on the $x$-coordinates of the points in $S$.
Each node $\mathbf{u} \in \mathcal{T}$ corresponds to a subset $S(\mathbf{u})$ of $x$-consecutive points in $S$, called the \textit{canonical subset} of $\mathbf{u}$.
At $\mathbf{u}$, there is an associated secondary tree $\mathcal{T}_\mathbf{u}$, which is a range tree built on the $y$-coordinates of the points in $S(\mathbf{u})$.
With an abuse of notation, for each node $\mathbf{v} \in \mathcal{T}_\mathbf{u}$, we still use $S(\mathbf{v})$ to denote the canonical subset of $\mathbf{v}$, which is a subset of $y$-consecutive points in $S(\mathbf{u})$.
As in \cite{gupta2014data}, for each (non-leaf) primary node $\mathbf{u} \in \mathcal{T}$, we fix a vertical line $l_\mathbf{u}$ such that the points in the canonical subset of the left (resp., right) child of $\mathbf{u}$ are to the left (resp., right) of $l_\mathbf{u}$.
Similarly, for each (non-leaf) secondary node $\mathbf{v}$, we fix a horizontal line $l_\mathbf{v}$ such that the points in the canonical subset of the left (resp., right) child of $\mathbf{v}$ are above (resp., below) $l_\mathbf{v}$.
Let $\mathbf{v} \in \mathcal{T}_\mathbf{u}$ be a secondary node.
Then at $\mathbf{v}$ we have two lines $l_\mathbf{v}$ and $l_\mathbf{u}$, which partition $\mathbb{R}^2$ into four quadrants.
We denote by $S_1(\mathbf{v}),\dots,S_4(\mathbf{v})$ the subsets of $S(\mathbf{v})$ contained in these quadrants; see Figure~\ref{figure:splitting1} for the correspondence.
In order to solve the problem, we need to store some additional data structures at the nodes of the tree (called \textit{sub-structures}).
At each secondary node $\mathbf{v}$, we store four $\mathcal{Q}$-RCP data structures $\mathcal{A}(S_1(\mathbf{v})),\dots,\mathcal{A}(S_4(\mathbf{v}))$ (Theorem~\ref{thm-quadrant}).

Now let us explain what we can do by using this 2D range tree (with the sub-structures).
Let $R = [x_1,x_2] \times [y_1,y_2] \in \mathcal{R}$ be a query rectangle.
We first find in $\mathcal{T}$ the \textit{splitting} node $\mathbf{u} \in \mathcal{T}$ corresponding to the range $[x_1,x_2]$, which is by definition the LCA of all the leaves whose corresponding points are in $[x_1,x_2] \times \mathbb{R}$.
Then we find in $\mathcal{T}_\mathbf{u}$ the splitting node $\mathbf{v} \in \mathcal{T}_\mathbf{u}$ corresponding to the range $[y_1,y_2]$.
If either of the splitting nodes does not exist or is a leaf node, then $|S \cap R| \leq 1$ and nothing should be reported.
So assume $\mathbf{u}$ and $\mathbf{v}$ are non-leaf nodes.
By the property of splitting node, we have $S \cap R = S(\mathbf{v}) \cap R$, and the lines $l_\mathbf{u}$ and $l_\mathbf{v}$ both intersect $R$.
Thus, $l_\mathbf{u}$ and $l_\mathbf{v}$ decompose $R$ into four smaller rectangles $R_1,\dots,R_4$; see Figure~\ref{figure:splitting2} for the correspondence.
By construction, we have $S(\mathbf{v}) \cap R_i = S_i(\mathbf{v}) \cap R_i$.
In order to find the closest-pair in $S \cap R$, we first try to compute the closest-pair in $S \cap R_i$ for all $i \in \{1,\dots,4\}$.
This can be done by querying the sub-structures stored at $\mathbf{v}$.
Indeed, $S \cap R_i = S(\mathbf{v}) \cap R_i = S_i(\mathbf{v}) \cap R_i = S_i(\mathbf{v}) \cap Q_i$, where $Q_i$ is the quadrant obtained by removing the two sides of $R_i$ that coincide with $l_\mathbf{u}$ and $l_\mathbf{v}$.
Therefore, we can query $\mathcal{A}(S_i(\mathbf{v}))$ with  $Q_i$ to find the closest-pair in $S \cap R_i$.
Once the four closest-pairs are computed, we take the shortest one (i.e., the one of the smallest length) among them and denote it by $\phi$.

\begin{figure}[h]
    \centering
	\includegraphics[width=6.5cm]{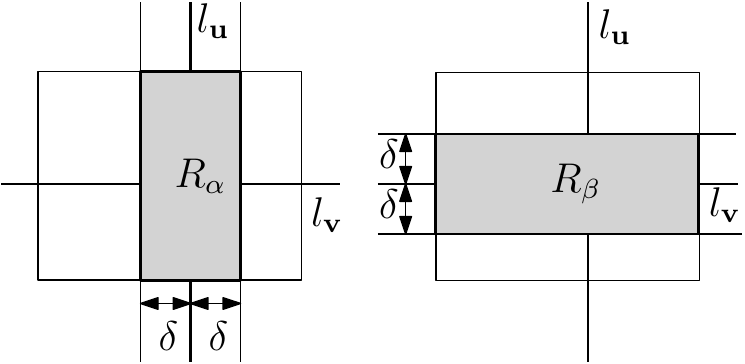}
	\caption{Illustrating the rectangles $R_\alpha$ and $R_\beta$.}
	\label{figure:2strips}
\end{figure}

Clearly, $\phi$ is not necessarily the closest-pair in $S \cap R$ as the two points in the closest-pair may belong to different $R_i$'s.
However, as we will see, with $\phi$ in hand, finding the closest-pair in $S \cap R$ becomes easier.
Suppose $l_\mathbf{u}: x = \alpha$ and $l_\mathbf{v}: y = \beta$, where $x_1 \leq \alpha \leq x_2$ and $y_1 \leq \beta \leq y_2$.
Let $\delta$ be the length of $\phi$.
We define $P_\alpha = [\alpha-\delta,\alpha+\delta] \times \mathbb{R}$ (resp., $P_\beta = \mathbb{R} \times [\beta-\delta,\beta+\delta]$) and $R_\alpha = R \cap P_\alpha$ (resp., $R_\beta = R \cap P_\beta$); see Figure~\ref{figure:2strips}.
We have the following key observation.
\begin{lemma} \label{lem-1of3}
	The closest-pair in $S \cap R$ is the shortest one among $\{\phi,\phi_\alpha,\phi_\beta\}$, where $\phi_\alpha$ \textnormal{(}resp., $\phi_\beta$\textnormal{)} is the closest-pair in $S \cap R_\alpha$ \textnormal{(}resp., $S \cap R_\beta$\textnormal{)}.
\end{lemma}
\textit{Proof.}
Let $\phi^* = (a^*,b^*)$ be the closest-pair in $S \cap R$.
Since $\phi,\phi_\alpha,\phi_\beta$ are all point-pairs in $S \cap R$, it suffices to show that $\phi^* \in \{\phi,\phi_\alpha,\phi_\beta\}$.
If $\phi^* = \phi$, we are done.
So assume $\phi^* \neq \phi$.
Then $a^*$ and $b^*$ must be contained in different $R_i$'s.
It follows that the segment $[a^*,b^*]$ intersects either $l_\mathbf{u}$ or $l_\mathbf{v}$.
Note that the length of $\phi^*$ is at most $\delta$ (recall that $\delta$ is the length of $\phi$), which implies $|a^*.x - b^*.x| \leq \delta$ and $|a^*.y - b^*.y| \leq \delta$.
If $[a^*,b^*]$ intersects $l_\mathbf{u}$, then $a^*,b^* \in P_\alpha$ (because $|a^*.x - b^*.x| < \delta$).
Thus, $a^*,b^* \in R_\alpha$ and $\phi^* = \phi_\alpha$.
Similarly, if $[a^*,b^*]$ intersects $l_\mathbf{v}$, we have $\phi^* = \phi_\beta$.
As a result, $\phi^* \in \{\phi,\phi_\alpha,\phi_\beta\}$.
\hfill $\Box$
\medskip

\noindent
Due to the above lemma, it now suffices to compute $\phi_\alpha$ and $\phi_\beta$.
Note that $R_\alpha$ and $R_\beta$ are rectangles, so computing $\phi_\alpha$ and $\phi_\beta$ still requires rectangle RCP queries.
Fortunately, there are some additional properties which make it easy to search for the closest-pairs in $S \cap R_\alpha$ and $S \cap R_\beta$.
For a set $A$ of points in $\mathbb{R}^2$ and $a,b \in A$, we define the $x$-\textit{gap} (resp., $y$-\textit{gap}) between $a$ and $b$ in $A$ as the number of the points in $A \backslash \{a,b\}$ whose $x$-coordinates (resp., $y$-coordinates) are in between $a.x$ and $b.x$ (resp., $a.y$ and $b.y$).
\begin{lemma} \label{lem-adjacent}
    There exists a constant integer $k$ such that the $y$-gap \textnormal{(}resp., $x$-gap\textnormal{)} between the two points of $\phi_\alpha$ \textnormal{(}resp., $\phi_\beta$\textnormal{)} in $S \cap R_\alpha$ \textnormal{(}resp., $S \cap R_\beta$\textnormal{)} is at most $k$.
\end{lemma}
\textit{Proof.}
We only need to consider $\phi_\alpha$.
Let $k=100$.
Suppose $\phi_\alpha = (a,b)$.
We denote by $w$ the left-right width of $R_\alpha$, i.e., the distance between the left and right boundaries of $R_\alpha$.
By the construction of $R_\alpha$, we have $w \leq 2 \delta$.
We consider two cases: $|a.y-b.y| \geq 2w$ and $|a.y-b.y| < 2w$.
Suppose $|a.y-b.y| \geq 2w$.
Assume there are more than $k$ points in $(S \cap R_\alpha) \backslash \{a,b\}$ whose $y$-coordinates are in between $a.y$ and $b.y$.
Then we can find, among these points, two points $a'$ and $b'$ such that $|a'.y-b'.y| \leq |a.y-b.y|/k$.
Since $a',b' \in R_\alpha$, we have $|a'.x-b'.x| \leq w \leq |a.y-b.y|/2$.
It follows that
\begin{equation*}
    \text{dist}(a',b') \leq |a'.x-b'.x| + |a'.y-b'.y| < |a.y-b.y| \leq \text{dist}(a,b),
\end{equation*}
which contradicts the fact that $\phi_\alpha$ is the closest-pair in $S \cap R_\alpha$.
Next, suppose $|a.y-b.y| < 2w$.
Then $|a.y-b.y| < 4 \delta$.
Consider the rectangle $R^* = R_\alpha \cap (\mathbb{R} \times [a.y,b.y])$.
Note that $(S \cap R^*) \backslash \{a,b\}$ consists of exactly the points in $(S \cap R_\alpha) \backslash \{a,b\}$ whose $y$-coordinates are in between $a.y$ and $b.y$.
Therefore, it suffices to show $|S \cap R^*| \leq k$.
Let $R_i^* = R^* \cap R_i$ for $i \in \{1,\dots,4\}$.
Since $R_i^* \subseteq R_i$, the pairwise distances of the points in $S \cap R_i^*$ are at least $\delta$.
Furthermore, the left-right width of each $R_i^*$ is at most $\delta$ and the top-bottom width of each $R_i^*$ is at most $|a.y-b.y|$ (which is smaller than $4 \delta$).
Therefore, a simple argument using Pigeonhole principle shows that $|S \cap R_i^*| \leq 16 < k/4$.
As such, $|S \cap R^*| \leq k$.
\hfill $\Box$
\medskip

\noindent
We shall properly use the above lemma to help compute $\phi_\alpha$ and $\phi_\beta$.
At this point, our two solutions diverge.

\subsection{Preliminary: Extreme point data structures}
Before presenting our solutions, we introduce the so-called \textit{top/bottom extreme point} (TBEP) and \textit{left/right extreme point} (LREP) data structures.
For a query space $\mathcal{X}$ and a constant integer $k$, an $(\mathcal{X},k)$-TBEP (resp. $(\mathcal{X},k)$-LREP) data structure stores a given set $S$ of points in $\mathbb{R}^2$ and can report the $k$ topmost/bottommost (resp., leftmost/rightmost) points in $S \cap X$ for a query range $X \in \mathcal{X}$.
\begin{lemma} \label{lem-epds}
Let $k$ be a constant integer.
There exists a $(\mathcal{P}^\textnormal{v},k)$-TBEP data structure $\mathcal{K}^\textnormal{v}$ such that for any $S \subseteq \mathbb{R}^2$ of size $n$, $\mathsf{Space}(\mathcal{K}^\textnormal{v}(S)) = O(n)$ and $\mathsf{Qtime}(\mathcal{K}^\textnormal{v}(S)) = O(\log n)$.
Symmetrically, there also exists a $(\mathcal{P}^\textnormal{h},k)$-LREP data structure $\mathcal{K}^\textnormal{h}$ satisfying the same bounds.
\end{lemma}
\textit{Proof.}
Let $S \subseteq \mathbb{R}^2$ be a dataset of size $n$.
The $(\mathcal{P}^\text{v},k)$-TBEP data structure instance $\mathcal{K}^\text{v}(S)$ is a standard 1D range tree $\mathcal{T}$ built on the $x$-coordinates of the points in $S$.
By the construction of a range tree, each node $\mathbf{u} \in \mathcal{T}$ corresponds to a subset $S(\mathbf{u})$ of $x$-consecutive points in $S$, called the \textit{canonical subset} of $\mathbf{u}$.
The leaves of $\mathcal{T}$ one-to-one correspond to the points in $S$.
At each node $\mathbf{u} \in \mathcal{T}$, we store the $k$ topmost and $k$ bottommost points in $S(\mathbf{u})$; we denote the set of these $2k$ points by $K(\mathbf{u})$.
The overall space cost of the range tree (with the stored points) is clearly $O(n)$, as $k$ is a constant.
To answer a query $P = [x_1,x_2] \times \mathbb{R} \in \mathcal{P}^\text{v}$, we first find the $t = O(\log n)$ canonical nodes $\mathbf{u}_1,\dots,\mathbf{u}_t \in \mathcal{T}$ corresponding to the range $[x_1,x_2]$.
This is a standard range-tree operation, which can be done in $O(\log n)$ time.
We compute $K = \bigcup_{i=1}^t K(\mathbf{u}_i)$ in $O(\log n)$ time.
We then use selection to find the $k$ topmost and $k$ bottommost points in $K$; this can be done in $O(\log n)$ time since $|K| = 2kt = O(\log n)$.
These $2k$ points are just the $k$ topmost and $k$ bottommost points in $S \cap P$.
The $(\mathcal{P}^\text{h},k)$-LREP data structure $\mathcal{K}^\text{h}$ is constructed in a symmetric way.
\hfill $\Box$
\begin{lemma} \label{lem-epdsrand}
Let $l$ be a vertical \textnormal{(}resp., horizontal\textnormal{)} line and $k$ be a constant integer.
There exists a $(\mathcal{P}_l,k)$-TBEP \textnormal{(}resp., $(\mathcal{P}_l,k)$-LREP\textnormal{)} data structure $\mathcal{K}_l$ such that for $S \propto \prod_{i=1}^n I_i$ where $I_1,\dots,I_n$ are distinct vertical \textnormal{(}resp., horizontal\textnormal{)} aligned segments, $\mathbb{E}[\mathsf{Space}(\mathcal{K}_l(S))] = O(\log n)$ and $\mathbb{E}[\mathsf{Qtime}(\mathcal{K}_l(S))] = O(\log\log n)$.
\end{lemma}

\subsection{First solution} \label{sec-rect1}
We now introduce our first solution, which achieves the desired worst-case bounds.
Let $k$ be the constant integer in Lemma~\ref{lem-adjacent}.
In our first solution, besides the 2D range tree presented before, we build additionally two 1D range trees $\mathcal{T}'$ and $\mathcal{T}''$ on $S$, where $\mathcal{T}'$ (resp., $\mathcal{T}''$) is built on $y$-coordinates (resp., $x$-coordinates).
For $\mathbf{u}' \in \mathcal{T}'$ (resp., $\mathbf{u}'' \in \mathcal{T}''$), we still use $S(\mathbf{u}')$ (resp., $S(\mathbf{u}'')$) to denote the canonical subset of $\mathbf{u}'$ (resp., $\mathbf{u}'' \in \mathcal{T}''$).
At each node $\mathbf{u}' \in \mathcal{T}'$, we store a $\mathcal{P}$-RCP data structure $\mathcal{B}(S(\mathbf{u}'))$ (Theorem~\ref{thm-strip}) and a $(\mathcal{P}^\text{v},k)$-TBEP data structure $\mathcal{K}^\text{v}(S(\mathbf{u}'))$ (Lemma~\ref{lem-epds}).
Similarly, at each node $\mathbf{u}'' \in \mathcal{T}''$, we store a $\mathcal{P}$-RCP data structure $\mathcal{B}(S(\mathbf{u}''))$ (Theorem~\ref{thm-strip}) and a $(\mathcal{P}^\text{h},k)$-LREP data structure $\mathcal{K}^\text{h}(S(\mathbf{u}''))$ (Lemma~\ref{lem-epds}).

We now explain how to compute $\phi_\alpha$ and $\phi_\beta$.
Suppose $R_\alpha = [x_\alpha,x_\alpha'] \times [y_\alpha,y_\alpha']$.
Let $P_x = [x_\alpha,x_\alpha'] \times \mathbb{R}$ and $P_y = \mathbb{R} \times [y_\alpha,y_\alpha']$.
To compute $\phi_\alpha$, we first find in $\mathcal{T}'$ the $t = O(\log n)$ canonical nodes $\mathbf{u}_1',\dots,\mathbf{u}_t' \in \mathcal{T}'$ corresponding to the range $[y_\alpha,y_\alpha']$.
Then $\bigcup_{i=1}^t S(\mathbf{u}_i') = S \cap P_y$, and each $S(\mathbf{u}_i')$ is a set of $y$-consecutive points in $S \cap P_y$.
Furthermore, $S \cap R_\alpha = \bigcup_{i=1}^t S(\mathbf{u}_i') \cap P_x$.
We query the sub-structures $\mathcal{B}(S(\mathbf{u}_1')),\dots,\mathcal{B}(S(\mathbf{u}_t'))$ with $P_x$ to find the closest-pairs $\phi_1,\dots,\phi_t$ in $S(\mathbf{v}_1) \cap P_x,\dots,S(\mathbf{v}_t) \cap P_x$, respectively.
We also query $\mathcal{K}^\text{v}(S(\mathbf{u}_1')),\dots,\mathcal{K}^\text{v}(S(\mathbf{u}_t'))$ with $P_x$ to obtain the $k$ topmost and bottommost points in $S(\mathbf{u}_1') \cap P,\dots,S(\mathbf{u}_t') \cap P$, respectively; we denote by $K$ the set of the $2kt$ reported points.
Then we find the closest-pair $\phi_K$ in $K$ using the standard divide-and-conquer algorithm.
We claim that $\phi_\alpha$ is the shortest one among $\{\phi_1,\dots,\phi_t,\phi_K\}$.
Suppose $\phi_\alpha = (a,b)$.
If the two points of $\phi_\alpha$ are both contained in some $S(\mathbf{u}_i')$, then clearly $\phi_\alpha = \phi_i$.
Otherwise, by Lemma~\ref{lem-adjacent} and the choice of $k$, the two points of $\phi_\alpha$ must belong to $K$ and hence $\phi_\alpha = \phi_K$.
It follows that $\phi_\alpha \in \{\phi_1,\dots,\phi_t,\phi_K\}$.
Furthermore, because the pairs $\phi_1,\dots,\phi_t,\phi_K$ are all contained in $R_\alpha$, $\phi_\alpha$ must be the shortest one among $\{\phi_1,\dots,\phi_t,\phi_K\}$.
Therefore, with $\phi_1,\dots,\phi_t,\phi_K$ in hand, $\phi_\alpha$ can be easily computed.
The pair $\phi_\beta$ is computed symmetrically using $\mathcal{T}''$.
Finally, taking the shortest one among $\{\phi,\phi_\alpha,\phi_\beta\}$, the query $R$ can be answered.

The 2D range tree together with the two 1D range trees $\mathcal{T}'$ and $\mathcal{T}''$ forms an $\mathcal{R}$-RCP data structure, which is our first solution.
A straightforward analysis gives us the worst-case space cost and query time of this data structure.
\begin{theorem} \label{thm-rectwst}
There exists an $\mathcal{R}$-RCP data structure $\mathcal{D}_1$ such that for any $S \subseteq \mathbb{R}^2$ of size $n$, $\mathsf{Space}(\mathcal{D}_1(S)) = O(n \log^2 n)$ and $\mathsf{Qtime}(\mathcal{D}_1(S)) = O(\log^2 n)$.
\end{theorem}
\textit{Proof.}
We first analyze the space cost.
Let $\mathbf{v}$ be a secondary node of the 2D range tree.
By Theorem~\ref{thm-quadrant}, the space cost of the sub-structures stored at $\mathbf{v}$ is $O(|S(\mathbf{v})|)$.
Therefore, for a primary node $\mathbf{u} \in \mathcal{T}$ of the 2D range tree, the space cost of $\mathcal{T}_\mathbf{u}$ (with the sub-structures) is $O(|S(\mathbf{u})| \log |S(\mathbf{u})|)$.
As a result, the entire space cost of the 2D range tree is $O(n \log^2 n)$.
Let $\mathbf{u}' \in \mathcal{T}'$ be a node of the 1D range tree $\mathcal{T}'$.
By Theorem~\ref{thm-strip} and Lemma~\ref{lem-epds}, the space cost of the sub-structures stored at $\mathbf{u}'$ is $O(|S(\mathbf{u}')| \log |S(\mathbf{u}')|)$.
As such, the entire space cost of $\mathcal{T}'$ is $O(n \log^2 n)$.
For the same reason, the space cost of $\mathcal{T}''$ is $O(n \log^2 n)$, and hence the entire space cost of $\mathcal{D}_1$ is $O(n \log^2 n)$.

Then we analyze the query time.
When answering a query, we need to compute the pairs $\phi,\phi_\alpha,\phi_\beta$ in Lemma~\ref{lem-1of3}.
To compute $\phi$, we first find the splitting nodes $\mathbf{u} \in \mathcal{T}$ and $\mathbf{v} \in \mathcal{T}_\mathbf{u}$.
This is done by a top-down walk in $\mathcal{T}$ and $\mathcal{T}_\mathbf{u}$, which takes $O(\log n)$ time.
Then we query the sub-structures $\mathcal{A}(S_1(\mathbf{v})),\dots,\mathcal{A}(S_4(\mathbf{v}))$, which can be done in $O(\log n)$ time by Theorem~\ref{thm-quadrant}.
Thus, the time for computing $\phi$ is $O(\log n)$.
To compute $\phi_\alpha$, we first find the $t = O(\log n)$ canonical nodes $\mathbf{u}_1',\dots,\mathbf{u}_t' \in \mathcal{T}'$, which can be done in $O(\log n)$ time.
Then we query the sub-structures $\mathcal{B}(S(\mathbf{u}_1')),\dots,\mathcal{B}(S(\mathbf{u}_t'))$ and $\mathcal{K}^\text{v}(S(\mathbf{u}_1')),\dots,\mathcal{K}^\text{v}(S(\mathbf{u}_t'))$ to obtain the pairs $\phi_1,\dots,\phi_t$ and the set $K$ of $2kt$ points.
By Theorem~\ref{thm-strip} and Lemma~\ref{lem-epds}, this step can be done in $O(\log^2 n)$ time.
Finally, we compute the closest-pair $\phi_K$ in $K$ using the standard divide-and-conquer algorithm, which takes $O(\log n \log \log n)$ time since $|K| = O(\log n)$.
Thus, the time for computing $\phi_\alpha$ is $O(\log^2 n)$, so is the time for computing $\phi_\beta$.
As a result, the overall query time is $O(\log^2 n)$.
\hfill $\Box$
\medskip

\noindent
Our first solution itself already achieves the desired worst-case bounds, which simultaneously improves the results given in \cite{gupta2014data} and \cite{sharathkumar2007range}.

\subsection{Second solution} \label{sec-rect2}
We now introduce our second solution, which has the desired average-case space cost and an $O(\log n)$ query time (even in worst-case).
In our second solution, we only use the 2D range tree presented before, but we need some additional sub-structures stored at each secondary node.
Let $k$ be the constant integer in Lemma~\ref{lem-adjacent}.
Define $S_\blacktriangle(\mathbf{v}) = S_3(\mathbf{v}) \cup S_4(\mathbf{v})$ (resp., $S_\blacktriangledown(\mathbf{v}) = S_1(\mathbf{v}) \cup S_2(\mathbf{v})$) as the subset of $S(\mathbf{v})$ consisting of the points above (resp., below) $l_\mathbf{v}$.
Similarly, define $S_\blacktriangleleft(\mathbf{v})$ and $S_\blacktriangleright(\mathbf{v})$ as the subsets to the left and right of $l_\mathbf{u}$, respectively.
Let $\mathbf{v} \in \mathcal{T}_\mathbf{u}$ be a secondary node.
Besides $\mathcal{A}(S_1(\mathbf{v})),\dots,\mathcal{A}(S_4(\mathbf{v}))$, we store at $\mathbf{v}$ two $(\mathcal{P}_{l_\mathbf{u}},k)$-TBEP data structures $\mathcal{K}_{l_\mathbf{u}}(S_\blacktriangle(\mathbf{v})),\mathcal{K}_{l_\mathbf{u}}(S_\blacktriangledown(\mathbf{v}))$ (Lemma~\ref{lem-epdsrand}) and two $(\mathcal{P}_{l_\mathbf{v}},k)$-LREP data structures $\mathcal{K}_{l_\mathbf{v}}(S_\blacktriangleleft(\mathbf{v})),\mathcal{K}_{l_\mathbf{v}}(S_\blacktriangleright(\mathbf{v}))$ (Lemma~\ref{lem-epdsrand}).
Furthermore, we need a new kind of sub-structures called \textit{range shortest-segment} (RSS) data structures.
For a query space $\mathcal{X}$, an $\mathcal{X}$-RSS data structure stores a given set of segments in $\mathbb{R}^2$ and can report the shortest segment contained in a query range $X \in \mathcal{X}$.
For the case $\mathcal{X} = \mathcal{U}$, we have the following RSS data structure.
\begin{lemma} \label{lem-shortestseg}
There exists a $\mathcal{U}$-RSS data structure $\mathcal{C}$ such that for any set $G$ of $m$ segments, $\mathsf{Space}(\mathcal{C}(G)) = O(m^2)$ and $\mathsf{Qtime}(\mathcal{C}(G)) = O(\log m)$.
\end{lemma}
\textit{Proof.}
It suffices to design the $\mathcal{U}^\downarrow$-RSS data structure.
We first notice the existence of a $\mathcal{P}^\text{v}$-RSS data structure using $O(m)$ space with $O(\log m)$ query time.
Indeed, by applying the method in Section~\ref{sec-strip}, we immediately obtain this data structure (a segment here corresponds to a candidate pair in Section~\ref{sec-strip}).
With this $\mathcal{P}^\text{v}$-RSS data structure in hand, it is quite straightforward to design the desired $\mathcal{U}^\downarrow$-RSS data structure.
Let $G = \{\sigma_1,\dots,\sigma_m\}$ be a set of $m$ segments where $\sigma_i = [a_i,b_i]$.
Define $y_i = \max\{a_i.y,b_i.y\}$ and assume $y_1 \leq \cdots \leq y_m$.
Now build a (balanced) binary search tree with keys $y_1,\dots,y_m$.
We denote by $\mathbf{u}_i$ the node corresponding to $y_i$.
At $\mathbf{u}_i$, we store a $\mathcal{P}^\text{v}$-RSS data structure described above built on the subset $G_i = \{\sigma_1,\dots,\sigma_i\} \subseteq G$.
The overall space cost is clearly $O(m^2)$.
To answer a query $U = [x_1,x_2] \times (-\infty,y] \in \mathcal{U}^\downarrow$, we first use the binary search tree to find the maximum $y_i$ that is less than or equal to $y$.
This can be done in $O(\log m)$ time.
Let $P = [x_1,x_2] \times \mathbb{R} \in \mathcal{P}^\text{v}$.
Note that a segment $\sigma \in G$ is contained in $U$ iff $\sigma \in G_i$ and $\sigma$ is contained in $P$.
Thus, we can find the desired segment by querying the $\mathcal{P}^\text{v}$-RSS data structure stored at $\mathbf{u}_i$ with $P$, which takes $O(\log m)$ time.
\hfill $\Box$
\medskip

\noindent
We now define $\varPhi_\blacktriangle(\mathbf{v}) = \varPhi_{l_\mathbf{u}}(S_\blacktriangle(\mathbf{v}),\mathcal{U}^\downarrow)$, $\varPhi_\blacktriangledown(\mathbf{v}) = \varPhi_{l_\mathbf{u}}(S_\blacktriangledown(\mathbf{v}),\mathcal{U}^\uparrow)$, $\varPhi_\blacktriangleleft(\mathbf{v}) = \varPhi_{l_\mathbf{v}}(S_\blacktriangleleft(\mathbf{v}),\mathcal{U}^\rightarrow)$, and $\varPhi_\blacktriangleright(\mathbf{v}) = \varPhi_{l_\mathbf{v}}(S_\blacktriangleright(\mathbf{v}),\mathcal{U}^\leftarrow)$.
We can view $\varPhi_\blacktriangle(\mathbf{v}), \varPhi_\blacktriangledown(\mathbf{v}), \varPhi_\blacktriangleleft(\mathbf{v}), \varPhi_\blacktriangleright(\mathbf{v})$ as four sets of segments by identifying each point-pair $(a,b)$ as a segment $[a,b]$.
Then we apply Lemma~\ref{lem-shortestseg} to build and store at $\mathbf{v}$ four $\mathcal{U}$-RSS data structures $\mathcal{C}(\varPhi_\blacktriangle(\mathbf{v})), \mathcal{C}(\varPhi_\blacktriangledown(\mathbf{v})), \mathcal{C}(\varPhi_\blacktriangleleft(\mathbf{v})), \mathcal{C}(\varPhi_\blacktriangle(\mathbf{v}))$.

We now explain how to compute $\phi_\alpha$ and $\phi_\beta$.
Let us consider $\phi_\alpha$.
Recall that $\phi_\alpha$ is the closest-pair in $S \cap R_\alpha$, i.e., in $S(\mathbf{v}) \cap R_\alpha$.
Let $P$ be the $l_\mathbf{u}$-anchored strip obtained by removing the top/bottom bounding line of $R_\alpha$.
If the two points of $\phi_\alpha$ are on opposite sides of $l_\mathbf{v}$, then by Lemma~\ref{lem-adjacent} its two points must be among the $k$ bottommost points in $S_\blacktriangle(\mathbf{v}) \cap P$ and the $k$ topmost points in $S_\blacktriangledown(\mathbf{v}) \cap P$ respectively.
Using $\mathcal{K}_{l_\mathbf{u}}(S_\blacktriangle(\mathbf{v}))$ and $\mathcal{K}_{l_\mathbf{u}}(S_\blacktriangledown(\mathbf{v}))$, we report these $2k$ points, and compute the closest-pair among them by brute-force.
If the two points of $\phi_\alpha$ are on the same side of $l_\mathbf{v}$, then they are both contained in either $S_\blacktriangle(\mathbf{v})$ or $S_\blacktriangledown(\mathbf{v})$.
So it suffices to compute the closest-pairs in $S_\blacktriangle(\mathbf{v}) \cap R_\alpha$ and $S_\blacktriangledown(\mathbf{v}) \cap R_\alpha$.
Without loss of generality, we only need to consider the closest-pair in $S_\blacktriangle(\mathbf{v}) \cap R_\alpha$.
We denote by $U$ the 3-sided rectangle obtained by removing the bottom boundary of $R_\alpha$, and by $Q_1$ (resp., $Q_2$) the quadrant obtained by removing the right (resp., left) boundary of $U$.
We query $\mathcal{A}(S_1(\mathbf{v}))$ with $Q_1$, $\mathcal{A}(S_2(\mathbf{v}))$ with $Q_2$, and $\mathcal{C}(\varPhi_\blacktriangle(\mathbf{v}))$ with $U$.
Clearly, the shortest one among the three answers is the closest-pair in $S_\blacktriangle(\mathbf{v}) \cap R_\alpha$.
Indeed, the three answers are all point-pairs in $S_\blacktriangle(\mathbf{v}) \cap R_\alpha$.
If the two points of the closest-pair in $S_\blacktriangle(\mathbf{v}) \cap R_\alpha$ are both to the left (resp., right) of $l_\mathbf{u}$, $\mathcal{A}(S_1(\mathbf{v}))$ (resp., $\mathcal{A}(S_2(\mathbf{v}))$) reports it; otherwise, the closest-pair crosses $l_\mathbf{u}$, and $\mathcal{C}(\varPhi_\blacktriangle(\mathbf{v}))$ reports it.
Now we see how to compute $\phi_\alpha$, and $\phi_\beta$ can be computed symmetrically.
Finally, taking the shortest one among $\{\phi,\phi_\alpha,\phi_\beta\}$, the query $R$ can be answered.

A straightforward analysis shows that the overall query time is $O(\log n)$ even in worst-case.
The worst-case space cost is not near-linear, as the $\mathcal{U}$-RSS data structure $\mathcal{C}$ may occupy quadratic space by Lemma~\ref{lem-shortestseg}.
Interestingly, we can show that the average-case space cost is in fact $O(n \log n)$.
The crucial thing is to bound the average-case space of the sub-structures stored at the secondary nodes.
The intuition for bounding the average-case space of the $\mathcal{Q}$-RCP and TBEP/LREP sub-structures comes directly from the average-case performance of our $\mathcal{Q}$-RCP data structure (Theorem~\ref{thm-quadrant}) and TBEP/LREP data structure (Lemma~\ref{lem-epdsrand}).
However, to bound the average-case space of the $\mathcal{U}$-RSS sub-structures is much more difficult.
By our construction, the segments stored in these sub-structures are 3-sided candidate pairs that cross a line.
As such, we have to study the expected number of such candidate pairs in a random dataset.
To this end, we recall Lemma~\ref{lem-probij}.
Let $l$ be a vertical line, and $S \propto \prod_{i=1}^n I_i$ be a random dataset drawn from vertical aligned segments $I_1,\dots,I_n$ as in Lemma~\ref{lem-probij}.
Suppose we build a $\mathcal{U}$-RSS data structure $\mathcal{C}(\varPhi)$ on $\varPhi = \varPhi_l(S,\mathcal{U}^\downarrow)$.
Using Lemma~\ref{lem-probij}, a direct calculation gives us $\mathbb{E}[|\varPhi_l(S,\mathcal{U}^\downarrow)|] = O(\log^2 n)$.
Unfortunately, this is not sufficient for bounding the average-case space of $\mathcal{C}(\varPhi)$, because $\mathbb{E}[\mathsf{Space}(\mathcal{C}(\varPhi))] = O(\mathbb{E}[|\varPhi_l(S,\mathcal{U}^\downarrow)|^2])$ and in general $\mathbb{E}[|\varPhi_l(S,\mathcal{U}^\downarrow)|^2] \neq \mathbb{E}^2[|\varPhi_l(S,\mathcal{U}^\downarrow)|]$.
Therefore, we need a bound for $\mathbb{E}[|\varPhi_l(S,\mathcal{U}^\downarrow)|^2]$, which can also be obtained using Lemma~\ref{lem-probij}, but requires nontrivial work.
\begin{lemma} \label{lem-3sidedcand^2}
Let $l$ be a vertical \textnormal{(}resp., horizontal\textnormal{)} line and $S \propto \prod_{i=1}^n I_i$ where $I_1,\dots,I_n$ are distinct vertical \textnormal{(}resp., horizontal\textnormal{)} aligned segments.
Then for $\mathcal{X} \in \{\mathcal{U}^\downarrow,\mathcal{U}^\uparrow\}$ \textnormal{(}resp., $\mathcal{X} \in \{\mathcal{U}^\leftarrow,\mathcal{U}^\rightarrow\}$\textnormal{)} , $\mathbb{E}[|\varPhi_l(S,\mathcal{X})|] = O(\log^2 n)$ and $\mathbb{E}[|\varPhi_l(S,\mathcal{X})|^2] = O(\log^4 n)$.
\end{lemma}
Now we are ready to prove the bounds of our second solution.
\begin{theorem} \label{thm-rectavg}
There exists an $\mathcal{R}$-RCP data structure $\mathcal{D}_2$ such that \\
$\bullet$ For any $S \subseteq \mathbb{R}^2$ of size $n$, $\mathsf{Qtime}(\mathcal{D}_2(S)) = O(\log n)$. \\
$\bullet$ For a random $S \propto R^n$ where $R$ is the unit square or more generally an arbitrary axes-parallel rectangle, $\mathbb{E}[\mathsf{Space}(\mathcal{D}_2(S))] = O(n \log n)$.
\end{theorem}

\subsection{Combining the two solutions} \label{sec-rectcomb}
We now combine the two data structures $\mathcal{D}_1$ (Theorem~\ref{thm-rectwst}) and $\mathcal{D}_2$ (Theorem~\ref{thm-rectavg}) to obtain a \textit{single} data structure $\mathcal{D}$ that achieves the desired worst-case and average-case bounds simultaneously.
For a dataset $S \subseteq \mathbb{R}^2$ of size $n$, if $\mathsf{Space}(\mathcal{D}_2(S)) \geq n \log^2 n$, we set $\mathcal{D}(S) = \mathcal{D}_1(S)$, otherwise we set $\mathcal{D}(S) = \mathcal{D}_2(S)$.
The worst-case bounds of $\mathcal{D}$ follows directly, while to see the average-case bounds of $\mathcal{D}$ requires a careful analysis using Markov's inequality.
\begin{theorem} \label{thm-rectangle}
There exists an $\mathcal{R}$-RCP data structure $\mathcal{D}$ such that \\
$\bullet$ For any $S \subseteq \mathbb{R}^2$ of size $n$, $\mathsf{Space}(\mathcal{D}(S)) = O(n \log^2 n)$ and $\mathsf{Qtime}(\mathcal{D}(S)) = O(\log^2 n)$. \\
$\bullet$ For a random $S \propto R^n$ where $R$ is the unit square or more generally an arbitrary axes-parallel rectangle, $\mathbb{E}[\mathsf{Space}(\mathcal{D}(S))] = O(n \log n)$ and $\mathbb{E}[\mathsf{Qtime}(\mathcal{D}(S))] = O(\log n)$.
\end{theorem}
\textit{Proof.}
As mentioned above, our data structure $\mathcal{D}$ is obtained by combining $\mathcal{D}_1$ (Theorem~\ref{thm-rectwst}) and $\mathcal{D}_2$ (Theorem~\ref{thm-rectavg}) as follows.
For any $S \subseteq \mathbb{R}^2$ of size $n$, if $\mathsf{Space}(\mathcal{D}_2(S)) \geq n \log^2 n$, we set $\mathcal{D}(S) = \mathcal{D}_1(S)$, otherwise we set $\mathcal{D}(S) = \mathcal{D}_2(S)$.
We claim that $\mathcal{D}$ satisfies the desired bounds.
Let $S \subseteq \mathbb{R}^2$ be a dataset of size $n$.
It is clear from the construction that $\mathsf{Space}(\mathcal{D}(S)) = O(n \log^2 n)$.
Also, $\mathsf{Qtime}(\mathcal{D}(S)) = O(\log^2 n)$, since $\mathsf{Qtime}(\mathcal{D}_1(S)) = O(\log^2 n)$ and $\mathsf{Qtime}(\mathcal{D}_2(S)) = O(\log n)$.
To analyze the average-case performance of $\mathcal{D}$, let $S \propto R^n$ for an axes-parallel rectangle $R$.
Define $E$ as the event $\mathsf{Space}(\mathcal{D}_2(S)) \geq n \log^2 n$ and $\neg E$ as the complement of $E$ ($\neg E$ is the event $\mathsf{Space}(\mathcal{D}_2(S)) < n \log^2 n$).
Since $\mathbb{E}[\mathsf{Space}(\mathcal{D}_2(S))] = O(n \log n)$, we have $\Pr[E] = O(1/\log n)$ by Markov's inequality.
To bound the average-case space cost, we observe
\begin{equation*}
    \mathbb{E}[\mathsf{Space}(\mathcal{D}(S))] = \Pr[E] \cdot \mathbb{E}[\mathsf{Space}(\mathcal{D}_1(S))\ |\ E] + \Pr[\neg E] \cdot \mathbb{E}[\mathsf{Space}(\mathcal{D}_2(S))\ |\ \neg E].
\end{equation*}
Note that $\Pr[E] \cdot \mathbb{E}[\mathsf{Space}(\mathcal{D}_1(S))\ |\ E] = O(n \log n)$, since $\mathsf{Space}(\mathcal{D}_1(S)) = O(n \log^2 n)$ and $\Pr[E] = O(1/\log n)$.
Also, $\Pr[\neg E] \cdot \mathbb{E}[\mathsf{Space}(\mathcal{D}_2(S))\ |\ \neg E] \leq \mathbb{E}[\mathsf{Space}(\mathcal{D}_2(S))] = O(n \log n)$.
Thus, $\mathbb{E}[\mathsf{Space}(\mathcal{D}(S))] = O(n \log n)$.
To bound the average-case query time, let $T_i$ be the worst-case query time of $\mathcal{D}_i$ built on a dataset of size $n$, for $i \in \{1,2\}$.
Then $\mathbb{E}[\mathsf{Qtime}(\mathcal{D}(S))] \leq \Pr[E] \cdot T_1 + \Pr[\neg E] \cdot T_2$.
Since $T_1 = O(\log^2 n)$, $T_2 = O(\log n)$, $\Pr[E] = O(1/\log n)$, we have $\mathbb{E}[\mathsf{Qtime}(\mathcal{D}(S))] = O(\log n)$.
\hfill $\Box$

\section{Halfplane query} \label{sec-halfplane}
We consider the RCP problem for halfplane queries, i.e., the $\mathcal{H}$-RCP problem.
In order to solve the $\mathcal{H}$-RCP problem, it suffices to consider the $\mathcal{H}^\uparrow$-RCP problem.
Let $S \subseteq \mathbb{R}^2$ be the dataset of size $n$.

We shall apply the standard duality technique \cite{de2000computational}.
A non-vertical line $l: y=ux+v$ in $\mathbb{R}^2$ is dual to the point $l^*=(u, -v)$ and a point $p = (s, t) \in \mathbb{R}^2$ is dual to the line $p^*: y=sx-t$.
A basic property of duality is that $p \in l^\uparrow$ (resp., $p \in l^\downarrow$) iff $l^* \in (p^*)^\uparrow$ (resp., $l^* \in (p^*)^\downarrow$).
To make the exposition cleaner, we distinguish between \textit{primal space} and \textit{dual space}, which are two copies of $\mathbb{R}^2$.
The dataset $S$ and query ranges are assumed to lie in the primal space, while their dual objects are assumed to lie in the dual space.
Duality allows us to transform the $\mathcal{H}^\uparrow$-RCP problem into a point location problem as follows.
Let $H = l^\uparrow \in \mathcal{H}^\uparrow$ be a query range.
The line $l$ bounding $H$ is dual to the point $l^*$ in the dual space; for convenience, we also call $l^*$ the dual point of $H$.
If we decompose the dual space into ``cells'' such that the query ranges whose dual points lie in the same cell have the same answer, then point location techniques can be applied to solve the problem directly.
Note that this decomposition must be a polygonal subdivision $\varGamma$ of $\mathbb{R}^2$, which consists of vertices, straight-line edges, and polygonal faces (i.e., cells).
This is because the cell-boundaries must be defined by the dual lines of the points in $S$.
In order to analyze the space cost and query time, we need to study the complexity $|\varGamma|$ of $\varGamma$.
An $O(n^2)$ trivial upper bound for $|\varGamma|$ follows from the fact that the subdivision formed by the $n$ dual lines of the points in $S$ has an $O(n^2)$ complexity. 
Surprisingly, using additional properties of the problem, we can show that $|\varGamma| = O(n)$, which is a key ingredient of our result in this section.

\begin{figure}[h]
    \begin{center}
        \includegraphics[width=5cm]{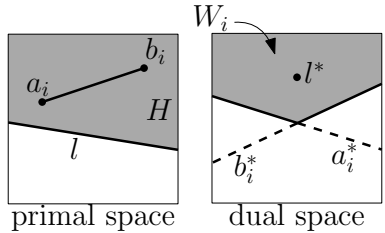}
    \end{center}
    \caption{Illustrating the upward-open wedge $W_i$.}
    \label{figure:openwedge}
\end{figure}
Suppose $\varPhi(S,\mathcal{H}^\uparrow) = \{\phi_1,\dots,\phi_m\}$ where $\phi_i = (a_i,b_i)$ and $\phi_1,\dots,\phi_m$ are sorted in increasing order of their lengths.
It was shown in \cite{abam2009power} that $m = O(n)$, and the candidate pairs do not cross each other (when identified as segments), i.e., the segments $[a_i,b_i]$ and $[a_j,b_j]$ do not cross for any $i \neq j$.
The non-crossing property of the candidate pairs is important and will be used later for proving Lemma~\ref{lem-hlfplinear}.
With this in hand, we now consider the subdivision $\varGamma$.
Let $H = l^\uparrow \in \mathcal{H}^\uparrow$ be a query range.
By the property of duality, $\phi_i$ is contained in $H$ iff $l^* \in (a_i^*)^\uparrow$ and $l^* \in (b_i^*)^\uparrow$, i.e., $l^*$ is in the upward-open \textit{wedge} $W_i$ generated by the lines $a_i^*$ and $b_i^*$ (in the dual space); see Figure~\ref{figure:openwedge}.
As such, the closest-pair in $S \cap H$ to be reported is $\phi_\eta$ for $\eta = \min \{i: l^* \in W_i\}$.
Therefore, $\varGamma$ can be constructed by successively overlaying the wedges $W_1,\dots,W_m$ (similarly to what we see in Section~\ref{sec-quadrant}).
Formally, we begin with a trivial subdivision $\varGamma_0$ of $\mathbb{R}^2$, which consists of only one face, the entire plane.
Suppose $\varGamma_{i-1}$ is constructed, which has an \textit{outer face} $F_{i-1}$ equal to the complement of $\bigcup_{j=1}^{i-1} W_j$ in $\mathbb{R}^2$.
Now we construct a new subdivision $\varGamma_i$ by ``inserting'' $W_i$ to $\varGamma_{i-1}$.
Specifically, $\varGamma_i$ is obtained from $\varGamma_{i-1}$ by decomposing the outer face $F_{i-1}$ via the wedge $W_i$; that is, we decompose $F_{i-1}$ into several smaller faces: one is $F_{i-1} \backslash W_i$ and the others are the connected components of $F_{i-1} \cap W_i$.
Note that $F_{i-1} \backslash W_i$ is the complement of $\bigcup_{j=1}^{i} W_j$, which is connected (as one can easily verify) and becomes the outer face $F_i$ of $\varGamma_i$.
In this way, we construct $\varGamma_1,\dots,\varGamma_m$ in order, and it is clear that $\varGamma_m = \varGamma$.
The linear upper bound for $|\varGamma|$ follows from the following technical result.
\begin{lemma} \label{lem-hlfplinear}
    $|\varGamma_i| - |\varGamma_{i-1}| = O(1)$ for $i \in \{1,\dots,m\}$.
    In particular, $|\varGamma| = O(m)$.
\end{lemma}
\begin{figure}[htbp]
	\begin{center}
		\begin{subfigure}{0.38\linewidth}
		    \includegraphics[width=150pt]{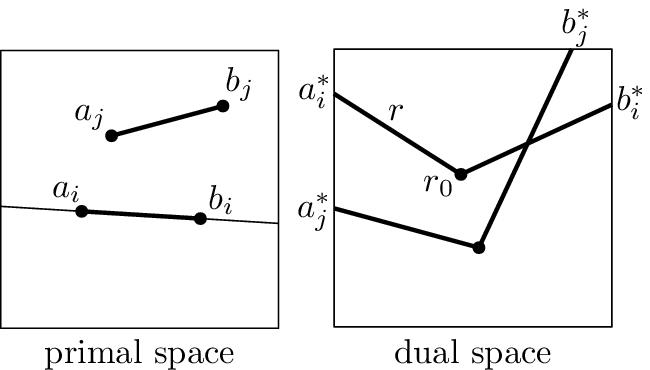}
		    \caption{An illustration of Case 1 \ \ \ \ \ \ \ \ \ \ \ \ \ \ \ \ \ \ \ \ }
			\label{figure:halfspace case 1}
		\end{subfigure}
		\hspace{0.2cm}
		\begin{subfigure}{0.38\linewidth}
		    \includegraphics[width=150pt]{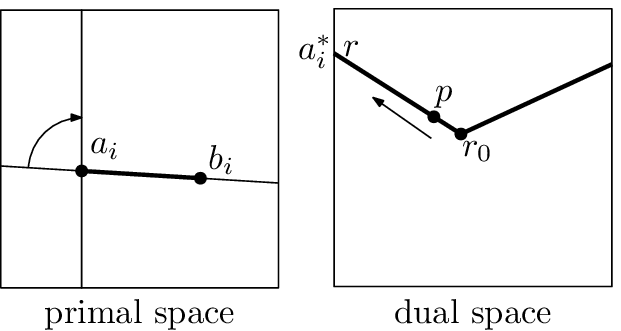}
		    \caption{Moving $p$ and rotating line $a_ib_i$ clockwise around $a_i$}
			\label{figure:halfspace case 2 1}
		\end{subfigure}
        \begin{subfigure}{0.38\linewidth}
		    \includegraphics[width=150pt]{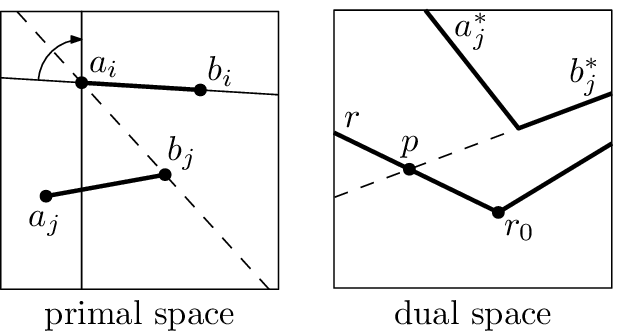}
		    \caption{An example for Case 2 where $r \cap W_j = \emptyset$}
			\label{figure:halfspace case 2 2}
		\end{subfigure}
		\hspace{0.2cm}
        \begin{subfigure}{0.38\linewidth}
		    \includegraphics[width=150pt]{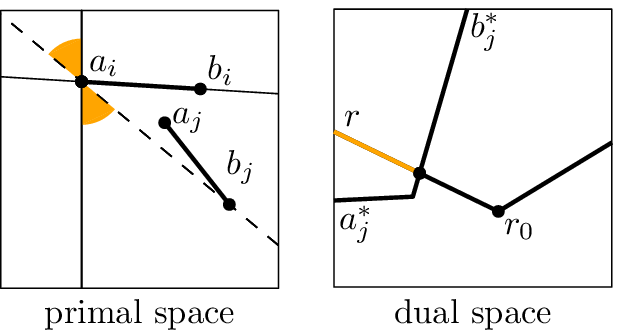}
		    \caption{An example for Case 2 where $r \cap W_j \neq \emptyset$}
			\label{figure:halfspace case 2 3}
		\end{subfigure}
        \begin{subfigure}{0.38\linewidth}
		    \includegraphics[width=150pt]{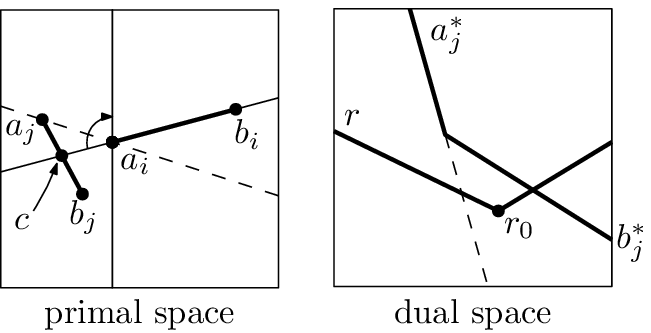}
		    \caption{An example for Case 3 where $a_jb_j$ is to the left and $r \cap W_j = \emptyset$}
			\label{figure:halfspace case 3 ajbj intersects aibi and to the left}
		\end{subfigure}
		\hspace{0.2cm}
        \begin{subfigure}{0.38\linewidth}
		    \includegraphics[width=150pt]{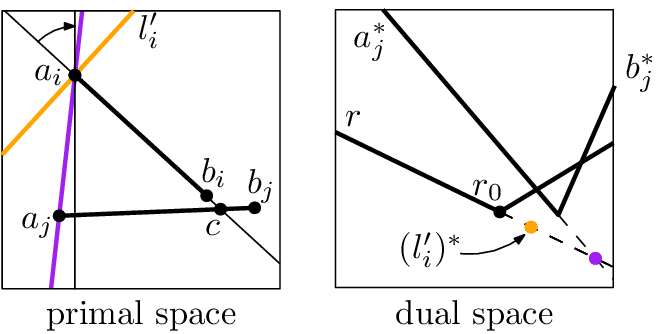}
		    \caption{An example for Case 3 where $(l_i')^* \not\in r$ and $r \cap W_j = \emptyset$}
			\label{figure:halfspace case 3, l does not exist in r}
		\end{subfigure}
        \begin{subfigure}{0.38\linewidth}
		    \includegraphics[width=150pt]{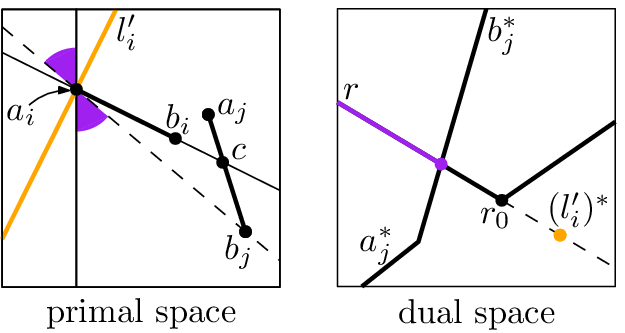}
		    \caption{An example for Case 3 where $(l_i')^* \not\in r$ and $r \cap W_j \neq \emptyset$ \\ \\}
			\label{figure:halfspace case 3, l does not exist in r, another}
		\end{subfigure}
		\hspace{0.2cm}
        \begin{subfigure}{0.38\linewidth}
		    \includegraphics[width=150pt]{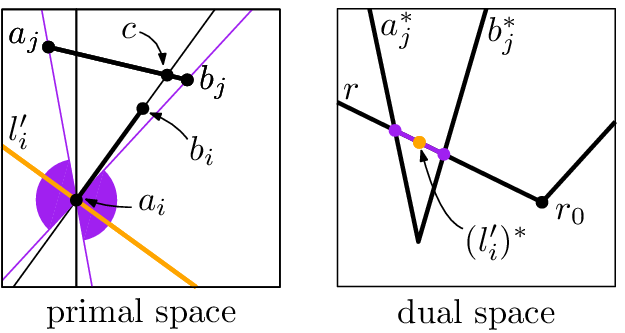}
		    \caption{An example for Case 3 where $(l_i')^* \in r$ and $r \cap W_j$ contains $(l_i')^*$ but does not contain $r_0$ or the infinite end of $r$}
			\label{figure:halfspace case 3, l in r}
		\end{subfigure}
	\caption{Illustrating the various cases in Lemma~\ref{lem-hlfplinear}.}
	\end{center}
\end{figure}
\textit{Proof.}
Let $F_i$ be the outer face of $\varGamma_i$, and $\partial W_i$ be the boundary of the wedge $W_i$ (which consists of two rays emanating from the intersection point of $a_i^*$ and $b_i^*$).
We first note that, to deduce that $|\varGamma_i| - |\varGamma_{i-1}| = O(1)$, it suffices to show that the number of the connected components of $\partial W_i \cap F_{i-1}$ is constant.
This is because every connected component of $\partial W_i \cap F_{i-1}$ contributes to $\varGamma_i$ exactly one new face, a constant number of new vertices, and a constant number of new edges.
Indeed, we only need to check one branch of $\partial W_i$ (i.e., one of the two rays of $\partial W_i$), say the ray contained in $a_i^*$ (we denote it by $r$).
We will show that $r \cap F_{i-1}$ has $O(1)$ connected components.
Without loss of generality, we may assume that $a_i$ is to the left of $b_i$.
Then each point on $r$ is dual to a line in the primal space, which goes through the point $a_i$ with the segment $[a_i, b_i]$ above it.
Note that $r \cap F_{i-1} = r \backslash \bigcup_{j=1}^{i-1} W_j = r \backslash \bigcup_{j=1}^{i-1} (r \cap W_j)$, and each $r \cap W_j$ is a connected portion of $r$.
We consider each $j \in \{1,\dots,i-1\}$ and analyze the intersection $r \cap W_j$.
Let $l_i$ be the line through $a_i$ and $b_i$.
There are three cases to be considered separately: (1) $a_j,b_j \in l_i^\uparrow$, (2) $a_j,b_j \in l_i^\downarrow$, or (3) one of $a_j,b_j$ is in $l_i^\uparrow \backslash l_i$ (i.e., strictly above $l_i$) while the other is in $l_i^\downarrow \backslash l_i$ (i.e., strictly below $l_i$).
\smallskip

\textbf{[Case 1]}
In this case, $a_j,b_j \in l_i^\uparrow$.
The wedge $W_j$ must contain the initial point $r_0$ of $r$ (i.e., the intersection point of $a_i^*$ and $b_i^*$, which is the dual of the line $l_i$), because $r_0 \in (a_j^*)^\uparrow$ and $r_0 \in (b_j^*)^\uparrow$.
(See Figure~\ref{figure:halfspace case 1}.)
\smallskip

\textbf{[Case 2]}
In this case, $a_j,b_j \in l_i^\downarrow$.
We claim that either $r \cap W_j$ is empty or it contains the infinite end of $r$ (i.e., the point at infinity along $r$).
Imagine that we have a point $p$ moving along $r$ from $r_0$ to the infinite end of $r$.
Then $p$ is dual to a line in the primal space rotating clockwise around $a_i$ from the line $l_i$ to the vertical line through $a_i$; see Figure~\ref{figure:halfspace case 2 1}.
Note that $p \in r \cap W_j$ (in the dual space) only when $a_j,b_j \in (p^*)^\uparrow$ (in the primal space).
But $a_j,b_j \in l_i^\downarrow$ in this case.
When $p$ is moving, the region $l_i^\downarrow \cap (p^*)^\uparrow$ expands.
As such, one can easily see that $r \cap W_j$ must contain the infinite end of $r$ if it is nonempty.
(See Figure~\ref{figure:halfspace case 2 2}~and~\ref{figure:halfspace case 2 3}.)
\smallskip

\textbf{[Case 3]}
In this case, one point of $a_j,b_j$ is in $l_i^\uparrow \backslash l_i$ while the other is in $l_i^\downarrow \backslash l_i$.
Thus, the segment $[a_j, b_j]$ must intersect the line $l_i$.
However, as argued before, the segments $[a_j, b_j]$ and $[a_i, b_i]$ do not cross.
So the intersection point $c$ of $[a_j, b_j]$ and $l_i$ is either to the left of $a_i$ or to the right of $b_i$ (recall that $a_i$ is assumed to be to the left of $b_i$).

If $c$ is to the left of $a_i$, we claim that $r \cap W_j$ is empty.
Observe that the dual line of any point on $r$ is through $a_i$ and below $b_i$, meaning that it must be above $c$ (as $c$ is to the left of $a_i$).
In other words, the dual line of any point on $r$ is above at least one of $a_j,b_j$, and thus any point on $r$ is not contained in the wedge $W_j$, i.e., $r \cap W_j$ is empty.
(See Figure~\ref{figure:halfspace case 3 ajbj intersects aibi and to the left}.)

The subtlest case occurs when $c$ is to the right of $b_i$.
In such a case, we consider the line through $a_i$ perpendicular to $l_i$, which we denote by $l_i'$.
We first argue that both $a_j$ and $b_j$ must be on the same side of $l_i'$ as $b_i$.
Since $c$ is to the right of $b_i$, at least one of $a_j,b_j$ is on the same side of $l_i'$ as $b_i$.
However, we notice that $[a_j,b_j]$ cannot intersect $l_i'$, otherwise the length of $\phi_j$ is (strictly) greater than that of $\phi_i$, contradicting the fact that $j<i$ (recall that $\phi_1,\dots,\phi_m$ is sorted in increasing order of their lengths).
So the only possibility is that $a_j,b_j,b_i$ are on the same side of $l_i'$.
Now we further have two sub-cases. \\
$\bullet$
$l_i'$ has no dual point (i.e., $l_i'$ is vertical) or its dual point $(l_i')^*$ is not on the ray $r$.
In this case, consider a point $p$ moving along $r$ from $r_0$ to the infinite end of $r$.
Clearly, when $p$ moves, the region $(l_i')^\rightarrow \cap (p^*)^\uparrow$ expands.
Thus, either $r \cap W_j$ is empty or it contains the infinite end of $r$.
(See Figure~\ref{figure:halfspace case 3, l does not exist in r}~and~\ref{figure:halfspace case 3, l does not exist in r, another}.)\\
$\bullet$
$(l_i')^*$ is on $r$.
Then $r \cap W_j$ may be a connected portion of $r$ containing neither $r_0$ nor the infinite end of $r$.
However, as $b_i \in (l_i')^\uparrow$ in this case, we have $a_j,b_j \in (l_i')^\uparrow$ (recall that $a_j,b_j,b_i$ are on the same side of $l_i'$).
This implies that $r \cap W_j$ contains $(l_i')^*$.
(See Figure~\ref{figure:halfspace case 3, l in r}.)
\smallskip

In sum, we conclude that for any $j \in \{1,\dots,i-1\}$, the intersection $r \cap W_j$ might be \textbf{(i)} empty, or \textbf{(ii)} a connected portion of $r$ containing $r_0$, or \textbf{(iii)} a connected portion of $r$ containing the infinite end of $r$, or \textbf{(iv)} a connected portion of $r$ containing $(l_i')^*$ (if $(l_i')^*$ is on $r$).
As such, the union $\bigcup_{j=1}^{i-1} (r \cap W_j)$ can have at most three connected components, among which one contains $r_0$, one contains the infinite end of $r$, and one contains $(l_i')^*$.
Therefore, the complement of $\bigcup_{j=1}^{i-1} (r \cap W_j)$ in $r$, i.e., $r \cap F_{i-1}$, has at most two connected components.
This in turn implies that $\partial W_i \cap F_{i-1}$ has only a constant number of connected components, and hence $|\varGamma_i| - |\varGamma_{i-1}| = O(1)$.
Finally, since $|\varGamma_0| = O(1)$ and $m = O(n)$, we immediately have $|\varGamma| = |\varGamma_m| = O(m)$.
\hfill $\Box$
\medskip

\noindent
With the above result in hand, we can build an optimal point-location data structure for $\varGamma$ using $O(m)$ space with $O(\log m)$ query time to solve the RCP problem.
Since $m = O(n)$, we obtain an $\mathcal{H}$-RCP data structure using $O(n)$ space and $O(\log n)$ query time in worst-case.

Next, we analyze the average-case bounds of the above data structure.
In fact, it suffices to bound the expected number of the candidate pairs.
Surprisingly, we have the following poly-logarithmic bound.
\begin{lemma} \label{lem-halfcand}
    For a random dataset $S \propto R^n$ where $R$ is an axes-parallel rectangle, $\mathbb{E}[|\varPhi(S,\mathcal{H})|] = O(\log^2 n)$.
\end{lemma}
Now we can conclude the following.
\begin{theorem}\label{thm-halfplane}
    There exists an $\mathcal{H}$-RCP data structure $\mathcal{E}$ such that \\
    $\bullet$ For any $S \subseteq \mathbb{R}^2$ of size $n$, $\mathsf{Space}(\mathcal{E}(S)) = O(n)$ and $\mathsf{Qtime}(\mathcal{E}(S)) = O(\log n)$. \\
    $\bullet$ For a random $S \propto R^n$ where $R$ is the unit square or more generally an arbitrary axes-parallel rectangle, $\mathbb{E}[\mathsf{Space}(\mathcal{E}(S))] = O(\log^2 n)$ and $\mathbb{E}[\mathsf{Qtime}(\mathcal{E}(S))] = O(\log \log n)$.
\end{theorem}

\subsection{Preprocessing} \label{sec-preproc}
In this section, we show how to build the $\mathcal{H}$-RCP data structure in Theorem~\ref{thm-halfplane} in $O(n \log^2 n)$ time.
It suffices to consider the $\mathcal{H}^\uparrow$-RCP data structure described in Section~\ref{sec-halfplane}.
To build this data structure, the key step is to construct the subdivision $\varGamma$ of the dual $\mathbb{R}^2$ (see Section~\ref{sec-halfplane}).
Since $|\varGamma| = O(n)$, once $\varGamma$ is constructed, one can build in $O(n \log n)$ time the point-location data structure for $\varGamma$, and hence our $\mathcal{H}^\uparrow$-RCP data structure.

Let us first consider an easier task, in which $\varPhi(S,\mathcal{H}^\uparrow)$ is already given beforehand.
In this case, we show that $\varGamma$ can be constructed in $O(n \log n)$ time.
As in Section~\ref{sec-halfplane}, suppose $\varPhi(S,\mathcal{H}^\uparrow) = \{\phi_1,\dots,\phi_m\}$ where $\phi_1,\dots,\phi_m$ are sorted in increasing order of their lengths.
Recall that in Section~\ref{sec-halfplane} we defined the $m$ subdivisions $\varGamma_0,\dots,\varGamma_m$.
Our basic idea for constructing $\varGamma$ is to begin with $\varGamma_0$ and iteratively construct $\varGamma_i$ from $\varGamma_{i-1}$ by inserting the wedge $W_i$ dual to $\phi_i$.
In this process, a crucial thing is to maintain the outer face $F_i$ (or its boundary).
Note that the boundary $\partial F_i$ of $F_i$ (i.e., the upper envelope of $F_i$) is an $x$-monotone polygonal chain consisting of segments and two infinite rays; we call these kinds of chains \textit{left-right polylines} and call their pieces \textit{fractions}.
Naturally, a binary search tree can be used to store a left-right polyline; the keys are its fractions in the left-right order.
Therefore, we shall use a (balanced) BST $\mathcal{T}$ to maintain $\partial F_i$.
That is, at the end of the $i$-th iteration, we guarantee the left-right polyline stored in $\mathcal{T}$ is $\partial F_i$.
At each node of $\mathcal{T}$, besides storing the corresponding fraction, we also store the wedge $W_j$ which contributes this fraction.

Suppose we are now at the beginning of the $i$-th iteration.
We have $\varGamma_{i-1}$ in hand and $\mathcal{T}$ stores $\partial F_{i-1}$.
We need to ``insert'' the wedge $W_i$ to generate $\varGamma_i$ from $\varGamma_{i-1}$, and update $\mathcal{T}$.
To this end, the first step is to compute $\partial W_i \cap F_{i-1}$.
Now let us assume in advance that $\partial W_i \cap F_{i-1}$ is already computed in $O(\log |\mathcal{T}|)$ time; later we will explain how to achieve this.
With $\partial W_i \cap F_{i-1}$ in hand, to construct $\varGamma_i$ is fairly easy.
By the proof of Lemma~\ref{lem-hlfplinear}, $\partial W_i \cap F_{i-1}$ has $O(1)$ connected components.
We consider these components one-by-one.
Let $\xi$ be a component, which is an $x$-monotone polygonal chain with endpoints (if any) on $\partial F_{i-1}$ (indeed, $\xi$ consists of at most two pieces as it is a portion of $\partial W_i$).
For convenience, assume $\xi$ has a left endpoint $u$ and a right endpoint $v$.
Then $\xi$ contributes a new (inner) face to $\varGamma_i$, which is the region bounded by $\xi$ and the portion $\sigma$ of $\partial F_{i-1}$ between $u,v$.
We then use $\mathcal{T}$ to report all the fractions of $\partial F_{i-1}$ intersecting $\sigma$ in left-right order, using which the corresponding new face can be directly constructed.
The time cost for reporting the fractions is $O(\log |\mathcal{T}| + k)$, where $k$ is the number of the reported fractions; see Appendix~\ref{appx-implement} for implementation details.
After all the components are considered, we can construct $\varGamma_i$ by adding the new faces to $\varGamma_{i-1}$ (and adjusting the involved edges/vertices if needed).
As there are $O(1)$ components, the total time cost for constructing $\varGamma_i$ from $\varGamma_{i-1}$ is $O(\log |\mathcal{T}| + K_i)$, where $K_i$ is the total number of the fractions reported from $\mathcal{T}$.
But we can charge the reported fractions to the corresponding new faces, and the fractions charged to each face are at most as many as its edges.
Therefore, $\sum_{i=1}^m K_i = O(m)$, and this part of the time cost is amortized $O(\log m)$ for each iteration.
The remaining task is to update the left-right polyline $\mathcal{T}$ to $\partial F_i$.
In fact, the update can also be done in amortized $O(\log m)$ for each iteration (see Appendix~\ref{appx-implement} for implementation details).
As such, the overall time cost for constructing $\varGamma$ is $O(m \log m)$, and thus $O(n \log n)$.

We now explain the missing part of the above algorithm: computing $\partial W_i \cap F_{i-1}$ in $O(\log |\mathcal{T}|)$ time.
Let $r$ be the left ray of $\partial W_i$ and $r_0$ be the initial point of $r$ (i.e., the vertex of $W_i$).
It suffices to compute $r \cap F_{i-1}$.
Recall that $l_i$ is the line through $a_i,b_i$ and $l_i'$ is the line through $a_i$ perpendicular to $l_i$.
Assume $(l_i')^* \in r$ (the case that $(l_i')^* \not\in r$ is in fact easier).
The point $(l_i')^*$ partitions $r$ into a segment $s = [r_0,(l_i')^*]$ and a ray $r'$ emanating from $(l_i')^*$, where $r'$ is to the left of $s$.
By the proof of Lemma~\ref{lem-hlfplinear}, each wedge $W_j$ for $j \in \{1,\dots,i-1\}$ with $W_j \cap r \neq \emptyset$ satisfies at least one of the following: (1) $r_0 \in W_j$, (2) $(l_i')^* \in W_j$, (3) $W_j$ contains the infinite end of $r$.
Therefore, $r \cap F_{i-1}$ can have one or two connected components; if it has two components, one should be contained in $r'$ and the other should be contained in $s$.
As such, $r'$ contains at most one left endpoint and one right endpoint of (some component of) $r \cap F_{i-1}$, so does $s$.
We show that one can find these endpoints by searching in $\mathcal{T}$.
Suppose we want to find the left endpoint $z$ contained in $r'$ (assume it truly exists).
Let $\gamma$ be a fraction of $\partial F_{i-1}$ which is contributed by the wedge $W_j$ for $j \in \{1,\dots,i-1\}$.
It is easy to verify that $\gamma$ contains $z$ iff $\gamma$ intersects $r'$ and $W_j$ contains the infinite end of $r$.
Also, $\gamma$ is to the left of $z$ iff $\gamma \subseteq R$ and $W_j$ contains the infinite end of $r$, where $R$ is the region to the left of $(l_i')^*$ and above $r'$.
As such, one can simply search in $\mathcal{T}$ to find the fraction $\gamma$ containing $z$ in $O(\log |\mathcal{T}|)$ time, if $z$ truly exists.
(If $z$ does not exist, by searching in $\mathcal{T}$ we can verify its non-existence, as we can never find the desired fraction $\gamma$.)
The right endpoint contained in $r'$ and the left/right endpoints contained in $s$ can be computed in a similar fashion.
With these endpoints in hand, one can compute $r \cap F_{i-1}$ straightforwardly.
The other case that $(l_i')^* \not\in r$ is handled similarly and more easily, as in this case $r \cap F_{i-1}$ has at most one connected component.
Therefore, $r \cap F_{i-1}$ (and thus $\partial W_i \cap F_{i-1}$) can be computed in $O(\log |\mathcal{T}|)$ time.

Next, we consider how to construct $\varGamma$ if we are only given the dataset $S$.
It was shown in \cite{abam2009power} that one can compute in $O(n \log^2 n)$ time a set $\varPsi$ of pairs of points in $S$ such that $\varPhi(S,\mathcal{H}^\uparrow) \subseteq \varPsi$ and $|\varPsi| = O(n \log n)$.
We use that method to compute $\varPsi$, and suppose $\varPsi = \{\psi_1,\dots,\psi_M\}$ where $\psi_1,\dots,\psi_M$ are sorted in increasing order of their lengths.
The $m$ candidate pairs $\phi_1,\dots,\phi_m \in \varPhi(S,\mathcal{H}^\uparrow)$ are among $\psi_1,\dots,\psi_M$.
Let $i_1 < \cdots < i_m$ be indices such that $\phi_1 = \psi_{i_1},\dots,\phi_m = \psi_{i_m}$ (note that at this point we do not know what $i_1,\dots,i_m$ are).
We shall consider $\psi_1,\dots,\psi_M$ in order.
When considering $\psi_i$, we want to verify whether $\psi_i$ is a candidate pair or not.
If this can be done, the candidate pairs $\phi_1,\dots,\phi_m$ will be found in order.
Whenever a new candidate pair $\phi_k$ is found, we construct $\varGamma_k$ from $\varGamma_{k-1}$ in $O(\log m)$ time by the approach above.
Now assume $\psi_1,\dots,\psi_{i-1}$ are already considered, the candidate pairs in $\{\psi_1,\dots,\psi_{i-1}\}$ are recognized (say they are $\phi_1,\dots,\phi_{k-1}$), and $\varGamma_{k-1}$ is constructed.
We then consider $\psi_i$.
We need to see whether $\psi_i$ is a candidate pair, i.e., whether $\psi_i = \phi_k$.
Let $W$ be the corresponding wedge of $\psi_i$ in the dual $\mathbb{R}^2$.
Observe that $\psi_i = \phi_k$ iff $W \nsubseteq \bigcup_{j=1}^{k-1} W_j$.
Indeed, if $\psi_i = \phi_k$, then $W = W_k$ and hence $W \nsubseteq \bigcup_{j=1}^{k-1} W_j$ (for $\phi_k$ is a candidate pair).
Conversely, if $W \nsubseteq \bigcup_{j=1}^{k-1} W_j$, then their exists some halfplane $H \in \mathcal{H}^\uparrow$ such that $H$ contains $\psi_i$ and does not contain $\phi_1,\dots,\phi_{k-1}$.
Then the closest-pair in $S \cap H$ cannot be in $\{\psi_1,\dots,\psi_{i-1}\}$ but must be in $\varPsi$, hence it is nothing but $\psi_i$.
Based on this observation, we can verify whether $\psi_i = \phi_k$ as follows.
We assume $\psi_i = \phi_k$ and try to use it to construct $\varGamma_k$ from $\varGamma_{k-1}$ by our above approach.
If our assumption is correct, then $\varGamma_k$ is successfully constructed in $O(\log m)$ time.
Furthermore, in the process of constructing $\varGamma_k$, our approach allows us to find a point in $W \backslash \bigcup_{j=1}^{k-1} W_j$, which we call \textit{witness} point.
This witness point then evidences the correctness of our assumption.
On the other hand, if our assumption is wrong, the process can still terminate in $O(\log m)$ time, but we can never find such a witness point because $W \subseteq \bigcup_{j=1}^{k-1} W_j$.
In this case, we just discard $\psi_i$ and continue to consider $\psi_{i+1}$.
After considering all pairs in $\varPsi$, we recognize all the $m$ candidate pairs and $\varGamma = \varGamma_m$ is constructed.
Since $m = O(\log n)$ and $|\varPsi| = O(n \log n)$, the overall process takes $O(n \log^2 n)$ time.

\subsection{Application to the $\mathcal{H}$-RSS problem}
Interestingly, our approach for solving the $\mathcal{H}$-RCP problem can also be applied to the $\mathcal{H}$-RSS problem, and leads to an optimal $\mathcal{H}$-RSS data structure for interior-disjoint (i.e., non-crossing) segments.
This by-product is of indepedent interest.
\begin{theorem} \label{thm-halfrss}
    There exists an $\mathcal{H}$-RSS data structure $\mathcal{F}$ such that for any set $G$ of $n$ interior-disjoint \textnormal{(}i.e., non-crossing\textnormal{)} segments in $\mathbb{R}^2$, $\mathsf{Space}(\mathcal{F}(G)) = O(n)$, $\mathsf{Qtime}(\mathcal{F}(G)) = O(\log n)$, and $\mathcal{F}(G)$ can be built in $O(n \log n)$ time.
\end{theorem}
\textit{Proof.}
The data structure is basically identical to the $\mathcal{H}$-RCP data structure given in Section~\ref{sec-halfplane}.
Let $\sigma_1,\dots,\sigma_n$ be the interior-disjoint segments in $G$ sorted in increasing order of their lengths.
Suppose $W_i$ is the wedge dual to $\sigma_i$.
We successively overlay the wedges $W_1,\dots,W_n$ to create a subdivision $\varGamma$ of the dual space, as what we do in Section~\ref{sec-halfplane} for the candidate pairs.
A point-location data structure on $\varGamma$ is then our $\mathcal{H}$-RSS data structure for $G$.
Note that Lemma~\ref{lem-hlfplinear} can be applied to show $|\varGamma| = O(n)$, because when proving Lemma~\ref{lem-hlfplinear} we only used the facts that the candidate pairs do not cross each other and the wedges are inserted in increasing order of the lengths of their corresponding candidate pairs (here the segments $\sigma_1,\dots,\sigma_n$ are also non-crossing and sorted in increasing order of their lengths).
As such, the space cost of the data structure is $O(n)$ and the query time is $O(\log n)$.
In Section~\ref{sec-preproc}, we show that if the candidate pairs are already given, our $\mathcal{H}$-RCP data structure can be built in $O(n \log n)$ time.
It follows that our $\mathcal{H}$-RSS data structure can be built in $O(n \log n)$ time, as we are directly given the segments in this case.
\hfill $\Box$

\section{Conclusion and future work} \label{sec-conclusion}
In this paper, we revisited the range closest-pair (RCP) problem, which aims to preprocess a set $S$ of points in $\mathbb{R}^2$ into a data structure such that when a query range $X$ is specified, the closest-pair in $S \cap X$ can be reported efficiently.
We proposed new RCP data structures for various query types (including quadrants, strips, rectangles, and halfplanes).
Both worst-case and average-case analyses were applied to these data structures, resulting in new bounds for the RCP problem.
See Table~\ref{tab-results} for a comparison of our new results with the previous work.

We now list some open questions for future study.
First, as mentioned in Section~\ref{sec-contribution}, the preprocessing for our orthogonal RCP data structures remains open.
It is not clear how to build these data structures in sub-quadratic time.
Besides, the RCP problem for other query types is also open.
One important example is the disk query, which is usually much harder than the rectangle query and halfplane query in traditional range search.
For an easier version, we can focus on the case where the query disks have a fixed radius, or equivalently, the query ranges are \textit{translates} of a fixed disk.
Along this direction, one can also consider translation queries of some shape other than a disk.
For instance, if the query ranges are translates of a fixed rectangle, can we have more efficient data structures than our rectangle RCP data structure in Section~\ref{sec-rectangle}?
Finally, the RCP problem in higher dimensions is quite open.
To our best knowledge, the only known result for this is a simple data structure given in \cite{gupta2014data} constructed by explicitly storing all the candidate pairs, which only has guaranteed average-case performance.

\bibliography{my_bib.bib}

\newpage
\noindent
{\LARGE \textbf{Appendix}}
\appendix
\section{Missing proofs} \label{appx-prfs}
\subsection{Proof of Lemma~\ref{lem-kappa}}
Without loss of generality, we can assume $R = [0,\Delta] \times [0,\Delta']$ where $\Delta \leq \Delta'$.
We first observe some simple facts.
Let $D$ be a disc centered at a point in $R$ with radius $\delta$.
Then $\mathsf{Area}(D \cap R) \geq \delta^2/9$ if $\delta \leq \Delta$, and $\mathsf{Area}(D \cap R) \geq \delta\Delta/9$ if $\Delta \leq \delta \leq \Delta'$.
Furthermore, we always have $\mathsf{Area}(D \cap R) \leq 4\delta^2$ and $\mathsf{Area}(D \cap R) \leq 2\delta\Delta$.
With these facts in hand, we can begin our proof.
\smallskip

\textbf{[Upper bound]}
First, we prove the upper bound for $\mathbb{E}[\kappa^p(A)]$.
To this end, we need to study the distribution of the random variable $\kappa(A)$.
For convenience, we assume $m$ is even and sufficiently large.
We make the following claims. \\
\textbf{(1)} For any $\delta \geq 2\Delta$, we have $\Pr[\kappa(A) \geq \delta] \leq e^{-\delta/(72\Delta'/m^2)}$. \\
\textbf{(2)} For any $\delta \in (0,2\Delta]$, we have $\Pr[\kappa(A) \geq \delta] \leq e^{-\delta^2/(144\Delta\Delta'/m^2)}$.\\
To prove the claims, suppose the $m$ random points in $A$ are $a_1,\dots,a_m$.
For any $\delta>0$ and each $i \in \{2,\dots,m\}$, we define $E_{\delta,i}$ as the event that $\kappa(\{a_1,\dots,a_i\}) \geq \delta$.
Note that $E_{\delta,i}$ happens only if $E_{\delta,i-1}$ does, thus for any $\delta>0$ we can write
\begin{equation*}
    \Pr[\kappa(A) \geq \delta] = \Pr[E_{\delta,m}] = \Pr[E_{\delta,2}] \cdot \prod_{i=3}^m \Pr[E_{\delta,i}|E_{\delta,i-1}].
\end{equation*}
We consider the probability $\Pr[E_{\delta,i}|E_{\delta,i-1}]$ for $i \in \{3,\dots,m\}$.
Let $D_1,\dots,D_{i-1}$ denote the discs with radii $\delta/2$ centered at $a_1,\dots,a_{i-1}$ respectively, and $U = \bigcup_{j=1}^{i-1} D_i$.
If $E_{\delta,i-1}$ happens, then $D_1,\dots,D_{i-1}$ are disjoint and hence
\begin{equation*}
    \mathsf{Area}(U \cap R) = \sum_{j=1}^{i-1} \mathsf{Area}(D_j \cap R).
\end{equation*}
Now assume we have a lower bound $\mu$ for all $\mathsf{Area}(D_j \cap R)$, i.e., $\mathsf{Area}(D_j \cap R) \geq \mu$ for any $j \in \{1,\dots,i-1\}$.
Then when $E_{\delta,i-1}$ happens, we always have $\mathsf{Area}(U \cap R) \geq (i-1)\mu$.
This implies
\begin{equation*}
    \Pr[a_i \in U \cap R|E_{\delta,i-1}] \geq \frac{(i-1)\mu}{\mathsf{Area}(R)} = \frac{(i-1)\mu}{\Delta\Delta'}.
\end{equation*}
Note that $E_{\delta,i}$ happens only if $a_i \notin U \cap R$.
Therefore,
\begin{equation*}
    \Pr[E_{\delta,i}|E_{\delta,i-1}] \leq \Pr[a_i \notin U \cap R|E_{\delta,i-1}] = 1-\Pr[a_i \in U \cap R|E_{\delta,i-1}] \leq 1-\frac{(i-1)\mu}{\Delta\Delta'}.
\end{equation*}
For $i \geq m/2+1$, we have $\Pr[E_{\delta,i}|E_{\delta,i-1}] \leq 1-(m\mu)/(2\Delta\Delta')$.
Then
\begin{equation*}
    \Pr[\kappa(A) \geq \delta] \leq \prod_{i=m/2+1}^m \Pr[E_{\delta,i}|E_{\delta,i-1}] \leq \left(1-\frac{m\mu}{2\Delta\Delta'}\right)^{m/2}.
\end{equation*}
Using the fact $(1-x)^{(1/x)}<e^{-1}$ for any $x \in [0,1]$, we deduce
\begin{equation*}
    \Pr[\kappa(A) \geq \delta] \leq \left(1-\frac{m\mu}{2\Delta\Delta'}\right)^{m/2} = \left(1-\frac{m\mu}{2\Delta\Delta'}\right)^{\frac{2\Delta\Delta'}{m\mu} \cdot \frac{m^2 \mu}{4\Delta\Delta'}} \leq e^{m^2 \mu/(4 \Delta\Delta')}.
\end{equation*}
If $\delta \geq 2\Delta$, then $\mathsf{Area}(D_j \cap R) \geq \delta\Delta/18$ for any $j$ (as argued at the beginning of the proof), so we can set $\mu = \delta\Delta/18$.
The above inequality directly implies the claim \textbf{(1)}.
If $\delta \in (0,2\Delta]$, then $\mathsf{Area}(D_j \cap R) \geq \delta^2/36$ for any $j$ (as argued at the beginning of the proof), so we can set $\mu = \delta^2/36$.
The above inequality directly implies the claim \textbf{(2)}.
With the two claims in hand, we now prove the lemma.
We shall use the formula
\begin{equation*}
    \mathbb{E}[\kappa^p(A)] = \int_0^\infty \Pr[\kappa^p(A) \geq t]\ \text{d}t.
\end{equation*}
We consider two cases: $\Delta'/m^2 \geq \sqrt{\Delta \Delta'}/m$ and $\Delta'/m^2 < \sqrt{\Delta \Delta'}/m$.
Assume $\Delta'/m^2 \geq \sqrt{\Delta \Delta'}/m$, i.e., $\Delta \leq \Delta'/m^2$.
In this case, what we want is $\mathbb{E}[\kappa^p(A)] = O((\Delta'/m^2)^p)$ for any constant $p \geq 1$.
Let $\alpha = 72\Delta'/m^2$.
Then for any constant $p \geq 1$, we can write
\begin{equation*}
    \mathbb{E}[\kappa^p(A)] = \int_0^\infty \Pr[\kappa^p(A) \geq t]\ \text{d}t \leq \alpha^p + \int_{\alpha^p}^\infty \Pr[\kappa^p(A) \geq t]\ \text{d}t.
\end{equation*}
Set $q = 1/p$.
For $t \geq \alpha^p$, we have $t^q \geq \alpha > 2\Delta'/m^2 \geq 2\Delta$.
Therefore, by applying the claim \textbf{(1)} above we have
\begin{equation*}
    \Pr[\kappa^p(A) \geq t] = \Pr[\kappa(A) \geq t^q] \leq e^{-t^q/\alpha} = e^{-(t/\alpha^p)^q}.
\end{equation*}
It follows that
\begin{equation*}
    \mathbb{E}[\kappa^p(A)] \leq \alpha^p + \int_{\alpha^p}^\infty e^{-(t/\alpha^p)^q}\ \text{d}t = \alpha^p + \alpha^p \int_1^\infty e^{-t^q}\ \text{d}t.
\end{equation*}
The integration $\int_1^\infty e^{-t^q}\ \text{d}t$ converges, thus $\mathbb{E}[\kappa^p(A)] = O(\alpha^p) = O((\Delta'/m^2)^p)$.
Next, assume $\Delta'/m^2 < \sqrt{\Delta \Delta'}/m$, i.e., $\Delta > \Delta'/m^2$.
In this case, what we want is $\mathbb{E}[\kappa^p(A)] = O((\sqrt{\Delta \Delta'}/m)^p)$ for any constant $p \geq 1$.
We first claim that 
\begin{equation} \label{eq-int2Delta^p}
    \int_{(2\Delta)^p}^\infty \Pr[\kappa^p(A) \geq t]\ \text{d}t = O((\sqrt{\Delta\Delta'}/m)^p).
\end{equation}
Again, let $\alpha = 72\Delta'/m^2$.
By applying the claim \textbf{(1)} above, we have
\begin{equation*}
    \int_{(2\Delta)^p}^\infty \Pr[\kappa^p(A) \geq t]\ \text{d}t \leq \int_{(2\Delta)^p}^\infty e^{-(t/\alpha^p)^q}\ \text{d}t,
\end{equation*}
where $q = 1/p$.
Since $\Delta > \Delta'/m^2$, we further deduce
\begin{equation*}
    \int_{(2\Delta)^p}^\infty e^{-(t/\alpha^p)^q}\ \text{d}t \leq \int_{(\alpha/36)^p}^\infty e^{-(t/\alpha^p)^q}\ \text{d}t = \alpha^p \int_{(1/36)^p}^\infty e^{-t^q}\ \text{d}t = O(\alpha^p).
\end{equation*}
By the assumption $\Delta'/m^2 < \sqrt{\Delta \Delta'}/m$, we have $\alpha = O(\sqrt{\Delta\Delta'}/m)$, thus Equation~\ref{eq-int2Delta^p} holds.
With this in hand, we bound $\mathbb{E}[\kappa^p(A)]$ as follows.
Let $\beta = 12 \sqrt{\Delta\Delta'}/m$.
If $\beta \geq 2\Delta$, then
\begin{equation*}
    \mathbb{E}[\kappa^p(A)] \leq \beta^p + \int_{\beta^p}^\infty \Pr[\kappa^p(A) \geq t]\ \text{d}t \leq \beta^p + \int_{(2\Delta)^p}^\infty \Pr[\kappa^p(A) \geq t]\ \text{d}t = O((\sqrt{\Delta\Delta'}/m)^p).
\end{equation*}
The rightmost equality above follows from Equation~\ref{eq-int2Delta^p}.
If $\beta \leq 2\Delta$, then
\begin{equation*}
    \mathbb{E}[\kappa^p(A)] \leq \beta^p + \int_{\beta^p}^{(2\Delta)^p} \Pr[\kappa^p(A) \geq t]\ \text{d}t + \int_{(2\Delta)^p}^\infty \Pr[\kappa^p(A) \geq t]\ \text{d}t.
\end{equation*}
It suffices to show $\int_{\beta^p}^{(2\Delta)^p} \Pr[\kappa^p(A) \geq t]\ \text{d}t = O((\sqrt{\Delta\Delta'}/m)^p)$.
By the claim \textbf{(2)} above,
\begin{equation*}
    \int_{\beta^p}^{(2\Delta)^p} \Pr[\kappa^p(A) \geq t]\ \text{d}t = \int_{\beta^p}^{(2\Delta)^p} \Pr[\kappa(A) \geq t^q]\ \text{d}t \leq \int_{\beta^p}^{(2\Delta)^p} e^{-(t/\beta^p)^{2q}}\ \text{d}t.
\end{equation*}
Furthermore, we have
\begin{equation*}
    \int_{\beta^p}^{(2\Delta)^p} e^{-(t/\beta^p)^{2q}}\ \text{d}t \leq \int_{\beta^p}^\infty e^{-(t/\beta^p)^{2q}}\ \text{d}t = \beta^p \int_1^\infty e^{-t^{2q}}\ \text{d}t.
\end{equation*}
Since the integration $\int_1^\infty e^{-t^{2q}}\ \text{d}t$ converges, $\int_{\beta^p}^{(2\Delta)^p} \Pr[\kappa^p(A) \geq t]\ \text{d}t = O((\sqrt{\Delta\Delta'}/m)^p)$.
As such, $\mathbb{E}[\kappa^p(A)] = O((\sqrt{\Delta\Delta'}/m)^p)$.
This proves the upper bound for $\mathbb{E}[\kappa^p(A)]$.
\smallskip

\textbf{[Lower bound]}
To prove the lower bound for $\mathbb{E}[\kappa^p(A)]$ is much easier.
It suffices to show that $\mathbb{E}[\kappa^p(A)] = \Omega((\Delta'/m^2)^p)$ and $\mathbb{E}[\kappa^p(A)] = \Omega((\sqrt{\Delta\Delta'}/m)^p)$.
Let $i,j \in \{1,\dots,m\}$ such that $i \neq j$.
Set $\delta_1 = \Delta'/(2m^2)$.
We observe that $\Pr[\text{dist}(a_i,a_j) \leq \delta_1] \leq 1/m^2$.
Indeed, if $D$ is the disc centered at $a_i$ with radius $\delta_1$, then we always have $\mathsf{Area}(D \cap R) \leq 2\delta_1\Delta = \Delta\Delta'/m^2$ (as argued at the beginning of the proof), and hence
\begin{equation*}
    \Pr[\text{dist}(a_i,a_j) \leq \delta_1] = \Pr[a_j \in D \cap R] \leq \frac{\Delta\Delta'/m^2}{\mathsf{Area}(R)} = \frac{1}{m^2}.
\end{equation*}
By union bound, we have
\begin{equation*}
    \Pr[\kappa(A) \leq \delta_1] \leq \binom{m}{2} \cdot \frac{1}{m^2} < \frac{1}{2}.
\end{equation*}
Thus, $\Pr[\kappa(A) \leq \delta_1] \geq 1/2$ and $\mathbb{E}[\kappa^p(A)] \geq \delta_1^p/2 = \Omega((\Delta'/m^2)^p)$.
Similarly, we can show $\mathbb{E}[\kappa^p(A)] = \Omega((\sqrt{\Delta\Delta'}/m)^p)$.
Set $\delta_2 = \sqrt{\Delta\Delta'}/(4m)$.
We again observe that $\Pr[\text{dist}(a_i,a_j) \leq \delta_2] \leq 1/m^2$ for any distinct $i,j \in \{1,\dots,m\}$.
Indeed, if $D$ is the disc centered at $a_i$ with radius $\delta_2$, then we always have $\mathsf{Area}(D \cap R) \leq 4\delta_2^2 = \Delta\Delta'/m^2$ (as argued at the beginning of the proof), and hence
\begin{equation*}
    \Pr[\text{dist}(a_i,a_j) \leq \delta_2] = \Pr[a_j \in D \cap R] \leq \frac{\Delta\Delta'/m^2}{\mathsf{Area}(R)} = \frac{1}{m^2}.
\end{equation*}
Applying the same argument as above, we can deduce $\Pr[\kappa(A) \leq \delta_2] \geq 1/2$, which implies that $\mathbb{E}[\kappa^p(A)] \geq \delta_2^p/2 = \Omega((\sqrt{\Delta\Delta'}/m)^p)$.
This proves the lower bound for $\mathbb{E}[\kappa^p(A)]$.
\smallskip

The above proof straightforwardly applies to the special case in which $R$ is a segment.

\subsection{Proof of Lemma~\ref{lem-quadcand}} \label{appx-prfquadcand}
Without loss of generality, assume $R = [0,1] \times [0,\Delta]$.
It suffices to show $\mathbb{E}[|\varPhi(S,\mathcal{Q}^\swarrow)|] = O(\log^2 n)$.
Suppose the $n$ random points in $S$ are $a_1,\dots,a_n$.
Let $E_{i,j}$ be the event $(a_i,a_j) \in \varPhi(S,\mathcal{Q}^\swarrow)$, then by the linearity of expectation,
\begin{equation*}
    \mathbb{E}[|\varPhi(S,\mathcal{Q}^\swarrow)|] = \sum_{i=1}^{n-1} \sum_{j=i+1}^n \Pr[E_{i,j}].
\end{equation*}
Since $a_1,\dots,a_n$ play the same roles here, the probabilities on the right-hand side of the above equation should be the same and thus $\mathbb{E}[|\varPhi(S,\mathcal{Q}^\swarrow)|] = O(n^2 \cdot \Pr[E_{1,2}])$.
In order to bound $\Pr[E_{1,2}]$, we introduce some random variables.
Let $x_\text{max} = \max\{a_1.x,a_2.x\}$, $y_\text{max} = \max\{a_1.y,a_2.y\}$, $x_\text{min} = \min\{a_1.x,a_2.x\}$, $y_\text{min} = \min\{a_1.y,a_2.y\}$.
The quadrant $Q = (-\infty,x_\text{max}] \times (-\infty,y_\text{max}]$ is the \textit{minimal} southwest quadrant containing both $a_1$ and $a_2$, and clearly $E_{1,2}$ happens iff $(a_1,a_2)$ is the closest-pair in $S \cap Q$.
Define $\varLambda = \{i \geq 3: a_i \in Q\}$, which is a random subset of $\{3,\dots,n\}$, i.e., a random variable taking value from the power set of $\{3,\dots,n\}$.
We achieve the bound for $\Pr[E_{1,2}]$ through three steps.
\smallskip

\textbf{[Step 1]}
Let us first fix the values of $x_\text{max},y_\text{max},\varLambda$, and consider the corresponding conditional probability of $E_{1,2}$.
Formally, we claim that, for all $\tilde{x} \in (0,1]$, all $\tilde{y} \in (0,\Delta]$, and all nonempty $J \subseteq \{3,\dots,n\}$,
\begin{equation} \label{eq-condprob}
    \Pr\left[E_{1,2}\ |\ (x_\text{max} = \tilde{x}) \wedge (y_\text{max} = \tilde{y}) \wedge (\varLambda = J)\right] = O(1/|J|^2).
\end{equation}
For convenience, we use $C_{\tilde{x},\tilde{y},J}$ to denote the condition in the above conditional probability.
Assume $|J| = m$.
Let $\delta_x = x_\text{max}-x_\text{min}$ and $\delta_y = y_\text{max}-y_\text{min}$.
Note that under the condition $C_{\tilde{x},\tilde{y},J}$, $E_{1,2}$ happens only if $\delta_x \leq \kappa(S_J)$ and $\delta_y \leq \kappa(S_J)$ where $S_J = \{a_j: j \in J\}$.
Indeed, if $\delta_x > \kappa(S_J)$ or $\delta_y > \kappa(S_J)$, then $\text{dist}(a_1,a_2) > \kappa(S_J)$, which implies $E_{1,2}$ does not happen because all the random points in $S_J$ are contained in $Q$ under the condition $C_{\tilde{x},\tilde{y},J}$.
Therefore, it suffices to bound $\Pr[(\delta_x \leq \kappa(S_J)) \wedge (\delta_y \leq \kappa(S_J))\ |\ C_{\tilde{x},\tilde{y},J}]$.
Note that under the condition $C_{\tilde{x},\tilde{y},J}$, $Q$ is just $(-\infty,\tilde{x}] \times (-\infty,\tilde{y}]$.
Thus the condition $C_{\tilde{x},\tilde{y},J}$ is equivalent to saying that the maximum of the $x$-coordinates (resp., $y$-coordinates) of $a_1,a_2$ is $\tilde{x}$ (resp., $\tilde{y}$), all $a_j$ for $j \in J$ are contained in the rectangle $R' = [0,\tilde{x}] \times [0,\tilde{y}]$, and all $a_j$ for $j \in J$ for $j \in \{3,\dots,n\} \backslash J$ are in $R \backslash R'$.
As such, one can easily verify that, under the condition $C_{\tilde{x},\tilde{y},J}$, the distribution of the random number $\delta_x$ (resp., $\delta_y$) is the uniform distribution on the interval $[0,\tilde{x}]$ (resp., $[0,\tilde{y}]$), and the distributions of the $m$ random points in $S_J$ are the uniform distribution on $R'$; furthermore, these random numbers/points are independent of each other.
This says, if we consider a new random experiment in which we independently generate two random numbers $\delta_x',\delta_y'$ from the uniform distributions on $[0,\tilde{x}],[0,\tilde{y}]$ respectively (which correspond to $\delta_x,\delta_y$) and a random dataset $S' \propto (R')^m$ (which corresponds to $S_J$), then we have
\begin{equation*}
    \Pr[(\delta_x' \leq \kappa(S')) \wedge (\delta_y' \leq \kappa(S'))] = \Pr[(\delta_x \leq \kappa(S_J)) \wedge (\delta_y \leq \kappa(S_J))\ |\ C_{\tilde{x},\tilde{y},J}].
\end{equation*}
So it suffices to bound $\Pr[(\delta_x' \leq \kappa(S')) \wedge (\delta_y' \leq \kappa(S'))]$ in the new experiment; we denote by $\lambda$ this probability.
We apply the formula
\begin{equation*}
    \lambda = \int_0^\infty p(t)\cdot \Pr\left[(\delta_x' \leq t) \wedge (\delta_y' \leq t)\right]\ \text{d}t = \int_0^\infty p(t)\cdot \Pr[\delta_x' \leq t] \cdot \Pr[\delta_y' \leq t]\ \text{d}t,
\end{equation*}
where $p(\cdot)$ is the probability distribution function of $\kappa(S')$.
Since $\delta_x'$ (resp., $\delta_y'$) is drawn uniformly on the interval $[0,\tilde{x}]$ (resp., $[0,\tilde{y}]$), we have $\Pr[\delta_x' \leq t] = \min\{t/\tilde{x},1\}$ (resp., $\Pr[\delta_y' \leq t] = \min\{t/\tilde{y},1\}$).
Without loss of generality, we assume $\tilde{x} \leq \tilde{y}$.
Then we have
\begin{equation*}
    \Pr[\delta_x' \leq t] \cdot \Pr[\delta_y' \leq t] = \min\{t^2/(\tilde{x}\tilde{y}),t/\tilde{y},1\} \leq \min\{t^2/(\tilde{x}\tilde{y}),t/\tilde{y}\}.
\end{equation*}
It follows that
\begin{equation*}
    \lambda \leq \int_0^\infty p(t)\cdot \min\left\{\frac{t^2}{\tilde{x}\tilde{y}},\frac{t}{\tilde{y}}\right\}\text{d}t \leq \min\left\{\int_0^\infty \frac{p(t) t^2}{\tilde{x}\tilde{y}}\text{d}t, \int_0^\infty \frac{p(t)t}{\tilde{y}}\text{d}t \right\}.
\end{equation*}
Noting the fact that $\int_0^\infty p(t) t^2 \text{d}t = \mathbb{E}[\kappa^2(S')]$ and $\int_0^\infty p(t) t \text{d}t = \mathbb{E}[\kappa(S')]$, we have
\begin{equation*}
    \lambda \leq \min\left\{ \frac{\mathbb{E}[\kappa^2(S')]}{\tilde{x}\tilde{y}}, \frac{\mathbb{E}[\kappa(S')]}{\tilde{y}}\right\}.
\end{equation*}
Since $\tilde{x} \leq \tilde{y}$ by assumption, Lemma~\ref{lem-kappa} implies that $\mathbb{E}[\kappa(S')] = O(\max\{\sqrt{\tilde{x}\tilde{y}}/m,\tilde{y}/m^2\})$ and $\mathbb{E}[\kappa^2(S')] = O(\max\{\tilde{x}\tilde{y}/m^2,\tilde{y}^2/m^4\})$.
If $\sqrt{\tilde{x}\tilde{y}}/m \leq \tilde{y}/m^2$, then $\mathbb{E}[\kappa(S')]/\tilde{y} = O(1/m^2)$, otherwise $\mathbb{E}[\kappa^2(S')]/(\tilde{x}\tilde{y}) = O(1/m^2)$.
In either of the two cases, we have $\lambda = O(1/m^2)$.
Therefore, we obtain Equation~\ref{eq-condprob}.
For an arbitrary nonempty $J \subseteq \{3,\dots,n\}$, since Equation~\ref{eq-condprob} holds for all $\tilde{x} \in (0,1]$ and $\tilde{y} \in (0,\Delta]$, we can remove the conditions $x_\text{max} = \tilde{x}$ and $y_\text{max} = \tilde{y}$ from Equation~\ref{eq-condprob} to deduce $\Pr[E_{1,2}\ |\ \varLambda = J] = O(1/|J|^2)$ (note that although we miss the case $\tilde{x}=0$ or $\tilde{y}=0$, it does not matter since the events $x_\text{max} = 0$ and $y_\text{max} = 0$ happen with probability 0).
This further implies that $\Pr[E_{1,2}\ |\ |\varLambda| = m] = O(1/m^2)$ for all $m \in \{1,\dots,n-2\}$.
For $m=0$, we have $\Pr[E_{1,2}\ |\ |\varLambda| = m] = 1$.
\smallskip

\textbf{[Step 2]}
In order to apply the result achieved in Step 1 to bound $\Pr[E_{1,2}]$, we need to bound $\Pr[|\varLambda| = m]$ for $m = \{0,\dots,n-2\}$.
This is a purely combinatorial problem, because the random variable $|\varLambda|$ only depends on the orderings of the $x$-coordinates and $y$-coordinates of $a_1,\dots,a_n$.
The ordering of the $x$-coordinates ($x$-ordering for short) of $a_1,\dots,a_n$ can be represented as a permutation of $\{a_1,\dots,a_n\}$; so is the $y$-ordering.
Thus, in terms of the ordering of coordinates, there are $(n!)^2$ different configurations of $S$, each of which can be represented by a pair $(\pi,\pi')$ of permutations of $\{a_1,\dots,a_n\}$ where $\pi$ (resp., $\pi'$) represents the $x$-ordering (resp., $y$-ordering) of $a_1,\dots,a_n$; that is, if $\pi=(a_{i_0},\dots,a_{i_n})$ and $\pi'=(a_{i_0'},\dots,a_{i_n'})$, then $a_{i_0}.x < \cdots < a_{i_n}.x$ and $a_{i_0'}.y < \cdots < a_{i_n'}.y$ (we can ignore the degenerate case in which two random points have the same $x$-coordinates or $y$-coordinates, because the random points in $S$ have distinct coordinates with probability 1).
Note that every configuration occurs with the same probability $1/(n!)^2$.
If $S$ has the configuration $(\pi,\pi')$, then $\varLambda$ is just the subset of $\{3,\dots,n\}$ consisting of all $i$ such that $\mathsf{rk}_\pi(a_i) \leq \max\{\mathsf{rk}_\pi(a_1),\mathsf{rk}_\pi(a_2)\}$ and $\mathsf{rk}_{\pi'}(a_i) \leq \max\{\mathsf{rk}_{\pi'}(a_1),\mathsf{rk}_{\pi'}(a_2)\}$, where the function $\mathsf{rk}$ computes the \textit{rank} of an element in a permutation (i.e., the position of the element in the permutation).
Therefore, we can pass to a new random experiment in which we generate independently and uniformly the two permutations $\pi,\pi'$ of $\{a_1,\dots,a_n\}$ (i.e., uniformly generate a configuration of $S$), and study $\Pr[|\varLambda|=m]$ for $m \in \{0,\dots,n-2\}$ in the new experiment.
Let $r_i = \mathsf{rk}_\pi(a_i)$.
Fixing $m \in \{0,\dots,n-2\}$, we have the formula
\begin{equation} \label{eq-Lambda=k}
    \Pr[|\varLambda|=m] = \sum_{i=2}^n \Pr[\max\{r_1,r_2\}=i] \cdot \Pr[|\varLambda|=m\ |\ \max\{r_1,r_2\}=i].
\end{equation}
We first compute $\Pr[\max\{r_1,r_2\}=i]$.
By an easy counting argument, we see that, among the $n!$ permutations of $\{a_1,\dots,a_n\}$, there are exactly $2(i-1)(n-2)!$ permutations in which the maximum of the ranks of $a_1$ and $a_2$ is $i$.
Therefore, 
\begin{equation*}
    \Pr[\max\{r_1,r_2\}=i] = \frac{2(i-1)(n-2)!}{n!} = O\left(\frac{i}{n^2}\right).
\end{equation*}
We then consider $\Pr[|\varLambda|=m\ |\ \max\{r_1,r_2\}=i]$.
If $i<m+2$, the probability is 0, because $|\varLambda \cup \{1,2\}| \leq \max\{r_1,r_2\}$ by definition.
So assume $i \geq m+2$.
Suppose the permutation $\pi$ has already been generated and satisfies $\max\{r_1,r_2\}=i$.
Let $A = \{a_j:r_j \leq i\}$.
Note that $|A| = i$ and $a_1,a_2 \in A$.
Now we randomly generate the permutation $\pi'$ and observe the probability of $|\varLambda|=m$.
Clearly, $|\varLambda|=m$ iff $\max\{\mathsf{rk}_{\pi'|_A}(a_1),\mathsf{rk}_{\pi'|_A}(a_2)\} = m+2$, where $\pi'|_A$ is the permutation of $A$ induced by $\pi'$ (i.e., the permutation obtained by removing the points in $S \backslash A$ from $\pi'$).
Using the same counting argument as above, we have
\begin{equation*}
    \Pr[\max\{\mathsf{rk}_{\pi'|_A}(a_1),\mathsf{rk}_{\pi'|_A}(a_2)\} = m+2] = O\left(\frac{m+1}{i^2}\right).
\end{equation*}
Therefore, $\Pr[|\varLambda|=m\ |\ \max\{r_1,r_2\}=i] = O((m+1)/i^2)$.
Plugging in these results to Equation~\ref{eq-Lambda=k}, a direct calculation gives us $\Pr[|\varLambda|=m] = O((m+1)\log n/n^2)$.
\smallskip

\textbf{[Step 3]}
Using the results achieved in the previous steps, to bound $\Pr[E_{1,2}]$ is quite straightforward.
We apply the formula
\begin{equation*}
    \Pr[E_{1,2}] = \sum_{m=0}^{n-2} \Pr[|\varLambda| = m] \cdot \Pr[E_{1,2}\ |\ |\varLambda| = m].
\end{equation*}
We use the result achieved in Step 1 to bound $\Pr[E_{1,2}\ |\ |\varLambda| = m]$ and the result achieved in Step 2 to bound $\Pr[|\varLambda| = m]$.
Then a direct calculation gives us $\Pr[E_{1,2}] = O(\log^2n/n^2)$.
As such, $\mathbb{E}[|\varPhi(S,\mathcal{Q}^\swarrow)|] = O(\log^2 n)$, and hence $\mathbb{E}[|\varPhi(S,\mathcal{Q})|] = O(\log^2 n)$.

\subsection{Proof of Lemma~\ref{lem-probij}}
It suffices to consider the case in which $I_1,\dots,I_n$ are vertical aligned segments and $\mathcal{X} = \mathcal{U}^\downarrow$.
Without loss of generality, we may assume $I_i = x_i \times [0,1]$ where $x_1<\cdots<x_n$ are real numbers.
Fix $i,j \in \{1,\dots,n\}$ such that $i<j$.
We first define some random variables.
Let $y_k = a_k.y$ for all $k \in \{1,\dots,n\}$.
Define $y_\text{max} = \max\{y_i,y_j\}$ and $y_\text{min} = \min\{y_i,y_j\}$.
The 3-sided rectangle $U = [x_i,x_j] \times (-\infty,y_\text{max}]$ is the \textit{minimal} bottom-unbounded rectangle containing both $a_i$ and $a_j$, and clearly $(a_i,a_j) \in \varPhi(S,\mathcal{U}^\downarrow)$ iff $(a_i,a_j)$ is the closest-pair in $S \cap U$.
Define $\varLambda = \{k: i<k<j \text{ and } a_k \in U\}$, which is a random subset of $\{i+1,\dots,j-1\}$.
We bound $\Pr[(a_i,a_j) \in \varPhi(S,\mathcal{U}^\downarrow)]$ through three steps.
\smallskip

\textbf{[Step 1]}
Let us first fix the values of $y_\text{max}$ and $\varLambda$, and consider the corresponding conditional probability of the event $(a_i,a_j) \in \varPhi(S,\mathcal{U}^\downarrow)$.
Formally, we claim that, for all $\tilde{y} \in (0,1]$ and all nonempty $K \subseteq \{i+1,\dots,j-1\}$,
\begin{equation} \label{eq-condprob2}
    \Pr[(a_i,a_j) \in \varPhi(S,\mathcal{U}^\downarrow)\ |\ (y_\text{max} = \tilde{y}) \wedge (\varLambda = K)] = O(1/|K|^2).
\end{equation}
For convenience, we use $C_{\tilde{y},K}$ to denote the condition in the above conditional probability.
Assume $|K| = m$.
Let $\delta = y_\text{max} - y_\text{min}$ and $Y_K = \{y_k: k \in K\}$.
We first notice that, under the condition $C_{\tilde{y},K}$, $(a_i,a_j) \in \varPhi(S,\mathcal{U}^\downarrow)$ only if $\delta \leq \kappa(Y_K)$.
Indeed, if $\delta > \kappa(Y_K) = |y_{i'}-y_{j'}|$ for some distinct $i',j' \in K$, then $\text{dist}(a_{i'},a_{j'}) < \text{dist}(a_i,a_j)$ since $|x_i-x_j| \geq |x_{i'}-x_{j'}|$, which implies $(a_i,a_j) \notin \varPhi(S,\mathcal{U}^\downarrow)$.
Therefore, it suffices to bound $\Pr[\delta \leq \kappa(Y_K)\ |\ C_{\tilde{y},K}]$.
Note that under the condition $C_{\tilde{y},K}$, $U$ is just $[x_i,x_j] \times (-\infty,\tilde{y}]$.
Thus the condition $C_{\tilde{y},K}$ is equivalent to saying that the maximum of $y_i,y_j$ is $\tilde{y}$, all $y_k$ for $k \in K$ are in $[0,\tilde{y}]$, and all $y_k$ for $k \in \{i+1,\dots,j-1\} \backslash K$ are in $(\tilde{y},1]$.
As such, one can easily verify that, under the condition $C_{\tilde{y},K}$, the distribution of $\delta$ is the uniform distribution on $[0,\tilde{y}]$, and the distributions of the $m$ random numbers in $Y_K$ are also the uniform distribution on $[0,\tilde{y}]$; furthermore, these random numbers are independent of each other.
This says, if we consider a new random experiment in which we independently generate a random number $\delta'$ from the uniform distribution on $[0,\tilde{y}]$ (which corresponds to $\delta$) and a random dataset $Y' \propto [0,\tilde{y}]^m$ (which corresponds to $Y_K$), then we have
\begin{equation*}
    \Pr[\delta' \leq \kappa(Y')] = \Pr[\delta \leq \kappa(Y_K)\ |\ C_{\tilde{y},K}].
\end{equation*}
So it suffices to bound $\Pr[\delta' \leq \kappa(Y')]$ in the new experiment.
We apply the formula
\begin{equation*}
    \Pr[\delta' \leq \kappa(Y')] = \int_0^{\tilde{y}} p(t)\cdot\Pr[\delta' \leq t]\ \text{d}t,
\end{equation*}
where $p(\cdot)$ is the probability distribution function of $\kappa(Y')$.
Since $\delta'$ is drawn from the uniform distribution on $[0,\tilde{y}]$, $\Pr[\delta' \leq t] = t/\tilde{y}$ for $t \in [0,\tilde{y}]$.
Thus, 
\begin{equation*}
    \Pr[\delta' \leq \kappa(Y')] = \left(\int_0^{\tilde{y}} p(t)t\ \text{d}t\right)/\tilde{y} = \mathbb{E}[\kappa(Y')]/\tilde{y}.
\end{equation*}
By Lemma~\ref{lem-kappa} (segment case), $\mathbb{E}[\kappa(Y')] = \tilde{y}/m^2$.
This implies $\Pr[\delta' \leq \kappa(Y')] = O(1/m^2)$, which proves Equation~\ref{eq-condprob2}.
For an arbitrary nonempty $K \subseteq \{i+1,\dots,j-1\}$, since Equation~\ref{eq-condprob2} holds for all $\tilde{y} \in (0,1]$, we can remove the condition $y_\text{max} = \tilde{y}$ from Equation~\ref{eq-condprob2} to deduce $\Pr[(a_i,a_j) \in \varPhi(S,\mathcal{U}^\downarrow)\ |\ \varLambda = K] = O(1/|K|^2)$.
This further implies $\Pr[(a_i,a_j) \in \varPhi(S,\mathcal{U}^\downarrow)\ |\ |\varLambda| = m] = O(1/m^2)$ for all $m = \{1,\dots,j-i-1\}$.
For $m=0$, we have $\Pr[(a_i,a_j) \in \varPhi(S,\mathcal{U}^\downarrow)\ |\ |\varLambda| = m] = 1$.
\smallskip

\textbf{[Step 2]}
In order to apply the result achieved in Step 1 to bound the unconditional probability of $(a_i,a_j) \in \varPhi(S,\mathcal{U}^\downarrow)$, we need to bound $\Pr[|\varLambda| = m]$ for all $m \in \{0,\dots,j-i-1\}$.
This is a combinatorial problem, because the random variable $|\varLambda|$ only depends on the ordering of $y_i,\dots,y_j$.
There are $(j-i+1)!$ possible orderings, each of which can be represented by a permutation of $\{y_i,\dots,y_j\}$.
Every ordering occurs with the same probability $1/(j-i+1)!$.
For a permutation $\pi$ of $\{y_i,\dots,y_j\}$, we write $\lambda_\pi = \max\{\mathsf{rk}_\pi(y_i),\mathsf{rk}_\pi(y_j)\}$, where the function $\mathsf{rk}$ computes the rank of an element in a permutation.
Clearly, if the ordering is $\pi$, then $|\varLambda| = \lambda_\pi-2$.
As such, we can pass to a new random experiment in which we generate uniformly a permutation $\pi$ of $\{y_i,\dots,y_j\}$ and study $\Pr[|\varLambda| = m]$ for $m \in \{0,\dots,j-i-1\}$ in this new experiment.
Fixing $m \in \{0,\dots,j-i-1\}$, it follows that $|\varLambda| = m$ iff $\lambda_\pi = m+2$.
By an easy counting argument, we see that, among the $(j-i+1)!$ permutations of $\{y_i,\dots,y_j\}$, there are exactly $2(m+1)(j-i-1)!$ permutations in which the maximum of the ranks of $y_i$ and $y_j$ is $m+2$.
Therefore,
\begin{equation*}
    \Pr[|\varLambda| = m] = \Pr[\lambda_\pi = m+2] = \frac{2(m+1)(j-i-1)!}{(j-i+1)!} = O\left(\frac{m+1}{(j-i)^2}\right).
\end{equation*}
\smallskip

\textbf{[Step 3]}
Using the results achieved in the previous steps, the lemma can be readily proved.
We apply the formula
\begin{equation} \label{eq-sumup}
    \Pr[(a_i,a_j) \in \varPhi(S,\mathcal{U}^\downarrow)] = \sum_{m=0}^{j-i-1} \Pr[|\varLambda| = m] \cdot \Pr[(a_i,a_j) \in \varPhi(S,\mathcal{U}^\downarrow)\ |\ |\varLambda| = m].
\end{equation}
We use the result achieved in Step 1 to bound $\Pr[(a_i,a_j) \in \varPhi(S,\mathcal{U}^\downarrow)\ |\ |\varLambda| = m]$ and the result achieved in Step 2 to bound $\Pr[|\varLambda| = m]$.
Then a direct calculation gives us $\Pr[(a_i,a_j) \in \varPhi(S,\mathcal{U}^\downarrow)] = O(\log(j-i) / (j-i)^2)$.

\subsection{Proof of Lemma~\ref{lem-epdsrand}}
It suffices to consider the case in which $l$ is a vertical line.
Then $\mathcal{P}_l \subseteq \mathcal{P}^\text{v}$, so a $(\mathcal{P}^\text{v},k)$-TBEP data structure is naturally a $(\mathcal{P}_l,k)$-TBEP data structure.
In a dataset $S \subseteq \mathbb{R}^2$, we say a point $a \in S$ is a \textit{candidate} point if $a$ is one of the $k$ topmost/bottommost points in $S \cap P$ for some $P \in \mathcal{P}_l$.
We denote by $\varPsi(S)$ the subset of $S$ consisting of all candidate points in $S$.
Note that for any $P \in \mathcal{P}_l$, the $k$ topmost/bottommost points in $S \cap P$ are just the $k$ topmost/bottommost points in $\varPsi(S) \cap P$.
As such, answering a $(\mathcal{P}_l,k)$-TBEP query on $S$ is equivalent to answering a $(\mathcal{P}_l,k)$-TBEP query on $\varPsi(S)$.
Therefore, we can define our $(\mathcal{P}_l,k)$-TBEP data structure $\mathcal{K}_l$ as $\mathcal{K}_l(S) = \mathcal{K}^\text{v}(\varPsi(S))$, where $\mathcal{K}^\text{v}$ is the $(\mathcal{P}^\text{v},k)$-TBEP data structure defined in Lemma~\ref{lem-epds}.
Now let $S \propto \prod_{i=1}^n I_i$ where $I_1,\dots,I_n$ are distinct vertical aligned segments.
Assume $I_1,\dots,I_n$ are sorted from left to right, in which $I_1,\dots,I_t$ are to the left of $l$ and $I_t,\dots,I_n$ are to the right of $l$.
Denote by $a_i \in S$ the random point drawn on $I_i$.
We claim that $\mathbb{E}[|\varPsi(S)|] = O(\log n)$.
Let $i \in \{1,\dots,t\}$.
If $a_i \in \varPsi(S)$, then it must be one of the $k$ topmost/bottommost points in $S \cap P$ for some $P \in \mathcal{P}_l$.
Note that any $P \in \mathcal{P}_l$ with $a_i \in P$ contains $a_i,\dots,a_t$.
Therefore, if $a_i \in \varPsi(S)$, it must be one of the $k$ topmost/bottommost points among $a_i,\dots,a_t$.
Since the random points are generated independently, the probability that $a_i$ is one of the $k$ topmost/bottommost points among $a_i,\dots,a_t$ is exactly $\min\{2k/(t-i+1),1\}$.
As such, $\Pr[a_i \in \varPsi(S)] \leq 2k/(t-i+1)$.
Using the same argument, we see that for $i \in \{t+1,\dots,n\}$, $\Pr[a_i \in \varPsi(S)] \leq 2k/(i-t)$.
Now
\begin{equation*}
    \mathbb{E}[|\varPsi(S)|] = \sum_{i=1}^n \Pr[a_i \in \varPsi(S)] \leq \sum_{i=1}^t \frac{2k}{t-i+1} + \sum_{i=t+1}^n \frac{2k}{t-i} = O(k \log n).
\end{equation*}
Since $k$ is a constant, we have $\mathbb{E}[|\varPsi(S)|] = O(\log n)$.
Thus, 
\begin{equation*}
    \mathbb{E}[\mathsf{Space}(\mathcal{K}_l(S))] = \mathbb{E}[\mathsf{Space}(\mathcal{K}^\text{v}(\varPsi(S)))] = \mathbb{E}[|\varPsi(S)|] = O(\log n).
\end{equation*}
Also, $\mathbb{E}[\mathsf{Qtime}(\mathcal{K}_l(S))] = \mathbb{E}[\mathsf{Qtime}(\mathcal{K}^\text{v}(\varPsi(S)))] = O(\mathbb{E}[\log |\varPsi(S)|])$.
Note that $\mathbb{E}[\log x] \leq \log \mathbb{E}[x]$ for a positive random variable $x$, hence we have $\mathbb{E}[\mathsf{Qtime}(\mathcal{K}_l(S))] = O(\log \log n)$.
The case in which $l$ is a horizontal line is handled symmetrically.

\subsection{Proof of Lemma~\ref{lem-3sidedcand^2}}
It suffices to consider the case in which $I_1,\dots,I_n$ are vertical aligned segments and $\mathcal{X} = \mathcal{U}^\downarrow$.
Without loss of generality, assume $I_i = x_i \times [0,1]$ where $x_1<\cdots<x_n$ are real numbers.
We denote by $a_i \in S$ the random point drawn on $I_i$.
Also, suppose $a_1,\dots,a_t$ are to the left of $l$, while $a_{t+1},\dots,a_n$ are to the right of $l$.
Let $E_{i,j}$ be the event that $(a_i,a_j) \in \varPhi(A,\mathcal{U}^\downarrow)$.
Then we have the equation
\begin{equation*}
    \mathbb{E}[|\varPhi_l(S,\mathcal{U}^\downarrow)|] = \sum_{i=1}^t \sum_{j=t+1}^n \Pr[E_{i,j}].
\end{equation*}
By applying Lemma~\ref{lem-probij} and the fact
\begin{equation*}
    \sum_{i=1}^t \sum_{j=t+1}^n \frac{\log(j-i)}{(j-i)^2} \leq \sum_{p=1}^n p \cdot \frac{\log p}{p^2} = O(\log^2n),
\end{equation*}
we have $\mathbb{E}[|\varPhi_l(S,\mathcal{U}^\downarrow)|] = O(\log^2n)$.

To prove $\mathbb{E}[|\varPhi_l(S,\mathcal{U}^\downarrow)|^2] = O(\log^4 n)$ is much more difficult.
Define $\varPsi = \varPhi_l^2(S,\mathcal{U}^\downarrow)$, i.e., the Cartesian product of two copies of $\varPhi_l(S,\mathcal{U}^\downarrow)$.
Then $|\varPsi| = |\varPhi_l(S,\mathcal{U}^\downarrow)|^2$.
So it suffices to bound $\mathbb{E}[|\varPsi|]$.
Clearly, for $i,i' \in \{1,\dots,t\}$ and $j,j' \in \{t+1,\dots,n\}$, $((a_i,a_j),(a_{i'},a_{j'})) \in \varPsi$ iff $E_{i,j} \wedge E_{i',j'}$.
Therefore, we have
\begin{equation} \label{eq-EPsi}
    \mathbb{E}[|\varPsi|] = \sum_{i=1}^t \sum_{j=t+1}^n \sum_{i'=1}^t \sum_{j'=t+1}^n \Pr[E_{i,j} \wedge E_{i',j'}].
\end{equation}
However, $\Pr[E_{i,j} \wedge E_{i',j'}] \neq \Pr[E_{i,j}] \cdot \Pr[E_{i',j'}]$ in general, as the events $E_{i,j}$ and $E_{i',j'}$ are not independent.
We investigate $\Pr[E_{i,j} \wedge E_{i',j'}]$ by considering various cases.
\smallskip

\textbf{[Case 1]}
We first consider the easiest case in which $i=i'$ and $j=j'$.
In this case, $\Pr[E_{i,j} \wedge E_{i',j'}] = \Pr[E_{i,j}] = O(\log(j-i)/(j-i)^2)$ by Lemma~\ref{lem-probij}.
Then the sum of the terms $\Pr[E_{i,j} \wedge E_{i',j'}]$ satisfying $i=i'$ and $j=j'$ is $O(\log^2 n)$.
\smallskip

\textbf{[Case 2]}
We then consider the case in which $i \neq i'$ and $j \neq j'$.
Let $\delta = j-i$ and $\delta' = j'-i'$.
In this case, we claim that $\Pr[E_{i,j} \wedge E_{i',j'}] = O((\log\delta\log\delta')/(\delta\delta')^2)$.
To prove this, we may assume that $\delta'$ is sufficiently large, say $\delta' \geq 5$.
Indeed, when $\delta'<5$, what we want is $\Pr[E_{i,j} \wedge E_{i',j'}] = O(\log\delta/\delta^2)$, which is true as $\Pr[E_{i,j} \wedge E_{i',j'}] \leq \Pr[E_{i,j}] = O(\log\delta/\delta^2)$.
For the same reason, we may also assume $\delta \geq 5$.
Let $S_0$ (resp., $S_1$) be the subsets of $S$ consisting of $a_i,a_j$ (resp., $a_{i'},a_{j'}$) and all the random points in $S \backslash \{a_i,a_j,a_{i'},a_{j'}\}$ with even indices (resp., odd indices).
Clearly, $S = S_0 \cup S_1$ and $S_0 \cap S_1 = \emptyset$.
Define $F_0$ (resp., $F_1$) as the event $(a_i,a_j) \in \varPhi(S_0,\mathcal{U}^\downarrow)$ (resp., $(a_{i'},a_{j'}) \in \varPhi(S_1,\mathcal{U}^\downarrow)$).
Since $S_0$ and $S_1$ are subsets of $S$, $E_{i,j}$ (resp., $E_{i',j'}$) happens only if $F_0$ (resp., $F_1$) happens.
Besides, $F_0$ and $F_1$ are independent events, because $S_0 \cap S_1 = \emptyset$.
Thus,
\begin{equation*}
    \Pr[E_{i,j} \wedge E_{i',j'}] \leq \Pr[F_0 \wedge F_1] = \Pr[F_0] \cdot \Pr[F_1].
\end{equation*}
To bound $\Pr[F_0]$ and $\Pr[F_1]$, we use Lemma~\ref{lem-probij} again.
By construction, there are $\Theta(\delta)$ points in $S_0$ whose $x$-coordinates are in $[a_i.x,a_j.x]$ (recall the assumption $\delta \geq 5$).
Therefore, we have $\Pr[F_0] = O(\log \delta/ \delta^2)$ by Lemma~\ref{lem-probij}.
Similarly, $\Pr[F_1] = O(\log \delta'/ (\delta')^2)$.
Using the above inequality, we have $\Pr[E_{i,j} \wedge E_{i',j'}] = O((\log\delta\log\delta')/(\delta\delta')^2)$.
The sum of the terms $\Pr[E_{i,j} \wedge E_{i',j'}]$ satisfying $i \neq i'$ and $j \neq j'$ is $O(\log^4 n)$, as one can easily verify.
\smallskip

\textbf{[Case 3]}
The subtlest case is that $i = i'$ and $j \neq j'$, or symmetrically $i \neq i'$ and $j = j'$.
Assume $i=i'$ and $j>j'$.
Let $\delta = j-i$ and $\delta' = j'-i$.
We claim that $\Pr[E_{i,j} \wedge E_{i,j'}] = O(\log\delta/(\delta^2\delta'))$.
Again, we may assume $\delta$ and $\delta'$ are sufficiently large, say $\delta > \delta' \geq 5$.
Let $S_0$ (resp., $S_1$) be the subsets of $S$ consisting of $a_i,a_j$ (resp., $a_i,a_{j'}$) and all the random points in $S \backslash \{a_i,a_j,a_{j'}\}$ with even indices (resp., odd indices).
Clearly, $S = S_0 \cup S_1$ and $S_0 \cap S_1 = \{a_i\}$.
Define $F_0$ (resp., $F_1$) as the event $(a_i,a_j) \in \varPhi(S_0,\mathcal{U}^\downarrow)$ (resp., $(a_i,a_{j'}) \in \varPhi(S_1,\mathcal{U}^\downarrow)$).
As in Case 2, we have
\begin{equation*}
    \Pr[E_{i,j} \wedge E_{i,j'}] \leq \Pr[F_0 \wedge F_1].
\end{equation*}
However, $\Pr[F_0 \wedge F_1] \neq \Pr[F_0] \cdot \Pr[F_1]$ in general, because $F_0$ and $F_1$ are not independent (both of them depends on $a_i.y$).
To handle this issue, we observe that
\begin{equation*}
    \Pr[F_0 \wedge F_1] = \int_0^1 \Pr[F_0 \wedge F_1\ |\ a_i.y = t]\ \text{d}t,
\end{equation*}
since the distribution of $a_i.y$ is the uniform distribution on $[0,1]$.
Note that under the condition $a_i.y = t$, $F_0$ and $F_1$ are in fact independent.
Indeed, when $a_i.y$ is fixed, $F_0$ (resp., $F_1$) only depends on the $y$-coordinates of the random points in $S_0 \backslash \{a_i\}$ (resp., $S_1 \backslash \{a_i\}$).
Therefore, we can write
\begin{equation*}
    \Pr[F_0 \wedge F_1] = \int_0^1 \Pr[F_0\ |\ a_i.y = t] \cdot \Pr[F_1\ |\ a_i.y = t]\ \text{d}t,
\end{equation*}
We first consider $\Pr[F_1\ |\ a_i.y = t]$ for a fixed $t \in [0,1]$.
Let $S_1' = S_1 \cap \{a_i,\dots,a_{j'}\}$, i.e., $S_1'$ is the subset of $S_1$ consisting of all the points whose $x$-coordinates are in $[x_i,x_{j'}]$.
We notice that $F_1$ happens only if $a_{j'}$ is $y$-adjacent to $a_i$ in $S_1'$, i.e., there is no other point whose $y$-coordinate is in between $a_i.y$ and $a_{j'}.y$.
Indeed, if there exists $a \in S_1' \backslash \{a_i,a_{j'}\}$ such that $a.y$ is in between $a_i.y$ and $a_{j'}.y$, then $\text{dist}(a_i,a) < \text{dist}(a_i,a_{j'})$ and $a$ is in the minimal bottom-unbounded 3-sided rectangle containing $a_i,a_{j'}$, which implies $F_1$ does not happen.
We claim that, under the condition $a_i.y = t$, the probability that $a_{j'}$ is $y$-adjacent to $a_i$ in $S_1'$ is $O(1/\delta')$.
The $y$-coordinates of the random points in $S_1' \backslash \{a_i\}$ are independently drawn from the uniform distribution on $[0,1]$, so every point in $S_1' \backslash \{a_i\}$ has the same probability (say $p$) to be $y$-adjacent to $a_i$.
Let $r$ be the number of the points in $S_1' \backslash \{a_i\}$ that are $y$-adjacent to $a_i$, which is a random variable.
Then $\mathbb{E}[r] = p \cdot |S_1' \backslash \{a_i\}|$.
But we always have $r \leq 2$, since there can be at most two points $y$-adjacent to $a_i$.
In particular, $\mathbb{E}[r] \leq 2$ and $p = O(1/|S_1' \backslash \{a_i\}|)$.
By construction, we have $|S_1' \backslash \{a_i\}| = \Theta(\delta')$ (recall the assumption $\delta' \geq 5$).
It follows that $p = O(1/\delta')$, i.e., the probability that $a_{j'}$ is $y$-adjacent to $a_i$ in $S_1'$ is $O(1/\delta')$.
Using our previous argument, we have $\Pr[F_1\ |\ a_i.y = t] = O(1/\delta')$.
Therefore,
\begin{equation*}
    \Pr[F_0 \wedge F_1] = O(1/\delta') \cdot \int_0^1 \Pr[F_0\ |\ a_i.y = t]\ \text{d}t.
\end{equation*}
Note that $\int_0^1 \Pr[F_0\ |\ a_i.y = t]\ \text{d}t = \Pr[F_0]$.
By construction, there are $\Theta(\delta)$ points in $S_0$ whose $x$-coordinates are in between $a_i.x$ and $a_j.x$ (recall the assumption $\delta \geq 5$).
Thus, Lemma~\ref{lem-probij} implies $\Pr[F_0] = O(\log \delta/\delta^2)$.
Plugging in this to the equation above, we have $\Pr[F_0 \wedge F_1] = O(\log\delta/(\delta^2\delta'))$.
As a result, $\Pr[E_{i,j} \wedge E_{i,j'}] = O(\log\delta/(\delta^2\delta'))$.
The sum of the terms $\Pr[E_{i,j} \wedge E_{i',j'}]$ satisfying $i = i'$ and $j > j'$ is $O(\log^3 n)$, as one can easily verify.
For the same reason, the terms satisfying $i = i'$ and $j < j'$ also sum up to $O(\log^3 n)$.
The symmetric case that $i \neq i'$ and $j = j'$ is handled in the same fashion.
\smallskip

Combining all the cases, we conclude that $E[|\varPhi_l(S,\mathcal{U}^\downarrow)|^2] = E[|\varPsi|] = O(\log^4 n)$.

\subsection{Proof of Theorem~\ref{thm-rectavg}} \label{appx-prfrectavg}
The $\mathcal{R}$-RCP data structure $\mathcal{D}_2$ is described in Section~\ref{sec-rect2}.
\smallskip

\textbf{[Query time]}
We first analyze the (worst-case) query time.
When answering a query, we first find the splitting nodes $\mathbf{u}$ and $\mathbf{v}$ in the 2D range tree.
As argued in the proof of Theorem~\ref{thm-rectwst}, this can be done in $O(\log n)$ time.
Then we query the sub-structures stored at $\mathbf{v}$ to compute $\phi,\phi_\alpha,\phi_\beta$.
Note that all the sub-structures have $O(\log n)$ query time and we only need constant number of queries.
Therefore, this step takes $O(\log n)$ time, and hence the overall query time is also $O(\log n)$.
\smallskip

\textbf{[Average-case space cost]}
We now analyze the average-case space cost of $\mathcal{D}_2$.
Let $R$ be an axes-parallel rectangle and $S \propto R^n$.
We denote by $a_1,\dots,a_n$ the $n$ random points in $S$.
The data structure instance $\mathcal{D}_2(S)$ is essentially a 2D range tree built on $S$ with some sub-structures stored at secondary nodes.
Note that a 2D range tree built on a set of $n$ points in $\mathbb{R}^2$ has a fixed tree structure independent of the locations of the points.
This says, while $\mathcal{D}_2(S)$ is a random data structure instance depending on the random dataset $S$, the 2D range tree in $\mathcal{D}_2(S)$ has a deterministic structure.
As such, we can view $\mathcal{D}_2(S)$ as a fixed 2D range tree with random sub-structures.
Let $\mathcal{T}$ denote the primary tree of this 2D range tree and $\mathcal{T}_\mathbf{u}$ denote the secondary tree at the node $\mathbf{u} \in \mathcal{T}$, as in Section~\ref{sec-rectangle}.
To bound $\mathbb{E}[\mathsf{Space}(\mathcal{D}_2(S))]$, it suffices to bound the average-case space cost of the sub-structures stored at each secondary node.

For convenience of exposition, we introduce some notations.
Let $\mathbf{u} \in \mathcal{T}$ be a primary node.
Suppose the $n$ leaves of $\mathcal{T}$ are $\mathbf{lf}_1,\dots,\mathbf{lf}_n$ sorted from left to right.
Then the leaves in the subtree rooted at $\mathbf{u}$ must be $\mathbf{lf}_\alpha,\dots,\mathbf{lf}_\beta$ for some $\alpha,\beta \in \{1,\dots,n\}$ with $\alpha \leq \beta$
We then write $\mathsf{range}(\mathbf{u}) = [\alpha:\beta]$ and $\mathsf{size}(\mathbf{u}) = \beta-\alpha+1$.
Due to the construction of a 2D range tree, we always have $|S(\mathbf{u})| = \mathsf{size}(\mathbf{u})$ no matter what the random dataset $S$ is.
Furthermore, if $\mathsf{range}(\mathbf{u}) = [\alpha:\beta]$, then $S(\mathbf{u})$ contains exactly the points in $S$ with $x$-ranks $\alpha,\dots,\beta$ (we say a point has $x$-rank $i$ in $S$ if it is the $i$-th leftmost point in $S$).
Let $\mathbf{v} \in \mathcal{T}_\mathbf{u}$ be a secondary node.
We can define $\mathsf{range}(\mathbf{v})$ and $\mathsf{size}(\mathbf{v})$ in the same way as above (just by replacing $\mathcal{T}$ with $\mathcal{T}_\mathbf{u}$).
Also, we always have $|S(\mathbf{v})| = \mathsf{size}(\mathbf{v})$.
If $\mathsf{range}(\mathbf{v}) = [\alpha:\beta]$, then $S(\mathbf{v})$ contains exactly the points in $S(\mathbf{u})$ with $y$-ranks $\alpha,\dots,\beta$ (we say a point has $y$-rank $i$ in $S(\mathbf{u})$ if it is the $i$-th bottommost point in $S(\mathbf{u})$).
In what follows, we fix a secondary node $\mathbf{v} \in \mathcal{T}_\mathbf{u}$ and analyze the sub-structures stored at $\mathbf{v}$.
Let $\mathbf{u}'$ (resp., $\mathbf{v}'$) denote the left child of $\mathbf{u}$ (resp., $\mathbf{v}$).
Suppose $\mathsf{range}(\mathbf{u}) = [\alpha: \beta]$, $\mathsf{range}(\mathbf{u}') = [\alpha: \beta']$ (where $\beta'<\beta$), $\mathsf{range}(\mathbf{v}) = [\gamma: \xi]$, $\mathsf{range}(\mathbf{v}') = [\gamma: \xi']$ (where $\xi'<\xi$).

We want to use Theorem~\ref{thm-quadrant}, Lemma~\ref{lem-epdsrand}, Lemma~\ref{lem-3sidedcand^2} to bound the average-case space cost of the $\mathcal{Q}$-RCP, TBEP/LREP, $\mathcal{U}$-RSS sub-structures, respectively.
However, before applying these results, there is a crucial issue to be handled.
Recall that in Theorem~\ref{thm-quadrant}, Lemma~\ref{lem-epdsrand}, Lemma~\ref{lem-3sidedcand^2}, we assume the random dataset is generated either from the uniform distribution on a rectangle ($S \propto R^n$) or from the uniform distributions on a set of aligned segments ($S \propto \prod_{i=1}^n I_i$).
Unfortunately, here the underlying datasets of the sub-structures are $S_1(\mathbf{v}),\dots,S_4(\mathbf{v})$ and $S_\blacktriangle(\mathbf{v}),S_\blacktriangledown(\mathbf{v}),S_\blacktriangleleft(\mathbf{v}),S_\blacktriangleright(\mathbf{v})$; these random point-sets are neither (independently and uniformly) generated from a rectangle nor generated from aligned segments.
For instance, we cannot directly use Theorem~\ref{thm-quadrant} to deduce $\mathbb{E}[\mathsf{Space}(\mathcal{A}(S_1(\mathbf{v})))] = O(\log^2 |S_1(\mathbf{v})|)$, since $S_1(\mathbf{v})$ is not uniformly generated from a rectangle, and even its size $|S_1(\mathbf{v})|$ is not a fixed number ($|S_1(\mathbf{v})|$ varies with $S$).
The main focus of the rest of this proof is to handle this issue.

We first consider $S_1(\mathbf{v})$.
Note that $S_1(\mathbf{v}) = S(\mathbf{u}') \cap S(\mathbf{v}')$ by definition.
We want to bound $\mathbb{E}[\mathsf{Space}(\mathcal{A}(S_1(\mathbf{v})))]$.
Our basic idea is the following: reducing this expectation to conditional expectations in which $S_1(\mathbf{v})$ can be viewed as uniformly and independently generated from an axes-parallel rectangle so that Theorem~\ref{thm-quadrant} applies.
To this end, let $\varLambda = \{i: a_i \in S_1(\mathbf{v})\}$, which is a random subset of $[n] = \{1,\dots,n\}$, i.e., a random variable taking value from the power set of $[n]$.
A \textit{configuration} refers to a pair $(J,f)$ where $J \subseteq [n]$ and $f: [n] \backslash J \rightarrow R$ is a \textit{coordinate-wise injective} function, i.e., $f(i)$ and $f(i')$ have distinct $x$-coordinates and $y$-coordinates if $i \neq i'$.
For a configuration $(J,f)$, we define a corresponding event $E_{J,f}$ as
\begin{equation*}
    E_{J,f} = \left(\bigwedge_{i \in [n] \backslash J} (a_i = f(i))\right) \wedge (\varLambda = J).
\end{equation*}
We say $(J,f)$ is a \textit{legal} configuration if $E_{J,f}$ is a possible event.
We shall show that, if $(J,f)$ is a legal configuration, then under the condition $E_{J,f}$, the $|J|$ random points in $\{a_j: j \in J\}$ can be viewed as independently drawn from the uniform distribution on an axes-parallel rectangle.
Suppose $(J,f)$ is a legal configuration.
Let $F = \{f(i): i \in [n] \backslash J\}$, and $F' \subseteq F$ be the subset consisting of the points with $x$-ranks $\alpha,\dots,\beta-|J|$ in $F$.
Define $x_1$ as the $x$-coordinate of the $(\alpha-1)$-th leftmost point in $F$, $x_2$ as the $x$-coordinate of the $(n-\beta')$-th rightmost point in $F$, $y_1$ as the $y$-coordinate of the $(\gamma-1)$-th bottommost point in $F'$, $y_2$ as the $y$-coordinate of the $(\mathsf{size}(\mathbf{u})-\xi')$-th topmost point in $F'$.
Set $R' = [x_1,x_2] \times [y_1,y_2]$.
We claim that $E_{J,f}$ happens iff $a_i = f(i)$ for all $i \in [n] \backslash J$ and $a_j \in R'$ for all $j \in J$.
Since $(J,f)$ is a legal configuration, there exists at least one instance of $S$ making $E_{J,f}$ happen.
Let $S^*: \{a_i = a_i^*\}_{i \in [n]}$ be such an instance, where $a_i^* \in R$ indicates the location of $a_i$ in the instance $S^*$.
Then $a_i^* = f(i)$ for all $i \in [n] \backslash J$, hence $\{a_1^*,\dots,a_n^*\} = F \cup \{a_j^*: j \in J\}$.
Since the points in $\{a_j^*: j \in J\}$ belong to $S(\mathbf{u}')$ (for $S^*$ makes $\varLambda = J$), the $\alpha-1$ leftmost points in $F \cup \{a_j^*: j \in J\}$ (which correspond to the points in $S$ to the left of $S(\mathbf{u}')$) must be contained in $F$, and hence they are just the $\alpha-1$ leftmost points in $F$ (which we denote by $F_1$).
This implies $a_j^*.x \geq x_1$ for all $j \in J$.
Similarly, the $n-\beta'$ rightmost points in $F \cup \{a_j^*: j \in J\}$ (which correspond to the points in $S$ to the right of $S(\mathbf{u}')$) must be the $n-\beta$ rightmost points in $F$ (which we denote by $F_2$).
This implies $a_j^*.x \leq x_2$ for all $j \in J$.
Clearly, the points corresponding to $S(\mathbf{u})$ are exactly those in $F' \cup \{a_j^*: j \in J\}$.
Since the points in $\{a_j^*: j \in J\}$ belong to $S(\mathbf{v}')$ (for $S^*$ makes $\varLambda = J$), the $\gamma-1$ bottommost points in $F' \cup \{a_j^*: j \in J\}$ (which correspond to the points in $S(\mathbf{u})$ below $S(\mathbf{v}')$) must be contained in $F'$, and hence they are just the $\gamma-1$ bottommost points in $F'$ (which we denote by $F_1'$).
This implies $a_j^*.y \geq y_1$ for all $j \in J$.
Similarly, the $\mathsf{size}(\mathbf{u})-\xi'$ topmost points in $F' \cup \{a_j^*: j \in J\}$ (which correspond to the points in $S(\mathbf{u})$ above $S(\mathbf{v}')$) must be the $\mathsf{size}(\mathbf{u})-\xi'$ topmost points in $F'$ (which we denote by $F_2'$).
This implies $a_j^*.y \leq y_2$ for all $j \in J$.
Now we already see $a_j^* \in R'$ for all $j \in J$.
It follows that $E_{J,f}$ happens only if $a_i = f(i)$ for all $i \in [n] \backslash J$ and $a_j \in R'$ for all $j \in J$.
Furthermore, we note that $F_1 \cup F_2 \cup F_1' \cup F_2'$ corresponds to $S \backslash S_1(\mathbf{v})$.
Since $S^*$ makes $\varLambda = J$, we must have $F = F_1 \cup F_2 \cup F_1' \cup F_2'$ (this argument relies on the existence of such an instance $S^*$ making $E_{J,f}$ happen, i.e., it may fail if $(J,f)$ is not a legal configuration).
We then use this fact to show the ``if'' part.
Let $S^*: \{a_i = a_i^*\}_{i \in [n]}$ be an instance of $S$ satisfying $a_i^* = f(i)$ for all $i \in [n] \backslash J$ and $a_j^* \in R'$ for all $j \in J$.
Then $\{a_1^*,\dots,a_n^*\} = F \cup \{a_j^*: j \in J\}$.
We look at the subsets $F_1,F_2,F_1',F_2'$ of $F$.
Since $a_j^*.x \in [x_1,x_2]$ for all $j \in J$, $F_1$ (resp., $F_2$) contains exactly the $\alpha-1$ leftmost points (resp., $n-\beta'$ rightmost points) in $F \cup \{a_j^*: j \in J\}$, which correspond to the points to the left (resp., right) of $S(\mathbf{u}')$.
Similarly, since $a_j^*.y \in [y_1,y_2]$ for all $j \in J$, $F_1'$ (resp., $F_2'$) contains exactly the $\gamma-1$ bottommost points (resp., $\mathsf{size}(\mathbf{u})-\xi'$ topmost points) in $F' \cup \{a_j^*: j \in J\}$, which correspond to the points in $S(\mathbf{u})$ below (resp., above) $S(\mathbf{v})$.
Then $F = F_1 \cup F_2 \cup F_1' \cup F_2'$ corresponds to $S \backslash S_1(\mathbf{v})$.
The remaining points, which correspond to $S_1(\mathbf{v})$, are exactly those in $\{a_j^*: j \in J\}$.
Therefore, $\varLambda = J$ and $S^*$ makes $E_{J,f}$ happen.
Now we see that $E_{J,f}$ happens iff $a_i = f(i)$ for all $i \in [n] \backslash J$ and $a_j \in R'$ for all $j \in J$, i.e.,
\begin{equation*}
    E_{J,f} = \left(\bigwedge_{i \in [n] \backslash J} (a_i = f(i))\right) \wedge \left(\bigwedge_{j \in J} (a_j \in R')\right).
\end{equation*}
As such, under the condition $E_{J,f}$, the random points in $S_J = \{a_j: j \in J\}$ can be viewed as independently drawn from the uniform distribution on $R'$.
Applying Theorem~\ref{thm-quadrant}, we have
\begin{equation*}
    \mathbb{E}[\mathsf{Space}(\mathcal{A}(S_1(\mathbf{v})))\ |\ E_{J,f}] = \mathbb{E}[\mathsf{Space}(\mathcal{A}(S_J))\ |\ E_{J,f}] = O(\log^2 |J|).
\end{equation*}
Noting that $|J| \leq \mathsf{size}(\mathbf{v}') \leq \mathsf{size}(\mathbf{v})$ if $(J,f)$ is a legal configuration, we can deduce
\begin{equation} \label{eq-condEJf}
    \mathbb{E}[\mathsf{Space}(\mathcal{A}(S_1(\mathbf{v})))\ |\ E_{J,f}] = O(\log^2 \mathsf{size}(\mathbf{v})) \text{ for any } E_{J,f} \in \mathcal{E},
\end{equation}
where $\mathcal{E} = \{E_{J,f}: (J,f) \text{ is a legal configuration}\}$.
Using this result, we further show that $\mathbb{E}[\mathsf{Space}(\mathcal{A}(S_1(\mathbf{v})))] = O(\log^2 \mathsf{size}(\mathbf{v}))$.
Clearly, $\mathcal{E}$ is a collection of \textit{mutually disjoint} (or \textit{mutually exclusive}) events.
Furthermore, we notice that whenever $a_1,\dots,a_n$ have distinct $x$-coordinates and $y$-coordinates, some $E_{J,f} \in \mathcal{E}$ happens.
That says, $\mathcal{E}$ is a collection of \textit{almost collectively exhaustive} events in the sense that with probability 1 some $E_{J,f} \in \mathcal{E}$ happens.
Since the events in $\mathcal{E}$ are mutually disjoint and almost collectively exhaustive, $\mathbb{E}[\mathsf{Space}(\mathcal{A}(S_1(\mathbf{v})))] = O(\log^2 \mathsf{size}(\mathbf{v}))$ follows directly from the law of total expectation and Equation~\ref{eq-condEJf}.
Clearly, the same idea applies to bound $\mathbb{E}[\mathsf{Space}(\mathcal{A}(S_i(\mathbf{v})))]$ for all $i \in \{1,\dots,4\}$.

Next, we consider $S_\blacktriangle(\mathbf{v})$.
We want to bound $\mathbb{E}[\mathsf{Space}(\mathcal{K}_{l_\mathbf{u}}(S_\blacktriangle(\mathbf{v})))]$ and $\mathbb{E}[\mathsf{Space}(\mathcal{C}(\varPhi_\blacktriangle(\mathbf{v})))]$ where $\varPhi_\blacktriangle(\mathbf{v}) = \varPhi_{l_\mathbf{u}}(S_\blacktriangle(\mathbf{v}),\mathcal{U}^\downarrow)$ by definition.
The idea is totally the same as in the last paragraph: reducing to conditional expectations in which $S_\blacktriangle(\mathbf{v})$ can be viewed as independently generated from a set of (vertical) aligned segments so that Lemma~\ref{lem-epdsrand} and Lemma~\ref{lem-3sidedcand^2} apply.
We change the definition of $\varLambda$ in the last paragraph to $\varLambda = \{i: a_i \in S_\blacktriangle(\mathbf{v})\}$, and again define
\begin{equation*}
    E_{J,f} = \left(\bigwedge_{i \in [n] \backslash J} (a_i = f(i))\right) \wedge (\varLambda = J)
\end{equation*}
based on the new definition of $\varLambda$.
As we see in the last paragraph, it suffices to bound the conditional expectations $\mathbb{E}[\mathsf{Space}(\mathcal{K}_{l_\mathbf{u}}(S_\blacktriangle(\mathbf{v})))\ |\ E_{J,f}]$ and $\mathbb{E}[\mathsf{Space}(\mathcal{C}(\varPhi_\blacktriangle(\mathbf{v})))\ |\ E_{J,f}]$ for all legal configuration $(J,f)$.
Suppose $(J,f)$ is a legal configuration.
Let $F = \{f(i): i \in [n] \backslash J\}$, and $F' \subseteq F$ be the subset consisting of the points with $x$-ranks $\alpha,\dots,\beta-|J|$ in $F$.
Define $x_1$ as the $x$-coordinate of the $(\alpha-1)$-th leftmost point in $F$, $x_2$ as the $x$-coordinate of the $(n-\beta)$-th rightmost point in $F$, $y_1$ as the $y$-coordinate of the $(\gamma-1)$-th bottommost point in $F'$, $y_2$ as the $y$-coordinate of the $(\mathsf{size}(\mathbf{u})-\xi')$-th topmost point in $F'$.
Set $R' = [x_1,x_2] \times [y_1,y_2]$.
Using the same argument as in the last paragraph, one can easily verify that $E_{J,f}$ happens iff $a_i = f(i)$ for all $i \in [n] \backslash J$ and $a_j \in R'$ for all $j \in J$.
For an injective function $g: J \rightarrow (x_1,x_2)$, we further define
\begin{equation*}
    E_{J,f,g} = E_{J,f} \wedge \left(\bigwedge_{j \in J} (a_j.x = g(j))\right).
\end{equation*}
Now $E_{J,f,g}$ happens iff $a_i = f(i)$ for all $i \in [n] \backslash J$ and $a_j \in \{g(j)\} \times [y_1,y_2]$ for all $j \in J$.
Thus, under $E_{J,f,g}$, the $|J|$ random points in $S_J = \{a_j: j \in J\}$ can be viewed as independently drawn from the $|J|$ vertical aligned segments in $\{\{g(j)\} \times [y_1,y_2]: j \in J\}$.
To apply Lemma~\ref{lem-epdsrand} and Lemma~\ref{lem-3sidedcand^2}, we still need to consider one thing: the line $l_\mathbf{u}$.
The line $l_\mathbf{u}$ is a random vertical line depending on $S$.
However, we notice that under $E_{J,f,g}$, $l_\mathbf{u}$ is fixed.
Indeed, under $E_{J,f,g}$, $S(\mathbf{u})$ corresponds to $F' \cup \{a_j: j \in J\}$.
Thus, the $x$-coordinates of the points in $S(\mathbf{u})$ are fixed under $E_{J,f,g}$, and hence $l_\mathbf{u}$ is fixed.
As such, we are able to apply Lemma~\ref{lem-epdsrand} to deduce
\begin{equation*}
    \mathbb{E}[\mathsf{Space}(\mathcal{K}_{l_\mathbf{u}}(S_\blacktriangle(\mathbf{v})))\ |\ E_{J,f,g}] = \mathbb{E}[\mathsf{Space}(\mathcal{K}_{l_\mathbf{u}}(S_J))\ |\ E_{J,f,g}] = O(\log |J|) = O(\log \mathsf{size}(\mathbf{v})),
\end{equation*}
and apply Lemma~\ref{lem-3sidedcand^2} to deduce
\begin{equation*}
    \mathbb{E}[\mathsf{Space}(\mathcal{C}(\varPhi_\blacktriangle(\mathbf{v})))\ |\ E_{J,f,g}] = \mathbb{E}[|\varPhi_{l_\mathbf{u}}(S_J,\mathcal{U}^\downarrow)|^2] = O(\log^4 |J|) = O(\log^4 \mathsf{size}(\mathbf{v})).
\end{equation*}
Note that, if $E_{J,f}$ happens, then with probability 1 some $E_{J,f,g}$ happens.
Therefore, the collection $\mathcal{E} = \{E_{J,f,g}\}$, which consists of all $E_{J,f,g}$ where $(J,f)$ is a legal configuration and $g: J \rightarrow (x_1,x_2)$ is an injective function with range $(x_1,x_2)$ depending on $(J,f)$, is a collection of mutually disjoint and almost collectively exhaustive events.
By the law of total expectation, we immediately have $\mathbb{E}[\mathsf{Space}(\mathcal{K}_{l_\mathbf{u}}(S_\blacktriangle(\mathbf{v}))))] = O(\log \mathsf{size}(\mathbf{v}))$ and $\mathbb{E}[\mathsf{Space}(\mathcal{C}(\varPhi_\blacktriangle(\mathbf{v})))] = O(\log^2 \mathsf{size}(\mathbf{v}))$.
The expected space cost of the sub-structures built on $S_\blacktriangledown(\mathbf{v})$ can be bounded using the same argument.
Also, one can handle $S_\blacktriangleleft(\mathbf{v})$ and $S_\blacktriangleright(\mathbf{v})$ in a similar way.
The only difference is that, in the event $E_{J,f,g}$, the $g$ function should indicate the $y$-coordinates of the points in $\{a_j: j \in J\}$ instead of the $x$-coordinates.

Once we know that the expected space cost of all the substructures stored at $\mathbf{v}$ is poly-logarithmic in $\mathsf{size}(\mathbf{v})$, we can deduce that the expected space cost of each secondary tree $\mathcal{T}_\mathbf{u}$ (with the sub-structures) is $O(\mathsf{size}(\mathbf{u}))$.
As a result, $\mathbb{E}[\mathsf{Space}(\mathcal{D}_2(S))] = O(n \log n)$.

\subsection{Proof of Lemma~\ref{lem-halfcand}}
Without loss of generality, assume $R = [0,1] \times [0,\Delta]$.
It suffices to show $\mathbb{E}[|\varPhi(S,\mathcal{H}^\downarrow)|] = O(\log^2 n)$.
This can be further reduced to showing $\mathbb{E}[|\varPhi(S,\mathcal{H}')|] = O(\log^2 n)$ where
\begin{equation*}
    \mathcal{H}' = \{l^\downarrow: l \text{ is a non-vertical line whose slope is non-positive}\} \subseteq \mathcal{H}^\downarrow.
\end{equation*}
Suppose the $n$ random points in $S$ are $a_1,\dots,a_n$.
Let $E_{i,j}$ be the event that $(a_i,a_j) \in \varPhi(S,\mathcal{H}')$, and observe
\begin{equation*}
    \mathbb{E}[|\varPhi(S,\mathcal{H}')|] = \sum_{i=1}^{n-1} \sum_{j=i+1}^n \Pr[E_{i,j}].
\end{equation*}
Note that all $\Pr[E_{i,j}]$ in the above equation are the same, which implies $\mathbb{E}[|\varPhi(S,\mathcal{H}')|] = O(n^2 \cdot \Pr[E_{1,2}])$.
Thus, it suffices to bound $\Pr[E_{1,2}]$.
As in the proof of Lemma~\ref{lem-quadcand}, we define random variables $x_\text{max} = \max\{a_1.x,a_2.x\}$, $y_\text{max} = \max\{a_1.y,a_2.y\}$, $x_\text{min} = \min\{a_1.x,a_2.x\}$, $y_\text{min} = \min\{a_1.y,a_2.y\}$, $Q = (-\infty,x_\text{max}] \times (-\infty,y_\text{max}]$, and $\varLambda = \{i \geq 3: a_i \in Q\}$.
We also define $Q' = (-\infty,x_\text{max}/2] \times (-\infty,y_\text{max}/2]$ and $\varLambda' = \{i \geq 3: a_i \in Q'\}$.
We achieve the bound for $\Pr[E_{1,2}]$ through four steps.
\smallskip

\textbf{[Step~1]}
We begin with establishing the following key observation: for any $H \in \mathcal{H}'$, $a_1,a_2 \in H$ implies $Q' \subseteq H$.
To see this, let $H \in \mathcal{H}'$ and assume $a_1,a_2 \in H$.
If $\{a_1,a_2\} = \{(x_\text{min},y_\text{min}),(x_\text{max},y_\text{max})\}$, then $H$ contains the point $(x_\text{max},y_\text{max})$.
This implies that $H$ contains the point $(x_\text{max}/2,y_\text{max}/2)$ and hence contains $Q'$, because $H = l^\downarrow$ for a line $l$ of non-positive slope.
If $\{a_1,a_2\} = \{(x_\text{min},y_\text{max}),(x_\text{max},y_\text{min})\}$, then $H$ contains the 5-polygon $P$ whose vertices are $(0,0)$, $(x_\text{max},0)$, $(x_\text{max},y_\text{min})$, $(x_\text{min},y_\text{max})$, $(0,y_\text{max})$.
Note that $P$ contains the point $(x_\text{max}/2,y_\text{max}/2)$, which implies that $H$ also contains the point $(x_\text{max}/2,y_\text{max}/2)$ and hence contains $Q'$.
\smallskip

\textbf{[Step~2]}
Based on the observation in Step 1, we prove a result which is similar to Equation~\ref{eq-condprob} in the proof of Lemma~\ref{lem-quadcand}.
We claim that for all $\tilde{x} \in (0,1]$, all $\tilde{y} \in (0,\Delta]$, and all nonempty $J' \subseteq \{3,\dots,n\}$,
\begin{equation} \label{eq-E12halfp}
    \Pr[E_{1,2}\ |\ (x_\text{max} = \tilde{x}) \wedge (y_\text{max} = \tilde{y}) \wedge (\varLambda' = J')] = O(1/|J'|^2).
\end{equation}
The argument for proving this is similar to that for proving Equation~\ref{eq-condprob}.
We use $C_{\tilde{x},\tilde{y},J'}'$ to denote the condition in the above conditional probability.
Assume $|J'| = k$.
Let $\delta_x = x_\text{max} - x_\text{min}$ and $\delta_y = y_\text{max} - y_\text{min}$.
Since any halfplane $H \in \mathcal{H}'$ containing $a_1,a_2$ must contain $Q'$, $E_{1,2}$ happens only if $\delta_x \leq \kappa(S_{J'})$ and $\delta_y \leq \kappa(S_{J'})$, where $S_{J'} = \{a_j: j \in J'\}$.
So it suffices to bound $\Pr[(\delta_x \leq \kappa(S_{J'})) \wedge (\delta_y \leq \kappa(S_{J'}))\ |\ C_{\tilde{x},\tilde{y},J'}']$.
Under the condition $C_{\tilde{x},\tilde{y},J'}'$, $Q'$ is just $(-\infty,\tilde{x}/2] \times (-\infty,\tilde{y}/2]$.
Thus the condition $C_{\tilde{x},\tilde{y},J'}'$ is equivalent to saying that the maximum of the $x$-coordinates (resp., $y$-coordinates) of $a_1,a_2$ is $\tilde{x}$ (resp., $\tilde{y}$), all $a_j$ for $j \in J'$ are contained in the rectangle $R' = [0,\tilde{x}/2] \times [0,\tilde{y}/2]$, and all $a_j$ for $j \in \{3,\dots,n\} \backslash J'$ are contained in $R \backslash R'$.
As such, one can easily verify that, under the condition $C_{\tilde{x},\tilde{y},J'}'$, the distribution of the random number $\delta_x$ (resp., $\delta_y$) is the uniform distribution on the interval $[0,\tilde{x}]$ (resp., $[0,\tilde{y}]$) and the distributions of the $k$ random points in $S_{J'}$ are the uniform distribution on $R'$; furthermore, these random numbers/points are independent of each other.
This says, if we consider a new random experiment in which we independently generate two random numbers $\delta_x',\delta_y'$ from the uniform distributions on $[0,\tilde{x}],[0,\tilde{y}]$ respectively (which correspond to $\delta_x,\delta_y$) and a random dataset $S' \propto (R')^k$ (which corresponds to $S_{J'}$), then we have
\begin{equation*}
    \Pr[(\delta_x' \leq \kappa(S')) \wedge (\delta_y' \leq \kappa(S'))] = \Pr[(\delta_x \leq \kappa(S_{J'})) \wedge (\delta_y \leq \kappa(S_{J'}))\ |\ C_{\tilde{x},\tilde{y},J'}'].
\end{equation*}
So it suffices to bound $\Pr[(\delta_x' \leq \kappa(S')) \wedge (\delta_y' \leq \kappa(S'))]$ in the new experiment; we denote by $\lambda$ this probability.
We apply the formula
\begin{equation*}
    \lambda = \int_0^\infty p(t) \cdot \Pr[(\delta_x' \leq t) \wedge (\delta_y' \leq t)]\ \text{d}t = \int_0^\infty p(t) \cdot \Pr[\delta_x' \leq t] \cdot \Pr[\delta_y' \leq t]\ \text{d}t,
\end{equation*}
where $p(\cdot)$ is the probability distribution function of $\kappa(S')$.
Since $\delta_x'$ (resp., $\delta_y'$) is uniformly drawn from the interval $[0,\tilde{x}]$ (resp., $[0,\tilde{y}]$), we have $\Pr[\delta_x' \leq t] = \min\{t/\tilde{x},1\}$ (resp., $\Pr[\delta_y' \leq t] = \min\{t/\tilde{y},1\}$).
Without loss of generality, we assume $\tilde{x} \leq \tilde{y}$.
Then we have
\begin{equation*}
    \Pr[\delta_x' \leq t] \cdot \Pr[\delta_y' \leq t] = \min\{t^2/(\tilde{x}\tilde{y}),t/\tilde{y},1\} \leq \min\{t^2/(\tilde{x}\tilde{y}),t/\tilde{y}\}.
\end{equation*}
It follows that
\begin{equation*}
    \lambda \leq \int_0^\infty p(t) \cdot \min\{t^2/(\tilde{x}\tilde{y}),t/\tilde{y}\}\ \text{d}t = \min\left\{\int_0^\infty \frac{p(t)t^2}{\tilde{x}\tilde{y}} \text{d}t, \int_0^\infty \frac{p(t)t}{\tilde{y}} \text{d}t \right\}.
\end{equation*}
Noting the fact that $\int_0^\infty p(t)t^2 \text{d}t = \mathbb{E}[\kappa^2(S')]$ and  $\int_0^\infty p(t)t \text{d}t = \mathbb{E}[\kappa(S')]$, we have
\begin{equation*}
    \lambda \leq \min\left\{\frac{\mathbb{E}[\kappa^2(S')]}{\tilde{x}\tilde{y}}, \frac{\mathbb{E}[\kappa(S')]}{\tilde{y}}\right\}.
\end{equation*}
Since $\tilde{x} \leq \tilde{y}$ by assumption, Lemma~\ref{lem-kappa} implies that $\mathbb{E}[\kappa(S')] = O(\max\{\sqrt{\tilde{x}\tilde{y}}/k, \tilde{y}/k^2\})$ and $\mathbb{E}[\kappa^2(S')] = O(\max\{\tilde{x}\tilde{y}/k^2, \tilde{y}^2/k^4\})$.
If $\sqrt{\tilde{x}\tilde{y}}/k \leq \tilde{y}/k^2$, then $\mathbb{E}[\kappa(S')]/\tilde{y} = O(1/k^2)$, otherwise $\mathbb{E}[\kappa^2(S')]/(\tilde{x}\tilde{y}) = O(1/k^2)$.
In either of the two cases, we have $\lambda = O(1/k^2)$.
Therefore, we obtain Equation~\ref{eq-E12halfp}.
For an arbitrary nonempty $J' \subseteq \{3,\dots,n\}$, since Equation~\ref{eq-E12halfp} holds for all $\tilde{x} \in (0,1]$ and $\tilde{y} \in (0,\Delta]$, we can remove the conditions $x_\text{max} = \tilde{x}$ and $y_\text{max} = \tilde{y}$ from Equation~\ref{eq-E12halfp} to deduce $\Pr[E_{1,2}\ |\ \varLambda' = J'] = O(1/|J'|^2)$ (note that although we miss the case $\tilde{x} = 0$ or $\tilde{y} = 0$ for Equation~\ref{eq-E12halfp}, it does not matter since the events $x_\text{max} = 0$ and $y_\text{max} = 0$ happen with probability 0).
This further implies that $\Pr[E_{1,2}\ |\ |\varLambda'| = k] = O(1/k^2)$ for all $k \in \{1,\dots,n-2\}$.
For $k=0$, we have $\Pr[E_{1,2}\ |\ |\varLambda'| = k] = 1$.
\smallskip

\textbf{[Step~3]}
Let $m$ be a sufficiently large integer, and $m' = \lfloor m/8 \rfloor$.
Our goal in this step is to bound $\Pr[|\varLambda'| \leq m' \ |\ |\varLambda| = m]$.
Again, we reduce to conditional probability.
We claim that for all $\tilde{x} \in (0,1]$, all $\tilde{y} \in (0,\Delta]$, and all $J \subseteq \{3,\dots,n\}$ with $|J|=m$,
\begin{equation} \label{eq-|Lambda'|}
    \Pr[|\varLambda'| \leq m' \ |\ (x_\text{max} = \tilde{x}) \wedge (y_\text{max} = \tilde{y}) \wedge (\varLambda = J)] \leq e^{-m/32}.
\end{equation}
We use $C_{\tilde{x},\tilde{y},J}$ to denote the condition in the above conditional probability.
Under the condition $C_{\tilde{x},\tilde{y},J}$, $Q = (-\infty,\tilde{x}] \times (-\infty,\tilde{y}]$ and $Q' = (-\infty,\tilde{x}/2] \times (-\infty,\tilde{y}/2]$.
Since $\varLambda' \subseteq \varLambda$ by definition, we have, under the condition $C_{\tilde{x},\tilde{y},J}$,
\begin{equation*}
    |\varLambda'| = \sum_{j \in J} \mathbf{1}_{a_j \in Q'}, \text{ where } \mathbf{1}_{a_j \in Q'} = \left\{ \begin{array}{ll}
         1 & a_j \in Q' \\
         0 & a_j \notin Q'
    \end{array} \right.
    \text{is the indicator function.}
\end{equation*}
As we have seen when proving Equation~\ref{eq-condprob} in the proof of Lemma~\ref{lem-quadcand}, under the condition $C_{\tilde{x},\tilde{y},J}$, the $m$ random points in $S_J$ can be viewed as independently drawn from the uniform distribution on the rectangle $[0,\tilde{x}] \times [0,\tilde{y}]$.
Note that a random point drawn from the uniform distribution on $[0,\tilde{x}] \times [0,\tilde{y}]$ has probability $1/4$ to be contained in $Q'$.
Therefore, under the condition $C_{\tilde{x},\tilde{y},J}$, $\{\mathbf{1}_{a_j \in Q'}:j \in J\}$ is a set of i.i.d. random variables each of which equals to 1 with probability $1/4$ and equals to 0 with probability $3/4$.
It follows that $\mathbb{E}[|\varLambda'|\ |\ C_{\tilde{x},\tilde{y},J}] = m/4$.
By Hoeffding's inequality, we have
\begin{equation*}
    \Pr[m/4 - |\varLambda| \geq m/8 \ |\ C_{\tilde{x},\tilde{y},J}] \leq e^{-2(m/8)^2/m} = e^{-m/32},
\end{equation*}
which implies Equation~\ref{eq-|Lambda'|}.
Since Equation~\ref{eq-|Lambda'|} holds for all $\tilde{x} \in (0,1]$, all $\tilde{y} \in (0,\Delta]$, and all $J \subseteq \{3,\dots,n\}$ with $|J|=m$, we can deduce that $\Pr[|\varLambda'| \leq m' \ |\ |\varLambda| = m] \leq e^{-m/32}$.
\smallskip


\textbf{[Step~4]}
Finally, we try to bound $\Pr[E_{1,2}]$ using the results obtained in the previous steps.
We apply the formula
\begin{equation*}
    \Pr[E_{1,2}] = \sum_{k=0}^{n-2} \Pr[|\varLambda'| = k] \cdot \Pr[E_{1,2}\ |\ |\varLambda'| = k].
\end{equation*}
Since $\Pr[|\varLambda'| = k] = \sum_{m=k}^{n-2} (\Pr[|\varLambda| = m] \cdot \Pr[|\varLambda'| = k \ |\  |\varLambda| = m])$, we further deduce
\begin{equation} \label{eq-sumcik}
    \Pr[E_{1,2}] = \sum_{m=0}^{n-2} \left(\Pr[|\varLambda| = m] \cdot \sum_{k=0}^m g_{m,k}\right).
\end{equation}
where $g_{m,k} = \Pr[E_{1,2}\ |\ |\varLambda'| = k] \cdot \Pr[|\varLambda'| = k \ |\  |\varLambda| = m]$.
We claim that $\sum_{k=0}^m g_{m,k} = O(1/m^2)$ for all $m \in \{1,\dots,n-2\}$.
To prove this, we may assume $i$ is sufficiently large.
Set $m' = \lfloor m/8 \rfloor$.
Using the result of Step 3, we can deduce that
\begin{equation*}
    \sum_{k=0}^{m'} g_{m,k} \leq \sum_{k=0}^{m'} \Pr[|\varLambda'| = k \ |\  |\varLambda| = m] = \Pr[|\varLambda'| \leq m' \ |\  |\varLambda| = m] \leq e^{-m/32}.
\end{equation*}
On the other hand, by the choice of $m'$ and the result of Step 2, $\Pr[E_{1,2}\ |\ |\varLambda'| = k] = O(1/m^2)$ for all $k \in \{m'+1,\dots,m\}$.
As such, we have
\begin{equation*}
    \sum_{k=m'+1}^m g_{m,k} = \sum_{k=m'+1}^m O(1/m^2) \cdot \Pr[|\varLambda'| = k \ |\  |\varLambda| = m] = O(1/m^2).
\end{equation*}
It follows that
\begin{equation*}
    \sum_{k=0}^m g_{m,k} = \sum_{k=0}^{m'} g_{m,k} + \sum_{k=m'+1}^m g_{m,k} \leq e^{-m/32} + O(1/m^2) = O(1/m^2).
\end{equation*}
For $m=0$, we have the trivial bound $\sum_{k=0}^m g_{m,k} = g_{m,0} \leq 1$.
Thanks to Equation~\ref{eq-sumcik} and the bounds for $\sum_{k=0}^m g_{m,k}$, the only thing remaining for bounding $\Pr[E_{1,2}]$ is to bound $\Pr[|\varLambda| = m]$.
Recall that, in the proof of Lemma~\ref{lem-quadcand}, we have shown $\Pr[|\varLambda| = m] = O((m+1)\log n/n^2)$ for all $m \in \{0,\dots,n-2\}$.
Plugging in this and the bounds for $\sum_{k=0}^m g_{m,k}$ to Equation~\ref{eq-sumcik}, a direct calculation gives us $\Pr[E_{1,2}] = O(\log^2n/n^2)$.
As such, $\mathbb{E}[|\varPhi(S,\mathcal{H}')|] = O(\log^2 n)$ and thus $\mathbb{E}[|\varPhi(S,\mathcal{H})|] = O(\log^2 n)$.

\section{Implementation details of the preprocessing algorithm} \label{appx-implement}
We now discuss the implementation details in the preprocessing algorithm.
Let $\mathcal{T}$ be the BST currently storing $\partial F_{i-1}$.

The first thing we need to show is, given two points $u,v \in \partial F_{i-1}$, how to report in the left-right order the fractions of $\partial F_{i-1}$ intersecting $\sigma$, where $\sigma$ is the portion of $\partial F_{i-1}$ between $u,v$.
Clearly, we are reporting a set of consecutive fractions of $\partial F_{i-1}$.
We can find in $\mathcal{T}$ the node $\mathbf{u}$ corresponding to the leftmost fraction to be reported in $O(\log |\mathcal{T}|)$ time.
We then report this fraction.
After this, we simply apply a (in-order) traversal from $\mathbf{u}$ to report the other fractions in the left-right order.
Since the fractions to be reported are consecutive, it is easy to see that the time cost is $O(\log |\mathcal{T}|+k)$, where $k$ is the number of the reported fractions.

The second thing we need to show is how to update $\mathcal{T}$.
At the beginning of the $i$-th iteration, $\mathcal{T}$ stores $\partial F_{i-1}$, and we need to update it to $\partial F_i$.
Clearly, $\partial F_i$ is obtained by using the connected components of $\partial W_i \cap F_{i-1}$ to replace the corresponding portions of $\partial F_{i-1}$.
We consider the components of $\partial W_i \cap F_{i-1}$ one-by-one (there are constant number of components to be considered by the proof of Lemma~\ref{lem-hlfplinear}).
Let $\xi$ be a component, which must be an $x$-monotone polygonal chain consisting of at most two pieces.
For convenience, assume $\xi$ has a left endpoint $u$ and a right end point $v$.
It is clear that $u,v \in \partial F_{i-1}$.
We need to replace the portion $\sigma$ of $\partial F_{i-1}$ between $u,v$ with $\xi$; we call this a \textsf{Replace} operation.
To achieve this, we first report the fractions of $\partial F_{i-1}$ intersecting $\sigma$, by using the approach described above.
Suppose the reported fractions are $\gamma_1,\dots,\gamma_k$ sorted in the left-right order.
Then $u \in \gamma_1$ and $v \in \gamma_k$.
Clearly, the fractions $\gamma_2,\dots,\gamma_{k-1}$ should be removed, as they disappear after replacing $\sigma$ with $\xi$.
This can be done by deleting the corresponding nodes from $\mathcal{T}$ via $k-2$ BST-deletion operations.
Also, we need to modify $\gamma_1$ and $\gamma_k$: the portion of $\gamma_1$ (resp., $\gamma_k$) to the right (resp., left) of $u$ (resp., $v$) should be ``truncated''.
This can be done by directly updating the information stored in the two corresponding nodes.
Finally, $\xi$ should be inserted.
Each piece of $\xi$ becomes a new fraction, for which we create a new node storing the information of the fraction and insert it into $\mathcal{T}$ via a BST-insertion operation.
Now we analyze the time cost of this \textsf{Replace} operation.
Let $|\mathcal{T}|$ be the size of $\mathcal{T}$ before the operation.
The time cost for reporting is $O(\log |\mathcal{T}|+k)$.
Removing $\gamma_2,\dots,\gamma_{k-1}$ takes $O(k \log |\mathcal{T}|)$ time.
Modifying $\gamma_1,\gamma_k$ and Inserting $\xi$ takes $O(\log |\mathcal{T}|)$ time (note that $\xi$ has at most two pieces).
So the total time of this \textsf{Replace} operation is $O(k \log |\mathcal{T}|)$.
If $k \leq 2$, then the time cost is just $O(\log |\mathcal{T}|)$.
If $k>2$, we observe that there are $\Omega(k)$ nodes deleted from $\mathcal{T}$ in this \textsf{Replace} operation.
Note that the total number of the nodes deleted from $\mathcal{T}$ cannot exceed the total number of the nodes inserted.
Over the $m$ iterations, we have in total $O(m)$ \textsf{Replace} operations, each of which inserts $O(1)$ nodes into $\mathcal{T}$.
Therefore, one can delete at most $O(m)$ nodes from $\mathcal{T}$ in total.
It follows that the total time cost for all \textsf{Replace} operations is $O(m \log m)$, which is also the total time cost for updating $\mathcal{T}$.
In other words, $\mathcal{T}$ can be updated in amortized $O(\log m)$ time for each iteration.




\end{document}